\documentclass[a4paper,11pt]{article}
\usepackage[a4paper, left=2.5cm, right=2.5cm, top=2.5cm, bottom=2.5cm]{geometry}
\usepackage{epsfig}
\usepackage{amssymb}
\usepackage{amsmath}
\usepackage[ruled,vlined,linesnumbered]{algorithm2e}
\usepackage{alltt}
\usepackage{enumitem}
\usepackage{tabularx}
\usepackage{url}
\usepackage{pst-tree}
\usepackage{lineno}
\usepackage{longtable}
\usepackage{pdfpages}
\usepackage{graphicx}
\usepackage{balance}  % for  \balance command ON LAST PAGE  (only there!)
\usepackage[utf8]{inputenc}

\usepackage{pgfplots}
\usepackage{tikz}
\usepackage{tikz-3dplot}
\usepackage{soul} % SR
\usepackage{verbatim}
\usepackage{subfigure}
\usepackage{authblk}
\usepackage{multirow}
\usepackage{booktabs}
\usepackage[normalem]{ulem}
\usepackage{subcaption}

\pgfplotsset{compat=1.15}

\newcommand{\ic}[1]{\textit{#1}}
\newcommand{\secondo}{\textsc{Secondo}}

\newcommand{\Secondo}{\textsc{Secondo}}

\def\MM#1{\ifmmode#1 \else \mbox{$#1$} \fi}

\def\endbox{\hfill$\Box$}

\renewcommand{\emph}[1]{\textit{#1}}

\newcommand{\tc}[1]{\ensuremath{\underline{\smash{\mathit{#1}}}}}
\newcommand{\op}[1]{\textbf{#1}}

\newenvironment{query}{
\begin{alltt}\small}{\end{alltt}}

\newenvironment{numquery*}
{\begin{linenumbers*}\begin{query}}
{\end{query}\end{linenumbers*}}

\newtheorem{Def}{Definition}[section]
\newenvironment{definition}{\begin{Def}\rm}{\endbox\end{Def}}

\newtheorem{Exam}{Example}[section]
\newenvironment{example}{\begin{Exam}\rm}{\endbox\end{Exam}}

\newtheorem{Theorem}{Theorem}[section]

\newcommand{\q}{$\quad$}

\newcommand{\sothat}{\ |\ }

\newcommand{\IF}{{\bf if }}
\newcommand{\THEN}{{\bf then }}
\newcommand{\ELSE}{{\bf else }}

\newcommand{\ENDIF}{{\bf endif}}

\newcommand{\RETURN}{{\bf return }}

\newcommand{\SIDEEFFECT}{{\bf Side Effect}}

\begin{document}

\title{Exact Trajectory Similarity Search With N-tree: An Efficient Metric Index for kNN and Range Queries}
\date{}

\author[1]{Ralf Hartmut G\"{u}ting\thanks{Corresponding Author: rhg@fernuni-hagen.de}}
\author[2]{Suvam Kumar Das\thanks{suvam.das@unb.ca}}
\author[1]{Fabio Valdés\thanks{fabio.valdes@fernuni-hagen.de}} 
\author[2]{Suprio Ray\thanks{sray@unb.ca}}

\affil[1]{Fernuniversit\"{a}t in Hagen, Germany}
\affil[2]{University of New Brunswick, Fredericton, Canada}

\maketitle

%\tableofcontents

% Exact Trajectory Similarity Search With N-tree: A Scalable Metric Space Index for kNN and Range Queries

\begin{abstract}
Similarity search is the problem of finding in a collection of objects those that are similar to a given query object. It is a fundamental problem in modern applications and the objects considered may be as diverse as locations in space, text documents, images, twitter messages, or trajectories of moving objects. 

In this paper we are motivated by the latter application. Trajectories are recorded movements of mobile objects such as vehicles, animals, public transportation, or parts of the human body. We propose a novel distance function called \emph{DistanceAvg} to capture the similarity of such movements. To be practical, it is necessary to provide indexing for this distance measure.

Fortunately we do not need to start from scratch. A generic and unifying approach is metric space, which organizes the set of objects solely by a distance (similarity) function with certain natural properties. Our function \emph{DistanceAvg} is a metric. 

Although metric indexes have been studied for decades and many such structures are available, they do not offer the best performance with trajectories. In this paper we propose a new design, which outperforms the best existing indexes for \emph{kNN} queries and is equally good for range queries. It is especially suitable for expensive distance functions as they occur in trajectory similarity search. In many applications, \emph{kNN} queries are more practical than range queries as it may be difficult to determine an appropriate search radius. Our index provides exact result sets for the given distance function.
\end{abstract}

% Exact Trajectory Similarity Search With N-tree: A Scalable Metric Index for kNN and Range Queries

\section{Introduction}

Finding relevant objects in a large collection of objects is a fundamental database problem. Modern applications have to deal with huge collections of data over a wide variety of complex data types: audio or video data, text documents, photographs, twitter messages, social network profiles, recommendations, points of interest, spatio-textual data, moving object trajectories, to name only a few.  The question is how to specify what are relevant objects; what are we searching for.

One option is to somehow map the given objects into $k$-dimensional vectors, e.g. by extracting $k$ numeric features. After this transformation, query types and indexing techniques for spatial data are available, especially those tuned for high-dimensional spaces.

However, the most simple and natural approach to querying is to select one object from the given domain and ask for objects similar to it, that is, \emph{similarity search}. Similarity can be defined by a distance function. Distance is inverse to similarity; hence, an object is most similar to itself, with distance 0. 

In this paper, we are motivated by a specific application, similarity search for trajectories of moving objects. Such data are collected by GPS receivers at a massive scale in recent years, for example, the locations of mobile phone users in location-based services, vehicles in transportation planning, naval vessels, migrating animals, to name only a few. There exist two views or models of trajectories. The first is a sequence of (location, time stamp) pairs, ordered by time. The second views a trajectory as an approximation of a continuous function from time into space, that is, a continuous time dependent position. The approximation is represented as a sequence of line segments in the (2d, time) space. We refer to the two models as discrete and continuous trajectory data types, respectively \cite{SuLZZZ20}.

Distance (similarity) measures on trajectories are fundamental for many applications in trajectory data mining and analysis. The survey \cite{SuLZZZ20} compares 15 such distance functions. They may be classified into 4 categories, depending on (1) whether they consider only the spatial or also the temporal information, and (2) whether they are based on the discrete or continuous trajectory data model.

The distance function we propose in this paper is called \emph{DistanceAvg}.
It takes
two continuous trajectories as arguments. The basic idea is to consider these two movements as if they had started at the same time and then to observe the continuous distance function. The intended measure of similarity is the average of the continuous distance that can be computed as a piece-wise integral.

However, this works directly only when both trajectories have the same duration (temporal extent). To make this similarity measure applicable to arbitrary pairs of trajectories, we map them to the same duration by \emph{stretching} or \emph{shrinking}, i.e., applying a factor to the speeds to make them longer or shorter, respectively.
When distances in a larger set of trajectories are considered, they are all mapped to the same duration. This is necessary to let the distance function be a metric.
 
By nature, this is a spatio-temporal distance measure as it takes the speed of movements into account. However, it can also be used as a spatial distance measure by disregarding the original time stamps and replacing them by time stamps so that objects move at constant speed. In this way, effectively only the spatial information is taken into account.

In addition, we introduce related distance measures on approximations of trajectories.
A cylindrical approximation of trajectory $U$ is a sequence of slanted cylinders enclosing $U$. Each cylinder encloses a subsequence of segments of the original trajectory. The radii of cylinders can be chosen to obtain finer or coarser approximations. We establish bounds on the difference between the average distance on exact trajectories and the average distance of their approximations.

This allows one to implement a \emph{filter-and-refine} strategy: find trajectories within a certain range (distance) based on the cylinder approximations and then apply the exact distance function on the original trajectories. Since the number of so-called \emph{units} (cylinders) of the approximation can be much smaller than the number of units (segments) of the original trajectory, the evaluation of the distance function can be much cheaper. The expensive distance function on original trajectories then needs to be evaluated in the \emph{refine} step only for candidate pairs found on approximations.
 
We use the new distance function as an example of similarity measures in arbitrary domains. Given such a new distance function, we need indexing for it.

Such distance functions can in principle be constructed in arbitrary ways. However, if certain properties are known, they can be exploited in search.
A long and fruitful line of research has constructed indexes and search techniques \cite{ChenIndexingmetricSpaceSurveyPaper} based solely on the fact that the distance function is a \emph{metric}. A metric distance function $d$ fulfills (i) $d(x, y) \ge 0$, (ii) $d(x, x) = 0$, (iii) $d(x, y) = d(y, x)$, and (iv) $d(x, z) \leq d(x, y) + d(y, z)$. The last property is known as the \emph{triangle inequality}. This approach is known as the metric space approach to similarity search \cite{Zezula2006Metric}.
The beauty of this approach is that resulting index structures are automatically applicable to the wide diversity of data types mentioned above, as long as we can define a metric distance function for them. The crucial property for pruning is the triangle inequality.

The new distance function \emph{DistanceAvg} is a metric; so metric indexing is applicable.

In this paper, we propose an index structure for metric similarity search based on the concept of a Voronoi partitioning. The Voronoi diagram is a popular structure in computational geometry providing a distance-based partitioning of Euclidean space: Given a set of points $S$ in a $k$-dimensional space, space is partitioned into regions consisting of the points closest to each point in $S$.

A \emph{Voronoi partitioning} of a metric space can be defined similarly: given a set $S$ of objects with a metric distance function $d$, select a subset $C \subset S$ as \emph{centers} and assign each element of $S$ to its closest center in $C$. Let $c_{nn}(s, C)$ denote the center closest to $s$ in $C$ and let $P(u)$ denote the partition assigned to center $u$.

The purpose of partitioning is to be able to restrict in query processing attention to a subset of partitions. A crucial question is therefore how objects in different partitions can interact, that is, whether an object $s \in P(u)$ can be within distance $r$ from an object $t \in P(v)$. 

We can establish in Voronoi partitionings an important property that we call 
\emph{range distribution property}: For an object $q$ with $c_{nn}(q, C) = u$, only partitions $P(v)$ can have elements $t$ with $d(q, t) \leq r$ for which holds:
\begin{equation*}
d(q, v) \leq d(q, u) + 2r
\end{equation*}
Suppose we have a set of centers $C$ and two data sets $S$ and $T$ and we wish to find pairs of elements from $S \times T$ within distance $r$ of each other.
The range distribution property means that we can assign each element of $S$ to its closest center in $C$ and each element $t$ of $T$ to all centers in $C$ that are no further away from $t$ than the distance to its closest center plus $2r$. We can then find the matching pairs searching only within the pairs of partitions for the same centers.

The range distribution property is an excellent basis both for designing index structures and for designing parallel/distributed algorithms. In index structures, search in a node can be restricted to partitions with centers within the given range. In parallel computation, different processes or computing nodes can process pairs of partitions with the same center independently.

Our main focus in this paper is on presenting a new metric index, the \emph{N-tree}. However, we will also address parallel computation for  the construction of an N-tree index.

The proposed N-tree has parameters $k$ and $l$ for inner node size and leaf size, respectively. To build an N-tree for a dataset $S$, $k$ centers are selected (e.g. randomly) to form $C$ and $S$ is partitioned by $C$; if resulting partitions are larger than $l$, subtrees are created for them recursively. Hence so far an N-tree is simply a hierarchical Voronoi partitioning.

A simple range search algorithm for query object $q$ with radius $r$ would determine among the centers of a node the closest center and then use the range distribution property to determine the partitions that need to be searched recursively.
This works, but can be further optimized: to determine the closest center, all distances of $q$ to centers in $C$ need to be evaluated. These distance computations may be expensive.

To address this, a second major ingredient to the N-tree is pre-computation of all distances between centers of a node at tree construction time. An idea for range search with query object $q$ is to find the closest center $c_{nn}(q, C)$ and use it as a kind of proxy for $q$: the known distances from $c_{nn}(q, C)$ to other centers should be similar to the unknown (and expensive to calculate) distances from $q$ to these centers. Therefore we can use the known distances to prune away many of the expensive calculations. 

Moreover, we can apply an iterative approach already in finding the closest center: in each iteration, evaluate the distance to some center and then prune away many of the other centers that cannot be the closest center any more, based on this distance. Experiments show that this strategy is very effective.

Our contributions in this work are the following:

\begin{itemize}
\item We identify the range distribution property for Voronoi partitionings of metric space as a basis for designing index structures and parallel algorithms for similarity search.

\item We propose a new metric index structure, the N-tree. It is based on a hierarchical partitioning of metric space and of precomputation of all distances between centers in a node, enabling effective pruning techniques.

\item We provide efficient algorithms for range search and k-nearest-neighbor search.

\item We define a new metric distance function (similarity measure) called \emph{DistanceAvg} for exact trajectories of mobile objects and related bounds for their cylinder approximations, enabling a filter-and-refine technique. Moreover, it can be evaluated in linear time (i.e, $O(m+n)$ time for trajectories of sizes $m$ and $n$, respectively). It serves as an illustration for the usefulness of metric indexes such as an N-tree.

\item We provide an experimental evaluation of the N-tree and comparison with two of the best performing main-memory metric indexes, namely GNAT \cite{Brin1995Near} and MVPT \cite{Bozkaya1999Indexing}. In the survey~\cite{ChenIndexingmetricSpaceSurveyPaper}, MVPT is on rank 1 with respect to running time for range queries and \emph{kNN} queries, and GNAT is on rank 2 with respect to the number of distance computations and the running time for \emph{kNN} queries. We compare the three structures for different datasets and distance functions, including trajectories, images, text, and points. The N-tree performs equally well for range queries with small radius as the best structure, MVPT, and performs better at larger radii. For \emph{kNN} queries, it clearly outperforms the other two structures.

\item The N-tree exhibits a behaviour that, to our knowledge, is not present in other metric indexes: for increasing radius in range queries after some point the number of distance evaluations and the cost \emph{decreases}. We call this the \emph{U-turn effect}.

\item We provide an algorithm for parallel construction of an N-tree. This is useful for very expensive distance functions to speed up construction time. The technique includes a relational representation of N-trees which can be used to store a main-memory N-tree persistently and to rebuild it quickly.

\item The N-tree has been implemented in the DBMS prototype \Secondo\ and with it is freely available for download and experiments. 
\end{itemize}

Note that the N-tree as a metric index supports exact similarity search, returning for a given distance function the precise results sets for range and \emph{kNN} queries. Like \cite{ChenIndexingmetricSpaceSurveyPaper}, we do not compete with, or compare to, techniques that return only approximations of the result sets.

The rest of the paper is structured as follows. The new similarity measure for trajectories \emph{DistanceAvg} is introduced in Section \ref{sec:distance}.  Section~\ref{sec:preliminaries} explains the range distribution property. In Section~\ref{sec:ntree}, structure and construction and update algorithms of the N-tree are defined. Sections~\ref{sec:range} and \ref{sec:knn} are devoted to the algorithms for range queries and \emph{kNN} queries, respectively. Section~\ref{sec:evaluation} contains the experimental results, comparing the N-tree with GNAT and MVPT. Section~\ref{sec:distconst} addresses the parallel construction of an N-tree and techniques to save and restore it. Related work is covered in Section~\ref{sec:related}. Finally, conclusions are offered in Section~\ref{sec:conclusions}.

\section{A New Measure for Similarity of Trajectories - Exact and Approximate}
\label{sec:distance}

In this section we introduce a new distance measure to express similarity of trajectories. A trajectory represents the time-dependent position of an object, for example, a person, a vehicle, or an animal. Other representations of moving objects exist, for example, a moving region has a time dependent extent and can represent hurricanes, oil spills in the sea, or the growth of the Roman Empire.

\subsection{Background}

Trajectories can be considered in isolated applications, but the standard software environment to manage them are \emph{spatio-temporal} or \emph{moving objects databases} (MODs) \cite{GS05}. In such systems, trajectories are first class citizens and represented by their own data type called \tc{mpoint}, an abbreviation of \tc{moving}(\tc{point}). 
We have designed our new distance function in the context of such systems (e.g. \Secondo\ \cite{GAA+05}, Hermes \cite{PFGT08}, or MobilityDB \cite{ZSL20}) and their data model \cite{GBE+00, FGNS00}. It is related to an existing distance function
\begin{quote}
\op{distance}: \tc{mpoint} $\times$ \tc{mpoint} $\rightarrow$ \tc{mreal}
\end{quote}
providing the time dependent Euclidean distance of two moving point objects as a value of type \tc{mreal}. In this context, our similarity measure is an operation
\begin{quote}
\op{DistanceAvg}: \tc{mpoint} $\times$ \tc{mpoint} $\rightarrow$ \tc{real}
\end{quote}
Essentially, this is the average Euclidean distance between two moving point objects shifting them temporally to the same starting time and duration, computed as a piecewise integral.

In the sequel, first the exact integral-based distance computation is detailed. We then focus on an approximation of trajectories based on sequences of cylinders as well as lower and upper bound distances for such approximations.

\subsection{Exact Distance Function}
\label{sec:precisedistance}

In the DBMS \secondo{} \cite{GAA+05,GBD10}, a trajectory is represented by the data type \tc{mpoint}. An instance of this type corresponds to a sequence of so-called units (data type \tc{upoint}), each of which consists of a time interval and a start and end point. In such a trajectory, the units' time intervals are pairwise disjoint and ordered by time, but not necessarily consecutive. Gaps may especially arise from operations, e.g. reducing a trip to the times when it was inside a park.

For obtaining an exact distance between trajectories $U=\langle u_1,\dots,u_n\rangle$ and $V=\langle v_1,\dots,v_m\rangle$, where the $u_i$ and $v_j$ are units, we apply integral computation. It is necessary to shift temporally and stretch or shrink both trajectories to the same definition time interval, starting at instant $t$ and of duration $dur$ so that the distance function is guaranteed to fulfill the properties of a metric. 

This adjustment is conducted in Algorithm~\ref{algo:adjusttrajectory}. 
We refer to the start and end time and the start and end point of a unit $u$ by $u.start$, $u.end$, $u.p_0$, and $u.p_1$, respectively. Hence, $u$ is described as a tuple $(u.start, u.end, u.p_0, u.p_1)$.

\begin{algorithm}[!ht]
  \KwIn{Trajectory $U=\langle u_1, \dots, u_m\rangle$, instant $t$, duration $dur$}
  \KwOut{Trajectory $V$, starting at instant $t$ and of duration $dur$}
  $V := \langle\rangle$\;
  $\Delta := u_1.start - t$\;
  $f := \frac{u_m.end - u_1.start}{dur}$\; 
  \ForEach{$i \in \{1, \dots, m\}$}{
    $v_i := (t + (u_i.start - \Delta) \cdot f, t + (u_i.end - \Delta) \cdot f, 
      u_i.p_0, u_i.p_1)$\;
    append $v_i$ to $V$\;
  }  
  \ForEach{$i\in\{1,\dots,\ m-1\}$}{
    \If{$v_i.end < v_{i+1}.start$}{
      insert $(v_i.end, v_{i+1}.start, v_i.p_1, v_{i+1}.p_0)$ into $V$\;
    }
  }
  \Return $V$\;
  \caption{Adjust trajectory to a given definition time interval}
  \label{algo:adjusttrajectory}
\end{algorithm}

In Algorithm~\ref{algo:adjusttrajectory}, lines 4 - 6 perform the temporal shifting and stretching or shrinking and lines 7 - 9 close gaps. 
When distances are considered in a set of trajectories (e.g. in building an index), all trajectories must be mapped to the same definition time interval to let \emph{DistanceAvg} be a metric. Which common interval is chosen does not matter, for example, one can choose $t = 0$ and a duration of one hour.

After adjustment, the refinement partition $R(U,V)$ of $U$ and $V$ has to be determined, a standard technique in MODs \cite{LFGNS03}. Essentially, $R(U,V)$ is a sequence $\langle iv_1,\dots,iv_l\rangle$ of non-overlapping time intervals such that for every interval $iv_k$, no unit of $U$ or $V$ starts or ends inside it. The union of these intervals equals to the union of all time intervals of $U$ and $V$. Moreover, $R(U,V)$ contains the smallest possible number of intervals so that these two properties hold. Note that the number of time intervals of the refinement partition is at least $\max\{m,n\}$ and at most $m+n$.

\begin{figure}[ht]
  \centering
  \begin{tikzpicture}
    \pgfplotsset{%
      width=10cm, %\columnwidth
      height=4cm
    }
    \begin{axis}[axis x line=center, axis y line=none, xmin=0, xmax=1, ymin=0.15, ymax=1, tick label style={font=\footnotesize}, label style={font=\small}, legend style={font=\small}, axis x line = center, axis y line = center, every axis/.style={pin distance=1ex}, trim axis left, ticks=none, xlabel={time}, x label style={below=1mm}]
      \addplot[black,thick,domain=0.1:0.8] {0.7} node [pos=1.0, right] {$U$};
      \draw (axis cs:0.1,0.67) -- (axis cs:0.1,0.73);
      \draw (axis cs:0.3,0.67) -- (axis cs:0.3,0.73);
      \draw (axis cs:0.8,0.67) -- (axis cs:0.8,0.73);
      \addplot[black,thick,domain=0.1:0.8] {0.5} node [pos=1.0, right] {$V$};
      \draw (axis cs:0.1,0.47) -- (axis cs:0.1,0.53);
      \draw (axis cs:0.2,0.47) -- (axis cs:0.2,0.53);
      \draw (axis cs:0.5,0.47) -- (axis cs:0.5,0.53);
      \draw (axis cs:0.7,0.47) -- (axis cs:0.7,0.53);
      \draw (axis cs:0.8,0.47) -- (axis cs:0.8,0.53);
      \addplot[black,thick,domain=0.1:0.8] {0.3} node [pos=1.0, right] {$R(U,V)$};
      \draw (axis cs:0.1,0.27) -- (axis cs:0.1,0.33);
      \draw (axis cs:0.2,0.27) -- (axis cs:0.2,0.33);
      \draw (axis cs:0.3,0.27) -- (axis cs:0.3,0.33);
      \draw (axis cs:0.5,0.27) -- (axis cs:0.5,0.33);
      \draw (axis cs:0.7,0.27) -- (axis cs:0.7,0.33);
      \draw (axis cs:0.8,0.27) -- (axis cs:0.8,0.33);
    \end{axis}
  \end{tikzpicture}
  \caption{Example of a refinement partition}
  \label{fig:rp}
\end{figure}
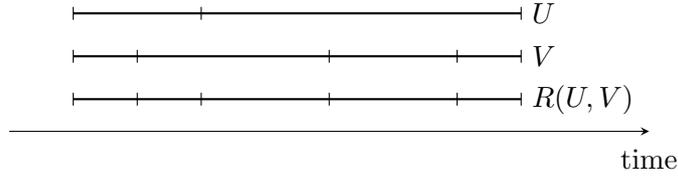

In Figure~\ref{fig:rp}, we illustrate the principle of a refinement partition. The example shows the division into time intervals for two trajectories $U$ and $V$ with~2 and 4~units, respectively, and for the corresponding refinement partition $R(U,V)$ having 5~time intervals. The computation cost for a refinement partition is in $O(n+m)$.

Let $U,V$ be adjusted trajectories and $R(U,V)=\langle iv_1,\dots,iv_l\rangle$ their refinement partition. Let $\langle u_1, \dots, u_l\rangle$ and $\langle v_1, \dots, v_l\rangle$ denote the sequences of units of $U$ and $V$, respectively, where each unit is reduced to the time interval(s) of the refinement partition.
We denote by $d(u_i, v_i)(t)$ the time dependent distance within the units $u_i, v_i$. Then the average distance can be computed as 

\[d(U, V) = \frac{\sum_{i=1}^{l} \int_{u_i.start}^{u_i.end} d(u_i, v_i)(t)}
{u_l.end - u_1.start}
\]

For every time interval of the refinement partition, the Euclidean distance function between the units $u$ and $v$ has to be determined. Both $u$ and $v$ can be considered as two linear functions of time describing the movement in $x$ direction ($u_x(t)$ and $v_x(t)$) and in $y$ direction ($u_y(t)$ and $v_y(t)$). Hence, we have time-dependent one-dimensional distance functions

\begin{equation*}
  f_x(u,v)(t)=\left|u_x(t) - v_x(t)\right| \mbox{ \ and \ } f_y(u,v)(t)=\left|u_y(t) - v_y(t)\right|,
\end{equation*}

\noindent{}respectively. The distance function for $u$ and $v$ thus equals the square root of a 2-degree polynomial, i.e., $f(t) = \sqrt{at^2+bt+c}$, where the coefficients $a$, $b$, and $c$ can be computed straightforwardly.

All these integral values are added and finally divided by the total duration of $U$ (or $V$, as they are guaranteed to be equal), in order to obtain an average distance value which is meaningful for any temporal duration.

It can be easily seen that this function is a metric. The required properties hold at every time instant because the measure relies on the Euclidean distance. Adding the values of the distance function for the whole temporal duration (which is done by the integral) and dividing this sum by that duration does not affect these properties.

\subsection{Distance of Approximations}
\label{sec:trajectoryapproximation}

In many cases, trajectories contain a large number of units, resulting in a considerable effort for computing precise distances between them. Approximations such as Douglas-Peucker \cite{R72,DP73} can be applied before the distance computation; however, the reliability of the result decreases. Particularly, it remains obscure whether and to what extent the precise distance is over- or underestimated.

Therefore, in the following we describe cylinder-based approximations   that allow us to control the distance between a trajectory and its approximation and so to establish lower and upper bounds on the distances of trajectories, based on their approximations. These bounds can be used in \emph{filter-and-refine} techniques for the evaluation of range queries: first, a superset of the results of a query is retrieved based on cheaper evaluations on approximations; second, the precise result set is determined from this superset using exact distance evaluation.

\subsubsection{Cylinder Unit Approximation}

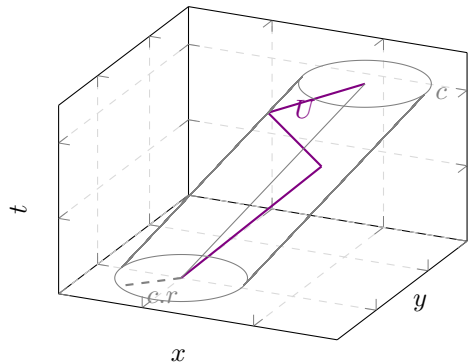
\begin{figure}[ht]
  \centering
  \begin{tikzpicture}
    \pgfplotsset{%
      height=6cm
    }
    \tikzstyle{every node}=[font=\small]
    \begin{axis}[grid, grid style={dashed,gray!30}, smooth, xmin=-1.5, xmax=3.5, ymin=-1.5, ymax=3.5, zmin=0, zmax=5,
        xlabel=$x$, xlabel style={below, anchor=north east,inner xsep=0pt}, xticklabels=\empty, ylabel=$y$, ylabel style={above,anchor=south,inner ysep=0pt}, yticklabels=\empty, zlabel=$t$, zticklabels=\empty]
    \addplot3[color=violet,thick] coordinates {(0,0,0) (2.1,0.9,3)};
    \addplot3[color=violet,thick] coordinates {(2.1,0.9,3) (1,1.2,4)} node[pos=0.7,above right] {$U$};
    \addplot3[color=violet,thick] coordinates {(1,1.2,4) (2.6,1.5,5)};
    \addplot3[color=gray] coordinates {(0,0,0) (2.6,1.5,5)};
    \addplot3[color=gray,domain=0:2*pi, samples = 60, samples y=0]({1.08*sin(deg(x))}, {1.08*cos(deg(x))}, {0});
    \addplot3[color=gray,domain=0:2*pi, samples = 60, samples y=0]({1.08*sin(deg(x))+2.6}, {1.08*cos(deg(x))+1.5}, {5}) node[right,pos=0.25] {$c$};
    \addplot3[color=gray,domain=0.25:4.7]({0.52*x-1.08},{0.3*x},{x});
    \addplot3[color=gray,domain=0.3:4.75]({0.52*x+1.08},{0.3*x},{x});
    \addplot3[color=gray,thick,dashed] coordinates {(0,0,0) (-0.75,-0.75,0)} node[below,pos=0.3] {$c.r$};
    \end{axis}
  \end{tikzpicture}
  \caption{Approximating a trajectory by a cylinder unit}
  \label{fig:cylinder}
\end{figure}

We introduce a cylinder unit of the data type \tc{cupoint} that extends the type \tc{upoint} by a radius indicating the tolerance value. A trajectory $U=\langle u_1,\dots,u_m\rangle$ (where the $u_i$ are \tc{upoint}s) can be approximated by a cylinder unit $u$ with start time $u_1.start$, end time $u_m.end$, start point $u_1.p_0$, and end point $u_m.p_1$. The radius $r$ of $u$ then equals the maximum distance between $u$ (conceived as a \tc{upoint}) and $U$. A cylinder unit $u$ can be written as a tuple $(u.start, u.end, u.p_0, u.p_1, u.r)$. The start and end time and the two points of $u$ together with the radius represent the smallest oblique cylinder that covers $U$, as illustrated in Figure~\ref{fig:cylinder}.

We define three distances on two cylinder units defined over the same time interval called $d, d_l$, and $d_u$, respectively. Here $d$ is the average distance of the central lines (axes of cylinders), $d_l$ a lower bound distance for the approximated trajectories, and $d_u$ an upper bound distance. They are defined as follows. Let $v, w$ be two cylinder units and $d(v, w)(t)$ the time dependent distance of their axes.

\[d(v, w) = \frac{\int_{v.start}^{v.end}d(v,w)(t)}{{v.end - v.start}} \]
\[d_l(v, w) = d(v, w) - v.r - w.r\]
\[d_u(v, w) = d(v, w) + v.r + w.r\]

\subsubsection{Cylinder Sequence Approximation}

However, for a longer trajectory, the radius of the approximating oblique cylinder can become very large, resulting in loose bounds on actual distances. The solution is to approximate a trajectory $U=\langle u_1,\dots,u_m\rangle$ by a sequence $C=\langle c_1,\dots,c_k\rangle$ of such cylinders (with $1\leq k\leq m$). A Douglas-Peucker approximation with radius $r$ conceptually yields a sequence of oblique cylinders with this radius enclosing the original trajectory. Technically, the result is just a regular trajectory. Let $dp(t, r)$ denote the approximation of trajectory $t$ with radius $r$.

What bounds can be established on the relative distances of approximated and precise trajectories? Let $c_1 = dp(t_1, r), c_2 = dp(t_2, r)$. The maximal time dependent distance between $t$ and $dp(t, r)$ is $r$. So the distance of two trajectories and the distance of their approximations can differ by at most $2r$. 

\[d(c_1, c_2) -2r \leq d(t_1, t_2) \leq d(c_1, c_2) + 2r \]
\[d(t_1, t_2) -2r \leq d(c_1, c_2) \leq d(t_1, t_2) + 2r \]

\begin{figure}[htb]
  \begin{center}
  \includegraphics[scale=0.8]{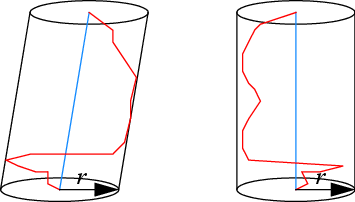}
  \end{center}
  \caption{ Two units of a cylinder sequence approximation}
  \label{fig:distances}
\end{figure}

This is illustrated for two units in Figure~\ref{fig:distances}. 

Suppose we want to perform a range query with radius $q$. Using

\[ d(c_1, c_2) \leq d(t_1, t_2) + 2r \] 

we have

\[ d(t_1, t_2) \leq q \Rightarrow d(c_1, c_2) \leq q + 2r \]

Hence, by performing a range query with radius $q + 2r$ on approximations we are sure to retrieve all objects whose trajectories have a radius up to $q$. However, the returned objects are not necessarily results of the query. We only know that 

\[ d(c_1, c_2) \leq q + 2r \Rightarrow d(t_1, t_2) \leq q + 2r + 2r = q + 4r \]

Hence we call the first query a \emph{filter step}; an additional \emph{refinement step} is needed checking the results of the filter step to ensure $d(t_1, t_2) \leq q$. This can be done by a further distance evaluation. However, for a subset of the filter result this is not necessary:

\[ d(c_1, c_2) \leq q - 2r \Rightarrow d(t_1, t_2) \leq q - 2r + 2r = q \]

Therefore, in the refinement step we first check $d(c_1, c_2) \leq q -2r$ and return objects fulfilling this directly; for the remaining objects we evaluate the more expensive $d(t_1, t_2) \leq q$.

In Figure~\ref{fig:distances} the two (red) trajectories are relatively far away from their approximation lines (blue). One can expect that often they will be close to the cylinder central lines and usually the average distance from center line to trajectory will be less than $r/2$. This is confirmed experimentally: In the New York Trips data set used in experiments (see Section~\ref{sec:evaluation}), in a 50 m approximation, only 164 out of 550841 approximations had an average distance of more than $r/2$, i.e., 25 meters.

We therefore slightly modify the construction of approximations: after running Douglas-Peucker with radius $r$ we check whether the resulting approximation has average distance less than $r/2$. If this is not the case, Douglas-Peucker is applied instead with radius $r/2$ to this trajectory. In this way we can guarantee, with a very small increase in the number of units of a few approximations, that the average distance between trajectory and approximation is less than $r/2$.

As a result, the difference between the distances of two trajectories and of their approximations, respectively, can only be $2 \cdot r/2 = r$. These smaller bounds are beneficial for the filter-and-refine technique:

\[ d(t_1, t_2) \leq q \Rightarrow d(c_1, c_2) \leq q + r \]
\[ d(c_1, c_2) \leq q - r \Rightarrow d(t_1, t_2) \leq q - r + r = q \]

\subsubsection{Filter-and-refine}
\label{sec:filterrefine}

Filter-and-refine techniques will be mainly used to accelerate index access in range queries, but they can also be used in a sequential scan without index. Algorithms \ref{alg:rangeScanFR} and \ref{alg:rangeSearchFR}
for both cases are shown below.

\begin{algorithm}[ht]
  \caption{\label{alg:rangeScanFR}\ic{rangeScanFR}$(T, s, q, r)$}
  \KwIn{$T$ -- a trajectory data set extended by cylinder unit approximations $cbb$ and cylinder sequence approximations $c$. The trajectory is in field $traj$\;
  \q\q $s$ -- a query trajectory object extended by the same approximations\;
  \q\q $q$ -- the query radius\;
  \q\q $r$ -- the radius of cylinder sequence approximations
  }
  \KwOut{all objects of $T$ within distance $q$ from $s$}
  \ForEach{$t \in T$}{
    \If {$d_l(s.cbb, t.cbb) \leq q$}
      {$d_c := d(s.c, t.c)$ \;
      \If {$d_c - r\leq q$}
        {\lIf {$d_c + r \leq q$} {\Return $t$}
          {\Else
            {\lIf {$d(s.traj, t.traj) \leq q$} {\Return $t$}}
          }
        }
      }
  }
\end{algorithm}

\begin{algorithm}[ht]
  \caption{\label{alg:rangeSearchFR}\ic{rangeSearchFR}$(T, I_c, s, q, r)$}
  \KwIn{$T$ -- a trajectory data set extended by cylinder sequence approximations $c$. The trajectory is in field $traj$\;
  \q\q $I_c$ -- an index over $T$ on field $c$. Queries on $I_c$ return objects from $T$\;
  \q\q $s$ -- a query trajectory object extended by cylinder sequence approximation $c$\;
  \q\q $q$ -- the query radius\;
  \q\q $r$ -- the radius of cylinder sequence approximations
  }
  \KwOut{all objects of $T$ within distance $q$ from $s$}
  $U := rangeSearch(I_c, s, q + r)$\;  
  \ForEach{$u \in U$}{
    \lIf {$d(s.c, u.c) \leq q - r$} {\Return $u$}
    \Else {\lIf {$d(s.traj, u.traj) \leq q$} {\Return $u$}}
  }
\end{algorithm}

\subsection{Final Remarks}

Finally, we note that the new similarity measure \emph{DistanceAvg} has the following properties:

\begin{itemize}
\item In contrast to other distance functions, e.g. Hausdorff or Frechet distance, it has a natural interpretation because the average Euclidean distance, expressed in meters, is a quantity that is easy to understand.
\item It has a linear runtime. A parallel scan through the two \tc{mpoint} values suffices. In the survey \cite{SuLZZZ20} all metric distance measures  are said to have a product time complexity\footnote{We note that this is not correct for STED, see below.}, i.e., $O(mn)$ for trajectories with  $m$ and $n$ elements (points, segments), respectively. The only exception is Euclidean distance on point sequences of equal length, which is not practical. The linear runtime of \emph{DistanceAvg} is advantageous especially for long trajectories (as confirmed in experiments, see Section~\ref{subsection:avg_dist_vs_hausdorff_eval}).

\item It can be used as a spatio-temporal or a purely spatial similarity measure (called \emph{sequence-only} in \cite{SuLZZZ20}). In the latter case, we convert the sequence of point locations into an \tc{mpoint} value by connecting points with movement at constant speed. If spatio-temporal trajectories (e.g. \tc{mpoint}) values are given that we wish to evaluate only spatially, we can adapt the time intervals of units in the same way.
\item There exist approximations by a single unit or a sequence of cylinder units with related distance bounds on approximated trajectories, enabling filter-and-refine techniques.
\end{itemize}

The new distance measure is related to the distance function defined in \cite{NP06} in the context of trajectory clustering, called spatio-temporal Euclidean distance (STED) in \cite{SuLZZZ20}. Indeed, STED is defined in the same way as \emph{DistanceAvg} computing the integral of the time dependent distance function. However, STED is restricted to trajectories defined over the same time interval; other applications are not considered. Our function may be considered as an extension of STED to handle arbitrary pairs of trajectories and to add related approximation distance measures.

Obviously, STED also works in linear time $O(m + n)$ \cite{NP06}, so the complexity $O(mn)$ in \cite{SuLZZZ20} is not correct.

\section{Preliminaries: Range Distribution on a Voronoi Partitioning}
\label{sec:preliminaries}

Suppose we are given a data set $S$ with a distance function $d$. We partition $S$ by selecting some of its elements as centers and assign each element of $S$ to its closest center. We call this a \emph{Voronoi partitioning}.
Let $P(u)$ denote the partition of $u$, the subset of $S$ assigned to center $u$.

Now let $u$ be a center and $s \in P(u)$. We want to perform a range query for $s$ with radius $r$ on $S$, retrieving all elements of $S$ within distance $r$ from $s$. The question is: \emph{which other partitions $P(v)$ may contain results?}

An answer to this question has several applications. A Voronoi partitioning can be used on the one hand to organize subtrees of a node in an index structure. On the other hand, the partitions can be assigned to different nodes of a distributed computing cluster. In the first case, for a given query point, all relevant subtrees need to be searched; in the latter case, the query point has to be sent to all relevant partitions in the cluster for  distributed processing.

Assuming $d$ is a metric distance function, an answer to this question has been given several times in the literature on metric indexing as well as in distributed similarity computation. In indexing, it is known as a pruning rule associated with generalized hyperplane partitioning, called \emph{double pivot filtering} in \cite{ChenIndexingmetricSpaceSurveyPaper} and attributed to \cite{Zezula2006Metric}. In distributed computation, it has been rediscovered at least in \cite{Sarma2014ClusterJoin} for similarity join and in \cite{GBN21} for density-based similarity clustering. Here, we review Theorem 6.1 and its proof from \cite{GBN21}.

For $s \in S$, let $N_{Eps}(s) = \{t \in S \sothat d(s, t) \leq Eps\}$. Here $Eps$ corresponds to radius $r$.

\begin{Theorem}
\label{th:range} Let $s, t \in S$ and $T \subset S$. Let $u, v \in T$ be the elements of $T$ with minimal distance to $s$ and $t$, respectively. Then $t \in N_{Eps}(s) \Rightarrow d(v, s) \leq d(u, s) + 2 \cdot Eps$.

Proof: Let $x$ be a location within $N_{Eps}(s)$ with equal distance to $u$ and $v$, that is, $d(u, x) = d(v, x)$. Such locations must exist, because $s$ is closer to $u$ and $t$ is closer to $v$. Then $d(u, x) \leq d(u, s) + Eps$. Further, $d(v, s) \leq d(v, x) + Eps = d(u, x) + Eps \leq d(u, s) + Eps + Eps = d(u, s) + 2 \cdot Eps$.
\end{Theorem}

\begin{figure}[thb]
  \begin{center}
  \includegraphics[scale=0.7]{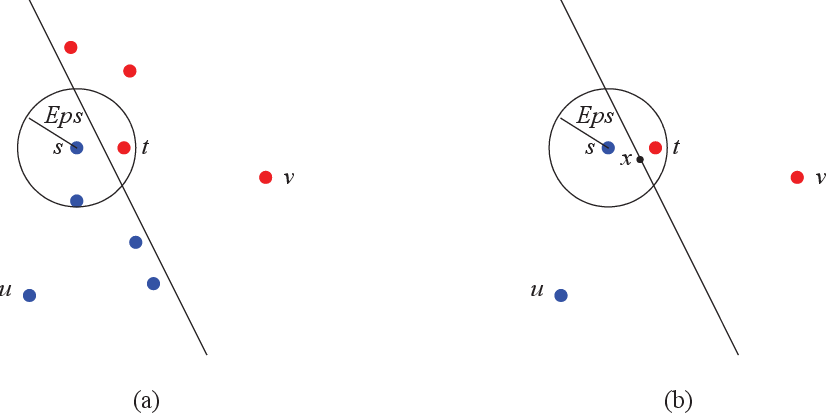}
  \end{center}
  \caption{Range search on a Voronoi partitioning (used with permission) \cite{GBN21}}
  \label{fig:neighbors}
\end{figure}

The setting of Theorem~\ref{th:range} is illustrated in Figure~\ref{fig:neighbors}~(a). Here $u$ and $v$ are partition centers; the blue objects are closer to $u$, the red objects are closer to $v$; hence $s \in P(u)$ and $t \in P(v)$. The slanted line represents equal distance between $u$ and $v$ (the generalized hyperplane). The proof is illustrated in Figure~\ref{fig:neighbors}~(b).

Theorem~\ref{th:range} says that when we perform a range query with radius $r$ from an object $s$ within a partition $P(u)$, we need to consider all partitions $P(v)$ such that $d(s, v) \leq d(s, u) + 2 r$. We call this the \emph{range distribution property} in this paper.

\section{The N-tree}
\label{sec:ntree}

Let $S$ be a set with a metric distance function $d$. An \emph{N-tree} over $S$ is an index structure supporting range search and \emph{kNN} search on $S$. In this paper we study it as a main memory index, but it is also suitable as a persistent disk based index.

\subsection{Structure}

An N-tree (\emph{neighborhood tree}) is a multiway tree like a B-tree or R-tree. It has two parameters: the \emph{degree} $k$ and the \emph{leaf size} $l$, $k \leq l$. 

The basic structure is quite simple. Let $C$ be a subset of $S$ of size $k$ called \emph{centers}. $S$ is partitioned by $C$, assigning each element of $S$ to its closest center in $C$. An internal node has a set of entries $(c_i, T(S_i))$ where $c_i$ is a center together with a pointer to a subtree $T(S_i)$ organizing the related subset $S_i$ of $S$. A leaf node just contains the subset. We start with a definition of this basic structure.

\begin{definition}
Let $S$ be a set with a metric distance function $d$. A \emph{basic N-tree over $S$} of degree $k$ and leaf size $l$ is defined as
\begin{eqnarray*}
T(S) & = &
  \begin{cases}

    ((c_1, T(S_1)), ..., (c_k, T(S_k))) & |S| > k
  \\
 \q \mbox{such that } S = \bigcup_{i \in \{1, ..., k\} } S_i, i \neq j \Rightarrow S_i \cap S_j = \emptyset,

 \\
 \q \forall i \in \{1, ..., k\}: \\
 \q\q c_i \in S, \\
 \q\q S_i = \{u \in S \sothat \forall j \in \{1, ..., k\}: d(u, c_i) \leq d(u, c_j)\} \\
     S	&  |S| \leq l
  \end{cases}
\end{eqnarray*}
\end{definition}

\begin{example} We construct an N-tree with parameters $k = 3,  l = 4$ for a collection of 2d points with Euclidean distance shown in Figure~\ref{fig:ntree} (a). 
\begin{figure}[thb]
\begin{center}
\begin{tabular}{ccc}
\includegraphics[scale=0.5]{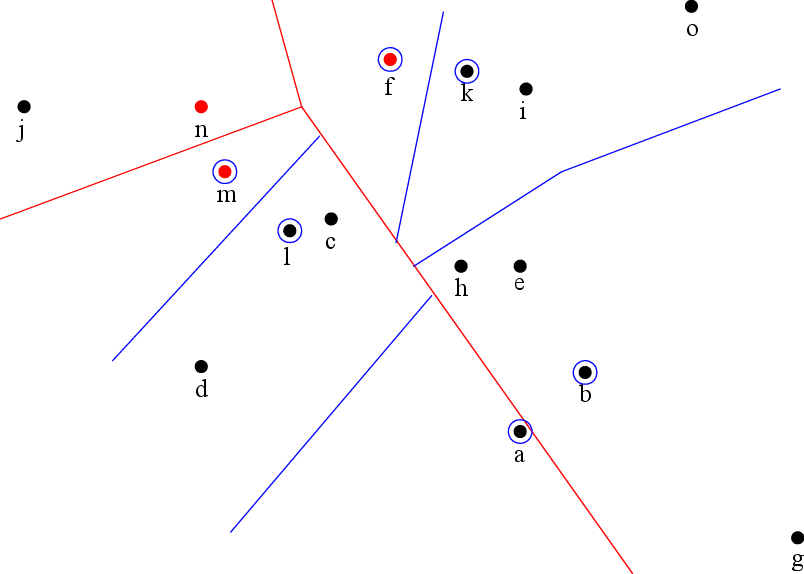}
& \q\q &
\includegraphics[scale=0.6]{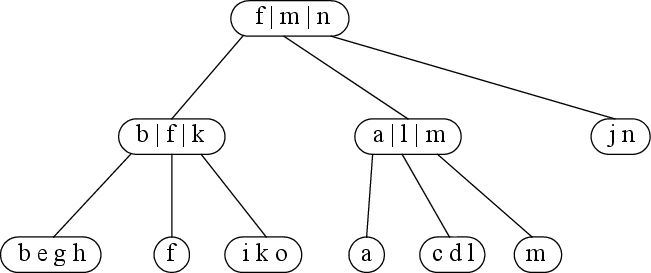} \\
% \vspace{0.5cm}\\
(a) &  & (b)
\end{tabular}
\end{center}
  \caption{ Constructing an N-tree: (a) a point set and its partitioning (b) the resulting N-tree}
  \label{fig:ntree}
\end{figure} 

At the top level we have a set of centers $\{f, m, n\}$ and partitions $P(f) = \{b, e, f, g, h, i, k, o\}$, $P(m) = \{a, c, d, l, m\}$, and $P(n) = \{j, n\}$. This partitioning is indicated by the red Voronoi diagram (red nodes and lines of equal distance between nodes). Partition $P(f)$ is recursively partitioned by centers $\{b, f, k\}$ into $P(b) = \{b, e, g, h\}$, $P(f) = \{f\}$, and $P(k) = \{i, k, o\}$. This level of partitioning is indicated by blue circles around nodes and blue partitioning lines. Similarly $P(m)$ is partitioned by set of centers $\{a, l, m\}$ into $P(a) = \{a\}$, $P(l) = \{c, d, l\}$, and $P(m) = \{m\}$. All resulting partitions are small enough (no more than $l=4$ elements), so no further partitioning is needed. The resulting N-tree is shown in Figure~\ref{fig:ntree} (b).
\end{example}

Note that we have separate parameters $k$ and $l$ to control the sizes of inner nodes and leaves, respectively. For a partition with $g$ elements, for $k < g < l$ we have a choice of splitting and creating a new internal node or assigning the partition to a leaf directly. In both cases we have a valid N-tree. The construction in Algorithm~\ref{alg:build} splits whenever the leaf size $l$ is exceeded.

The complete structure of an N-tree includes for each node some auxiliary information, namely
\begin{itemize}
\item all pairwise distances between centers,
\item two distinguished centers called \emph{pivots},
\item for each center, a vector of its distances to the two pivot elements,
\item for each center $c_i$ in an internal node, the \emph{radius} of its associated subtree $T(S_i)$, defined as the largest distance from $c_i$ to an element of $S_i$.
\end{itemize}
Such information is of course used for pruning in range search and \emph{kNN} search. The definition for the complete structure is:

\begin{definition}
Let $S$ be a set with a metric distance function $d$. An \emph{N-tree over $S$} of degree $k$ and leaf size $l$ is defined as
\begin{eqnarray*}
T(S) & = &
  \begin{cases}
    (((c_1, T(S_1), r_1), ..., (c_k, T(S_k), r_k)), D, \{p_1, p_2\}, PD) & |S| > k \\
    
\q \mbox{such that } \\
\q C = \{c_1, ..., c_k\}, S = \bigcup_{i \in \{1, ..., k\}} S_i, 
i \neq j \Rightarrow S_i \cap S_j = \emptyset,\\
\q \forall i \in \{1, ..., k\}: \\
\q\q c_i \in S, \\
\q\q S_i = \{u \in S \sothat \forall j \in \{1, ..., k\}: d(u, c_i) \leq d(u, c_j)\}, \\
\q\q r_i = \max_{v \in S_i} d(c_i, v), \\

     (S, D, \{p_1, p_2\}, PD)	&  |S| \leq l \\
     \q \mbox{such that } C = S\\
  \end{cases}\\
\end{eqnarray*}
where 

\[D = \{d_{ij} \sothat i, j \in \{1, ..., |C|\}, d_{ij} = d(c_i, c_j)\}\]
\[p_1, p_2 \in C\]
\[PD = \{v_i \sothat i \in \{1, ..., |C|\}, v_i = (d(c_i, p_1), d(c_i, p_2))\}  \]

\end{definition}
It is possible that a leaf has only one element; in this case $D$ and $PD$ are empty and the $p_i$ are undefined. This special case is omitted in the definition.
Note that there is no balancing condition; the subsets $S_i$ of a node may have different sizes and so the representing subtrees may have different depths.

\subsection{Construction}   

An N-tree is constructed for a set $S$ by selecting $k$ elements from $S$ as centers and then partitioning $S$ by these centers. For each partition, an N-tree is constructed recursively if it has more than $l$ elements (Algorithm~\ref{alg:build}).

\begin{algorithm}[ht]
  \caption{\label{alg:build}\ic{build}$(S, k, l)$}
  \KwIn{$S$ - a set of objects with a distance function $d$\;
  \q\q $k$ - an integer, the degree of the N-tree\;
  \q\q $l$ - an integer, the maximal leaf size}
  \KwOut{a node representing $S$}
  \If {$|S| \leq l$} {
  compute the auxiliary information $D, \{p1, p2\}, PD$\;
  \Return leaf($S, D, \{p1, p2\}, PD$)
  }
  \Else {
    $C$ := \emph{determineCenters}$(S, k)$\;
    $\{(c_1, S_1, r_1), ..., (c_k, S_k, r_k)\}$ := \emph{partition}$(S, C)$\;
    compute the auxiliary information $D, \{p1, p2\}, PD$\;
    \Return node($((c_1, build(S_1, k, l), r_1), ..., 
      (c_k, build(S_k, k, l), r_k)), D, \{p1, p2\}, PD$)
  }
\end{algorithm}

The two pivot elements are selected randomly.
Perhaps, the simplest  algorithm  for \emph{determineCenters} is random selection, which is expected to yield good results. In the report 
\cite{ntree_technical_report}, Section 6.2,
we compare it experimentally with three other algorithms for selecting centers. We found the \emph{Greedy} algorithm also used in GNAT \cite{Brin1995Near} to be the most suitable algorithm as construction is only a little more expensive than with random selection and the query times are a little better.  In contrast, the other two algorithms (called \emph{Base-prototypes selection} and \emph{Hull of Foci}, see \cite{ntree_technical_report}) had far higher construction times and worse query times than \emph{Greedy}. The \emph{Greedy} algorithm, to find $k$ centers, first selects randomly $3k$ candidates and then finds among them those $k$ that are farthest apart. We also use this algorithm in the experiments of this paper (except for those with \Secondo, where random partitioning is used).

The algorithm for \emph{partition} needs to determine for each element $u$ of $S$ the closest center in $C$. A straightforward implementation computes all distances $d(u, c_i)$. A better implementation uses the algorithm \emph{closestCenter} developed in Section~\ref{sec:closestCenter} in the context of range search, which prunes a lot of the expensive distance calculations.

\subsection{Updates}

The N-tree supports simple updates, i.e., inserting or deleting an element.

\subsubsection{Insert}

The insertion process is listed in Algorithm~\ref{alg:insert}.
When a new element $x$ is to be inserted into an existing N-tree originally built over a set $S$, we have to ensure that $x$ is assigned to the appropriate leaf node. At the beginning, we determine the center $c_x$ from the root node which is closest to $x$ among all its $k$ centers. This can be computed efficiently using the algorithm \emph{closestCenter} from Section~\ref{sec:closestCenter}. We also update the radius of the respective subtree if the new element is further away from $c_x$.
This procedure is repeated in every inner node until finally a leaf node is visited.
The notation \emph{node}$(c_j)$ refers to the root node of the subtree associated with center $c_j$.

\begin{algorithm}[ht]
  \caption{\emph{insert}$(T(S), k, l, x)$\label{alg:insert}}
  \KwIn{an N-tree $T(S)$ with parameters $k, l$, an element $x$}
  \KwOut{an N-tree $T(S \cup \{x\})$}
  $p := $ root of $T(S)$\; 
  $q := p$\;
  \While{$q$ is an inner node}{
    $(c_i, d_{min}) := \ic{closestCenter}(q.C, x, q.D, q.\{p_1, p_2\}, q.PD)$\;
    \lIf{$d_{min} > r_i$} {
      $r_i := d_{min}$\;
      }
    $q := node(c_i)$\;
  }
  /* $q$ is now a leaf node */ \\
  $q.S := q.S \cup \{x\}$\;
  \If{$\left| q.S \right| \leq l$}{
    update $q.D$ and $q.PD$ to include $x$\;
  }
  \Else{
    replace $q$ with new inner node $\ic{build}(q.S, k, l)    $\;
  }
  \Return $p$\;
\end{algorithm}

In the regular case, $x$ can simply be added to the leaf node and its distances to all other elements in the node as well as the distances to the pivots have to be computed for the sets $D$ and $PD$, respectively. However, if the number of elements is already $l$ before the insertion, the procedure becomes slightly more involved. In this case, the leaf node has to be replaced by a new inner node with $k$ new centers and leaf nodes containing the $l+1$ elements. The auxiliary information has to be recomputed accordingly (as shown in Algorithm~\ref{alg:build}).

\subsubsection{Delete}
The procedure of deleting an element $x$ from an existing N-tree, listed in Algorithm~\ref{alg:delete}, is similar to the insertion algorithm described in the previous subsection.

\begin{algorithm}[ht]
  \caption{\emph{delete}$(T(S), k, l, x)$\label{alg:delete}}
  \KwIn{an N-tree $T(S)$ with parameters $k, l$, an element $x$}
  \KwOut{an N-tree $T(S\setminus\{x\})$}
  $p := $ root of $T(S)$\;
  $q := p$\;
  \While{$q$ is an inner node}{
    $(c_i, d_{min}) := \ic{closestCenter}(q.C, x, q.D, q.\{p_1, p_2\}, q.PD)$\;
    $q := node(c_i)$\;
  }
  /* $q$ is now a leaf node */ \\
  \lIf{$x \notin q.S$}{
    \Return $p$\;
  }
  $q.S := q.S \setminus \{x\}$\;
  \If{$\left| q.S \right| > 0$}{
    remove auxiliary information concerning $x$ from $q.D$ and $q.PD$\;
  }
  \Else {
    $p :=$ parent node of $q$ with subtree $T(S_p)$\;
    $T'(S_p) :=$ \emph{build}$(S_p, k, l)$\;
    replace $T(S_p)$ with new inner node $T'(S_p)$\;
  }
  \Return $p$\;
\end{algorithm}

We follow the path of closest centers until a leaf node $q$ is found. If $x$ does not occur in $q$, we can be sure that it is not present in the whole tree and the latter remains unchanged. Otherwise, $x$ is deleted from the respective object set $q.S$ and the auxiliary information is reduced accordingly.

After removing $x$, the set $q.S$ and hence the leaf $q$ may be empty. In this case, the tree has to be restructured. Therefore we collect all objects corresponding to $q$'s parent node $p$   and replace its subtree by a new subtree constructed from the updated object set.

Note that the radii of subtrees along the path to the leaf containing $x$ are not changed. In case $x$ happens to be the most distant element in such a subtree, in principle this radius should be corrected. However, such a computation would be too expensive. The radius is used for pruning in range search and \emph{kNN} search; an occasionally slightly too high radius may lead to consider a subtree that is otherwise pruned; however, the results are still correct. 

We assume that insertions or deletions are relatively rare operations; in many applications they are not needed at all. For large amounts of updates the tree should be rebuilt from scratch. Note that the best competitors that we compare to in experiments, GNAT and MVPT, are static structures \cite{ChenIndexingmetricSpaceSurveyPaper}.

\section{Range Queries}
\label{sec:range}

In this section we develop algorithms for range queries; Section~\ref{sec:knn} addresses \emph{kNN} queries. The main idea is to use the precomputed distances available in nodes to avoid many expensive distance computations.

We briefly review the definitions of range queries and \emph{kNN} queries.

\begin{definition}
Let $S$ be a set, $q$ a query object, $d$ a distance function applicable to $S$ and $q$, $r \in \mathbb{R}$ a \emph{search radius}, and $k \in \mathbb{N}$. A \emph{range query} is
\[
range(S, q, r) = \{s \in S \sothat d(q, s) \leq r\}
\]
A \emph{\emph{kNN} query} (k-nearest-neighbors query) is
\[
kNN(S, q, k) = U \subseteq S \mbox{ such that } |U| = k \wedge \forall u \in U, \forall s \in S \setminus U: d(q, u) \leq d(q, s)
\]
\end{definition}

\subsection{Overview}
\label{sec:overview}

\begin{figure}[htb]
  \begin{center}
  \includegraphics[scale=0.5]{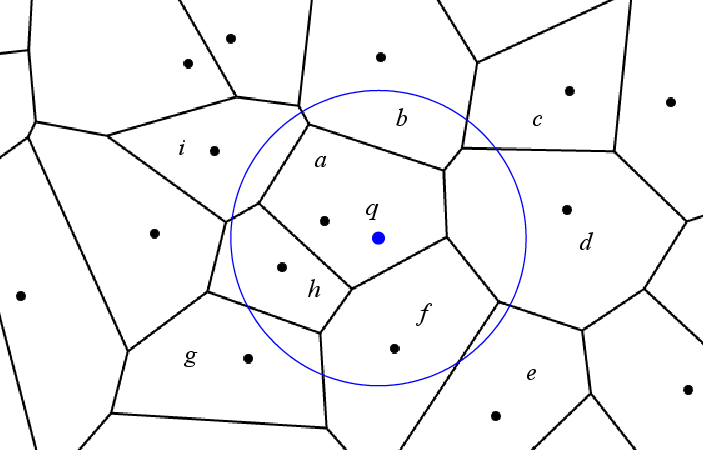}
  \end{center}
  \caption{ Range query with query point inside partitioning}
  \label{fig:rangequery}
\end{figure}

The problem of range search on the root node of an N-tree is illustrated in Figure~\ref{fig:rangequery}. The partitioning of set $S$ into partitions $((c_1, S_1), \dots (c_k, S_k))$, where each element of $S$ is assigned to its closest center, corresponds to a Voronoi partitioning of the metric space. Here we illustrate it by a Voronoi diagram in the 2D Euclidean space. There is one center $c_x$ in $C$ that is closest to the query object $q$; we say $q$ \emph{falls into} partition $S_x$. In Figure~\ref{fig:rangequery} this is partition $a$. There are other partitions that may contain objects within distance $r$ from $q$; these are partitions $b$ through $i$ in Figure~\ref{fig:rangequery}.

Hence the search on the root node needs to identify the partition into which $q$ falls as well as the other partitions that may contain objects within its query radius; the respective subtrees need to be searched recursively.

The latter are given by the range distribution property: Theorem~\ref{th:range} suggests that an object $p$ assigned to another partition with center $c_y$ can only have distance $d(q, p) \leq r$ if $d(q, c_y) \leq d(q, c_x) + 2r$.

As it is important for pruning
to know the distance $d(q, c_x)$, the algorithm for range search on the root node proceeds as follows:
\begin{enumerate}
\item Find the center $c_x$ closest to $q$;
\item Determine other centers $c_y$ fulfilling the condition $d(q, c_y) \leq d(q, c_x) + 2r$, using precomputed distances as much as possible. 
\item Recursively search all qualifying partitions.
\end{enumerate}

\begin{figure}[htb]
  \begin{center}
  \includegraphics[scale=0.7]{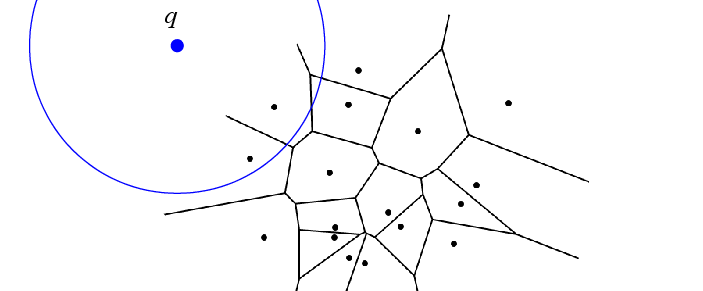}
  \end{center}
  \caption{ Range query with query point outside partitioning}
  \label{fig:rangequery2}
\end{figure}

At levels of the tree below the root node, however, the situation is not always the same as in the root node. For partition $S_x$ that $q$ falls into, it is the same, so we can search this child of the root node in the same way. However, for the other partitions $q$ lies ``outside''. Figure~\ref{fig:rangequery2} illustrates the partitioning of a child node of the root like $e$ in Figure~\ref{fig:rangequery} for which $q$ lies outside.

It turns out that in this case, the pruning criterion of Theorem~\ref{th:range} is not effective. It is therefore not a good strategy to find the closest center to $q$ first, also because the pruning of distance computations in finding the closest center is not effective. In the following subsections, we therefore address the following subproblems:

\begin{itemize}
\item Finding the closest center to a query point.
\item Range search for a query point inside a partition.
\item Range search for a query point outside a partition.
\item Overall algorithm for range search.
\end{itemize}

\subsection{Finding the Closest Center}
\label{sec:closestCenter}

The goal in designing an algorithm to find the closest center for a query point $q$ in a set of centers $C$ is to avoid as many of the expensive distance computations between the query point and a center as possible, using the precomputed distances between centers. The strategy is to consider all centers as candidates and then, in each step, to evaluate one distance $d(q, c_i)$ and to prune all centers $c_j$ based on their known distance to this center, $d_{ij} = d(c_i, c_j)$, that cannot be the closest center any more.

\subsubsection{Pruning Rules}

\begin{figure}[htb]
  \begin{center}
  \includegraphics[scale=0.5]{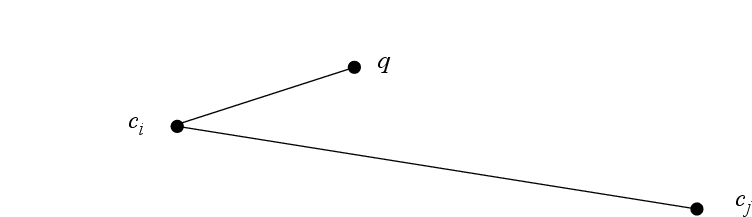}
  \end{center}
  \caption{ Simple pruning}
  \label{fig:simplepruning}
\end{figure}

\begin{figure}[htb]
  \begin{center}
  \includegraphics[scale=0.5]{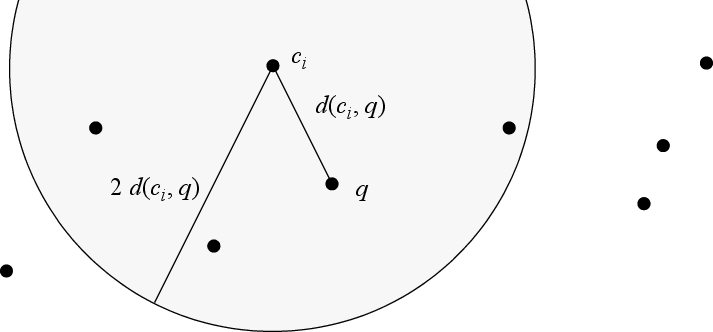}
  \end{center}
  \caption{ Simple pruning}
  \label{fig:simplepruning2}
\end{figure}

Two pruning rules can be determined. We call the first \emph{simple pruning}. It is illustrated in Figure~\ref{fig:simplepruning}. The distance $d(q, c_i)$ has just been evaluated; the known distances between centers are denoted as $d_{ij}$.
We have (triangle inequality for metric distance functions):
\[
|d_{ij} - d(q, c_i)| \leq d(q, c_j) \leq d_{ij} + d(q, c_i)
\]
Now assume $d_{ij} > 2d(q, c_i)$.
\[
2d(q, c_i) < d_{ij} \Rightarrow d(q, c_i) < d_{ij} - d(q, c_i) \leq d(q, c_j)
\]
Hence $c_i$ is closer to $q$ than $c_j$ and $c_j$ can be pruned from the set of candidates. This is illustrated in Figure~\ref{fig:simplepruning2}. All objects with distance larger than $2d(q, c_i)$ from $c_i$ can be pruned.

\begin{figure}[htb]
  \begin{center}
  \includegraphics[scale=0.5]{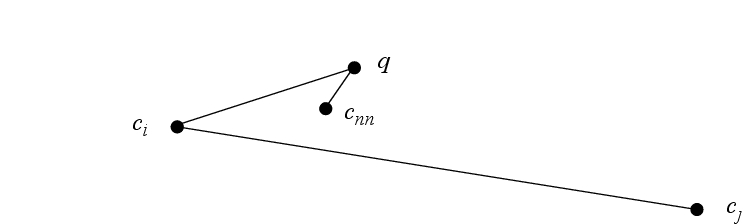}
  \end{center}
  \caption{ Mindist pruning}
  \label{fig:mindistpruning}
\end{figure}

A second pruning rule is called \emph{mindist pruning}. It is illustrated in Figure~\ref{fig:mindistpruning}. 

This rule uses not only the distance $d(q, c_i)$ just evaluated but also keeps track of the closest center $c_{nn}$ discovered so far and the related minimal distance $d_{min} = d(q, c_{nn})$. It is easy to see that in this case we can prune a center $c_j$ not only when $d_{ij}$ is too large, but also when it is too small. Obviously, any center $c_j$ can be pruned for which holds $d_{ij} < d(q, c_i) - d_{min}$ or $d_{ij} > d(q, c_i) + d_{min}$. Note that the second condition is the one for simple pruning when we substitute $d(q, c_i)$ for $d_{min}$. Hence this rule subsumes the first one.

\subsubsection{The Order of Candidates}

For pruning to be effective, it would be good to select first candidates $c_i$ for distance evaluation that are close to $q$, because then the radius for pruning is small and many points can be pruned early. 

\begin{figure}[htb]
  \begin{center}
  \includegraphics[scale=0.6]{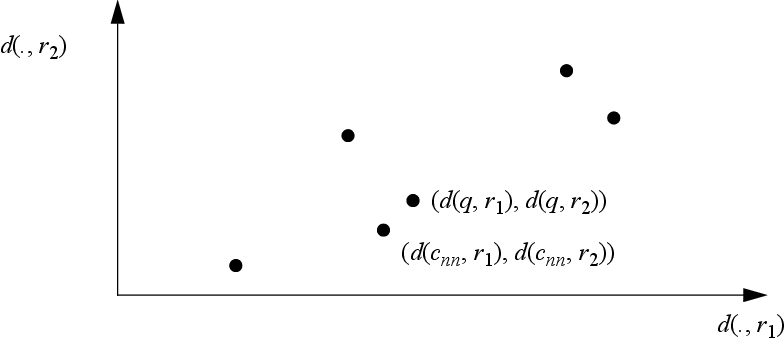}
  \end{center}
  \caption{ Mapping distances into a Euclidean space}
  \label{fig:pivotdistances}
\end{figure}

Consider evaluating distances from two selected elements of $C$,   $r_1$ and $r_2$, to (i) $q$ and (ii) to the nearest neighbor of $q$, $c_{nn}$. These distances should be similar. We can represent these distances as 2d vectors $(d(q, r_1), d(q, r_2))$ and $(d(c_{nn}, r_1), d(c_{nn}, r_2))$. We can visualize these vectors as shown in Figure~\ref{fig:pivotdistances}.

Hence it should be a good strategy to process candidates for distance evaluation with $q$ in the order of increasing distance of their vector to the vector of $q$. Reference points for mapping arbitrary distances into a Euclidean space are called \emph{pivots} in the literature. Instead of using two pivots, we might also use three or a higher number $n$, using Euclidean distance in an $n$-dimensional space. Experiments have shown that three pivots do not yield better results than two, so we use two pivots in our structure.

\subsubsection{Algorithm \emph{closestCenter}}

The algorithm \emph{closestCenter} can be formulated as shown in Algorithm~\ref{alg:closestCenter}. 

\begin{algorithm}[ht]
  \caption{\label{alg:closestCenter}\ic{closestCenter}$(C, q, D, \{p_1, p_2\}, PD)$}
  \KwIn{$C$ - the set of centers\;
  \q\q $q$ - a point\;
  \q\q $D$ - the set of pairwise distances $d_{ij}$ between centers in $C$\;
  \q\q $\{p_1, p_2\}$ - the two pivot elements of $C$\;
  \q\q $PD$ - the pivot distance vectors of $C$.}
  \KwOut{$c_{nn} \in C$ - the center with minimal distance to $q$\;
  \q\q $d_{min}$ - the distance between $q$ and $c_{nn}$}
  let $C = \{c_1, \dots, c_m\}$\;
  let $PD = \{v_1, \dots, v_m\}$\;
  $v_q := (d(q, p_1), d(q, p_2))$ (*)\;  
  $C' := \langle c_{i_1}, \dots, c_{i_m}\rangle \mbox{such that } \forall j, l \in \{1, \dots, m\}: j < l \Rightarrow dEuc(v_q, v_{i_j}) \leq dEuc(v_q, v_{i_l})$\;
  \label{line:order}
  $(c_{nn}, d_{min}) := (\bot, \infty)$\;
  \While {$\mathit{not}(\mathit{isempty}(C'))$} {
    $c_i := \mathit{first}(C'); C' := \mathit{rest}(C')$\;
    $u := d(q, c_i)$ (*); $DQ_i := u$\;
    \label{line:DQ}
    \If {$u < d_{min}$} {$c_{nn} := c_i; d_{min} := u$}
    $C' := \{c_j \in C \sothat u - d_{min} < d_{ij} <  u + d_{min}\}$
    }
  \Return $(c_{nn}, d_{min})$  
\end{algorithm}

Lines where expensive distance evaluations occur are marked as (*).

In Line~\ref{line:order}, $\langle x, y, z, \dots\rangle$ denotes the ordered sequence, or list, of elements $x, y, z, \dots$. In this algorithm, $dEuc$ denotes the 2d-distance of vectors in the Euclidean space. Hence the sequence of candidates is returned ordered by increasing distance of their 2d vectors from $v_q$, the 2d vector of the query point.

In Line~\ref{line:DQ}, after evaluating a distance $d(q, c_i)$ we store it in an array $DQ_i$. This avoids reevaluating the same distance in the algorithm \emph{rangeSearch1} presented below.

\subsection{Query Point Inside  Partitioning}

After determining the center $c_{nn}$ closest to $q$  and its distance $d_{min} = d(q, c_{nn})$, we can address the range search problem illustrated in Figure~\ref{fig:rangequery}. In inner nodes, we need to determine partitions that need to be searched recursively; in leaves we need to report centers within the query range $r$. The idea is to use the known distances $d_{min}$ and $d_{ij}$ to prune centers without performing expensive distance computations, as far as possible.

\subsubsection{Nearest Neighbor Pruning}
\label{sec:firstpruning}

A first pruning rule based on Theorem~\ref{th:range} suggests that a center $c_j$ must be considered (i.e., the search must traverse the related subtree) if $d(q, c_j) \leq d(q, c_{nn}) + 2r$.

\begin{figure}[htb]
  \begin{center}
  \includegraphics[scale=0.8]{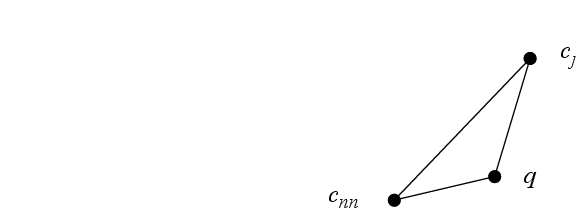}
  \end{center}
  \caption{ Range query}
  \label{fig:rangequeryA}
\end{figure}

Due to the triangle inequality we have:
\[	d(c_{nn}, c_j) \leq d(q, c_{nn}) + d(q, c_j)	\]
Therefore
\begin{eqnarray*}
d(q, c_j) \leq d(q, c_{nn}) + 2r  \Rightarrow  d(c_{nn}, c_j) & \leq &d(q, c_{nn}) + d(q, c_{nn}) + 2r \\
& = & 2 d(q, c_{nn}) + 2r
\end{eqnarray*}

Assuming that $c_{nn}$ is the center with index $i$, we can rewrite this as
\[
d(q, c_j) \leq d(q, c_{nn}) + 2r  \Rightarrow  d_{ij} \leq  2 \cdot d_{min} + 2r
\]
Hence we can retrieve all centers $c_j$ fulfilling $d_{ij} \leq  2 \cdot d_{min} + 2r$ and check whether they also fulfill $d(q, c_j) \leq d_{min} + 2r$. Other centers can be ignored. This requires an expensive distance calculation $d(q, c_j)$.

Let $T = \{c_j \in C \sothat d_{ij} \leq  2 \cdot d_{min} + 2r\}$. Can we infer for some elements of $T$ that $d(q, c_j) \leq d_{min} + 2r$ ? We know (triangle)
\begin{equation*}
                 d(q, c_j) \leq d_{min} + d_{ij}
\end{equation*} 

Therefore
\begin{equation*}
d_{ij} \leq 2r    \Rightarrow    d(q, c_j) \leq d_{min} + 2r
\end{equation*} 

In summary, we can retrieve all elements of $C$ within distance $2 \cdot d_{min} + 2r$ from $c_{nn}$. The elements fulfilling $d(q, p) \leq d_{min} + 2r$ must be among them. For those elements $c_j$ retrieved for which $d_{ij} \leq 2r$ holds, we do not need to evaluate the distance to $q$; they are guaranteed to fulfill $d(q, c_j) \leq d_{min} + 2r$. For the remaining elements, we need to check this condition.

\subsubsection{MaxDist Pruning}
\label{sec:maxdistpruning}

A second pruning rule uses the maximal distance of any element in the partition $S_j$ from center $c_j$ stored as the radius $r_j$. Considering a center $c_j$, we can distinguish the following cases:

\begin{enumerate}
\item $r$ is too small to reach $c_j$ (for a leaf) or any element in the partition $S_j$ (for an inner node). We can prune $c_j$ or $T(S_j)$.
\item $r$ is so large that it definitely includes $c_j$ or any element in $S_j$. We can report $c_j$ or $S_j$ without distance evaluations.
\item We need to evaluate the distance $d(q, c_j)$ (for a leaf) or search the subtree $T(S_j)$ (for an inner node).
\end{enumerate}

These cases can be determined as follows. Note that $d(q, c_i) = d_{min}$, the distance to the nearest neighbor of $q$.

\begin{figure}[thb]
\begin{center}
\begin{tabular}{ccc}
\includegraphics[scale=0.8]{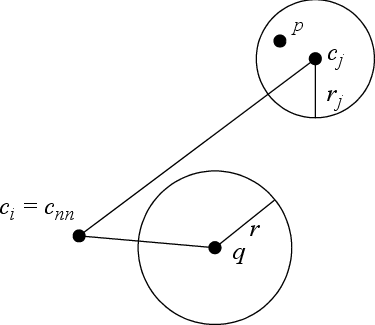}
& \q\q\q\q\q&
\includegraphics[scale=0.8]{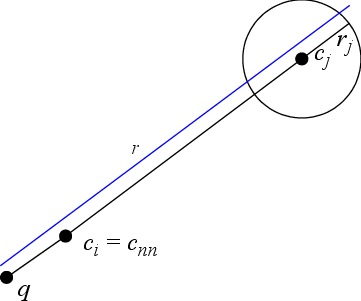} \\
(a) &  & (b)
\end{tabular}
\end{center}
  \caption{ MaxDist Pruning (a) small $r$ (b) large $r$}
  \label{fig:maxDist}
\end{figure}

\begin{enumerate}
\item $r < d_{ij} - d_{min} - r_j$ (Figure~\ref{fig:maxDist} (a))
\item $r \geq d_{ij} + d_{min} + r_j$ (Figure~\ref{fig:maxDist} (b))
\item Otherwise, distance or subtree needs to be evaluated.
\end{enumerate}
In case of a leaf, we can simply set $r_j = 0$.

\subsubsection{Algorithm \emph{rangeSearch1}}

The algorithm for range search for the ``inside'' case combines the two pruning rules. It is shown as Algorithm~\ref{alg:rangeSearch1}. The following notations are used:

\begin{verse}
   \emph{all}$(c_j) =$ return all elements of the subtree for center $c_j$.
\\ \q
\\ \emph{distance}$(q, c_j) =$
\\ \q \IF $DQ_j$ is defined \THEN \RETURN $DQ_j$ \ELSE \RETURN $d(q, c_j) (*)$ \ENDIF
\end{verse}

The array $DQ_i$ is defined in Algorithm~\ref{alg:closestCenter}; it stores distances already evaluated in that algorithm.

The algorithm returns separately results already found, the closest center, and other centers whose partitions need to be searched recursively. The partition for $c_{nn}$ will be processed again by this algorithm (for the ``inside'' case) whereas the other partitions will be processed by the algorithm of Section~\ref{sec:outside} (for the ``outside'' case).

\begin{algorithm}[ht]
  \caption{\label{alg:rangeSearch1}\ic{rangeSearch1}$(p, q, r)$}
  \KwIn{$p$ -- a node of the N-tree\;
  \q\q $q$ -- a query point\;
  \q\q $r$ -- the search radius}
  \KwOut{$c_i$ -- the closest center to $q$\;
  \q\q $Res$ -- elements within distance $r$ from $q$, to be returned\;
  \q\q $Search$ -- centers for subtrees to be searched}
  $(c_i, d_{min}) := \ic{closestCenter}(p.C, q, p.D, p.\{p_1, p_2\}, p.PD)$\;
  $Res := \emptyset$\;
  $Search := \emptyset$\;
  \If {$p$ is a leaf} {
    \ForEach{$c_j \in p.C$}{
      \lIf {$d_{ij} + d_{min} \leq r$} 
        {$Res := Res \cup \{c_j\}$}
      \Else {
        \If {$d_{ij} - d_{min} \leq r$} {
          \lIf {$\ic{distance}(q, c_j) \leq r$} {$Res := Res \cup \{c_j\}$}        
        }
      }
    } 
  }
  \Else { 
    \ForEach {$c_j \in p.C \setminus \{c_i\}$}{
      \lIf {$d_{ij} + d_{min} + r_j \leq r$}{$Res := Res \cup \ic{all}(c_j)$}
      \Else {
        \If {$(d_{ij} - d_{min} - r_j \leq r) \wedge (d_{ij} \leq 2d_{min} + 2r)$} {        
          \lIf {$d_{ij} \leq 2r$}  {$Search := Search \cup \{c_j\}$}
          \Else {
            \lIf {$\ic{distance}(q, c_j) \leq d_{min} + 2r$}  {$Search := Search \cup \{c_j\}$}
          }                
        }
      }    
    }
  }
  \Return $(c_i, Res, Search)$
\end{algorithm}

\begin{algorithm}[ht]
  \caption{\label{alg:rangeSearch2}\ic{rangeSearch2}$(p, q, r)$}
  \KwIn{$p$ -- a node of the N-tree\;
  \q\q $q$ -- a query point\;
  \q\q $r$ -- the search radius\;
  }
  \KwOut{the set of elements within distance $r$ from $q$}
let $C = \langle c_1, \dots, c_k\rangle$ be the centers of node $p$\;
$Res := \emptyset$\;
$C' := \emptyset$\;
$d_{min} := \infty$\;
\While {$\mathit{not}(\mathit{isempty}(C))$} {
  $c_i := \mathit{first}(C); C := \mathit{rest}(C)$\;
  $u := d(q, c_i)$ (*)\;
  \lIf {$u < d_{min}$}  {$d_{min} := u$}
  \If {$p$ is a leaf} {
    $Res := Res \cup \mathit{prune}(C, c_i, u, q, r, leaf)$\;
    \lIf {$r > u$} {$Res := Res \cup \{c_i\}$}
  }
  \Else {
    $Res := Res \cup \mathit{prune}(C, c_i, u, q, r, inner)$\;
    \If {$r > u + r_i$} {$Res := Res \cup \mathit{all}(c_i)$;}
    \ElseIf {$r > u - r_i$}
      {$C' := C' \cup \{(c_i, u)\}$;}
  }
}
$C' := \{(c, u) \in C' | u \leq d_{min} + 2r\}$\;
$Res := Res \cup \bigcup_{(c, u) \in C'} \mathit{rangeSearch2}(node(c), q, r)$\;
\Return $Res$
\end{algorithm}

\begin{algorithm}[ht]
  \caption{\label{alg:prune}\ic{prune}$(C, c_i, u, r, nodetype)$}
  \KwIn{
	$C$ - a set of candidate centers\;
	\q\q $c_i$ - a selected center with distance $u$ from query point $q$\;
	\q\q $u = d(q, c_i)$\;
	\q\q $r$ - the query radius\;
	\q\q $nodetype \in \{leaf, inner\}$\;
  {\SIDEEFFECT}: some subtrees or elements are removed from $C$ and their elements reported 
  \\ \q\q when appropriate
  }
  \KwOut{a set of elements within distance $r$ from $q$}
$Res := \emptyset$\;
\If {nodetype = leaf}{
  \For {$c_j \in C$}{
    \If {$r > u + d_{ij}$} 
    	{$Res := Res \cup \{c_j\}$; $C := C \setminus \{c_j\}$;} 
    \ElseIf {$r < |u - d_{ij}|$}
 	{$C := C \setminus \{c_j\}$;}
  }
}
\Else {
  \For {$c_j \in C$}{
    \If {$r > u + d_{ij} + r_j$}{$Res := Res \cup \mathit{all}(c_j)$; $C := C \setminus \{c_j\}$;}
    \ElseIf {$r < |u - d_{ij}| - r_j$}
      {$C := C \setminus \{c_j\}$;}
  }
}
\Return $Res$
\end{algorithm}

\subsection{Query Point Outside Partitioning}
\label{sec:outside}

We now address range searching from ``outside'' a partitioning as illustrated in Figure~\ref{fig:rangequery2}. In this case, it is not effective to determine the closest center initially. Instead, similar as in the algorithm for finding the closest center, we consider the elements of the set of centers sequentially and after each distance evaluation determine elements that can be pruned or reported.

\subsubsection{MaxDist Pruning}

Again we can use the maximal distance stored as radius $r_j$ for subtree $T(S_j)$. The cases to be considered are the same as in Section~\ref{sec:maxdistpruning}, namely

\begin{enumerate}
\item $r$ is so small that we can prune $c_j$ or $T(S_j)$.
\item $r$ is so large that we can report $c_j$ or $S_j$ without distance evaluations.
\item Evaluation is needed.
\end{enumerate}

We have the following cases. We assume $d(q, c_i)$ has just been evaluated and we consider center $c_j$.

\begin{figure}[thb]
\begin{center}
\begin{tabular}{ccc}
\includegraphics[scale=0.7]{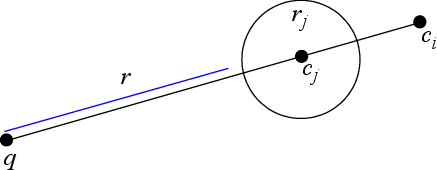}
& \q\q &
\includegraphics[scale=0.7]{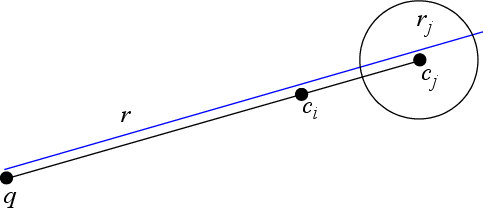} \\
(a) &  & (b)
\end{tabular}
\end{center}
  \caption{ MaxDist Pruning, Case 1: $d(q, c_i) \geq d_{ij}$. (a) small $r$ (b) large $r$}
  \label{fig:maxDistCase1}
\end{figure}

\begin{figure}[thb]
\begin{center}
\begin{tabular}{ccc}
\includegraphics[scale=0.7]{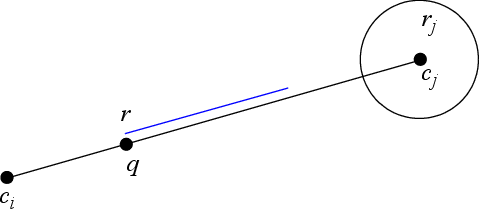}
& \q\q &
\includegraphics[scale=0.7]{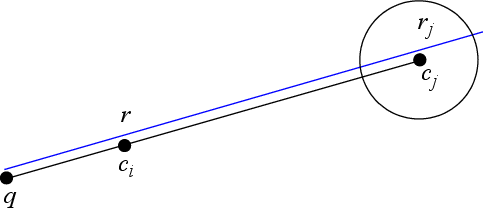} \\
(a) &  & (b)
\end{tabular}
\end{center}
  \caption{ MaxDist Pruning, Case 2: $d(q, c_i) < d_{ij}$. (a) small $r$ (b) large $r$}
  \label{fig:maxDistCase2}
\end{figure}

\begin{itemize}
\item Case 1: $d(q, c_i) \geq d_{ij}$
\begin{enumerate}
\item $r < d(q, c_i) - d_{ij} - r_j$. Subtree $c_j$ can be pruned. See Figure~\ref{fig:maxDistCase1} (a).
\item $r \geq d(q, c_i) + d_{ij} + r_j$. Subtree $c_j$ can be reported. See Figure~\ref{fig:maxDistCase1} (b).
\end{enumerate}
\item Case 2: $d(q, c_i) < d_{ij}$
\begin{enumerate}
\item $r < d_{ij} - d(q, c_i) - r_j$. Subtree $c_j$ can be pruned. See Figure~\ref{fig:maxDistCase2} (a).
\item $r \geq d_{ij} + d(q, c_i) + r_j$. Subtree $c_j$ can be reported. See Figure~\ref{fig:maxDistCase2} (b).
\end{enumerate}
\end{itemize}

\begin{figure}[htb]
  \begin{center}
  \includegraphics[scale=0.7]{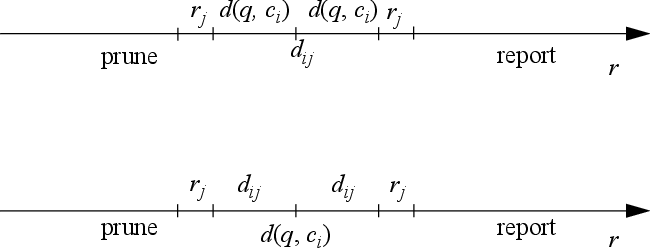}
  \end{center}
  \caption{ Pruning or reporting depending on radius}
  \label{fig:maxDistSummary}
\end{figure}

These results are illustrated in Figure~\ref{fig:maxDistSummary}. Depending on the radius, we can prune or report the subtree for $c_j$ without distance evaluation.

The results can be summarized as follows:

\begin{eqnarray}
r < | d_{ij} - d(q, c_i)| - r_j & \Rightarrow & \mbox{subtree $c_j$ can be pruned.}\\
r \geq d_{ij} + d(q, c_i) + r_j & \Rightarrow & \mbox{subtree $c_j$ can be reported.}
\end{eqnarray}

For centers in leaves we have the same analysis setting $r_j = 0$.

So the algorithm to process a node will sequentially consider its centers. For a given center $c_i$ it evaluates the distance $d(q, c_i)$ and prunes or reports all subtrees (elements) qualified by the conditions. The remaining centers are processed in the following steps.

\subsubsection{Nearest Neighbor Pruning}

When MaxDist pruning is finished, the distances from $q$ to all remaining centers have been computed. Hence at this time also the center $c_{nn}$ with minimal distance $d_{min}$ is known. Therefore the nearest neighbor pruning condition can be applied. 

Consider the non-pruned centers (call this set $C'$) in inner nodes. If for $c_j \in C'$ the condition

\begin{equation*}
d(q, c_j) \leq d_{min} + 2r
\end{equation*}

does not hold, then we can prune partition $S_j$ because no element of this partition can be within distance $r$ from $q$. If it were close enough, it would have been assigned to partition $S_i$ for $c_i = c_{nn}$ instead.

\subsubsection{Algorithm \emph{rangeSearch2}}

The algorithm can be formulated as shown in Algorithm~\ref{alg:rangeSearch2}. It uses a subalgorithm \emph{prune} shown as Algorithm ~\ref{alg:prune}. The notation
\emph{node}$(c_j)$ refers to the root node of the subtree associated with center $c_j$.

% algorithm rangeSearch2

The statement marked with (*) is the only one where an expensive distance evaluation is performed.

% algorithm prune

\subsection{Range Search Algorithm}

The overall algorithm \emph{rangeSearch} is 
% called
invoked on the root node of the N-tree. It starts by
applying Algorithm \emph{rangeSearch1} on the root node for the ``inside'' case. On lower levels it uses either \emph{rangesearch1} or \emph{rangeSearch2} depending on whether the query point is inside or outside the partitioning. The overall algorithm is shown as Algorithm~\ref{alg:rangeSearch}.

\begin{algorithm}[ht]
  \caption{\label{alg:rangeSearch}\ic{rangeSearch}$(p, q, r)$}
  \KwIn{$p$ -- a node of the N-tree\;
  \q\q $q$ -- a query point\;
  \q\q $r$ -- the search radius
  }
  \KwOut{the set of elements within distance $r$ from $q$}
  $Res := \emptyset$\;
  $(c_{nn}, Result, Search) := \ic{rangeSearch1}(p, q, r)$\;
  \If {$p$ is a leaf} { $Res := Result$; }
  \Else {
    $Res := Res \cup \ic{rangeSearch}(node(c_{nn}), q, r)$\;
    $Res := Res \cup Result$\;
    $Res := Res \cup \bigcup_{c \in Search} \ic{rangeSearch2}(node(c), q, r)$\;
  }
  \Return $Res$
\end{algorithm}

\section{k-Nearest-Neighbor Queries}
\label{sec:knn}

Apart from range search, another important query addressed by metric indexes is to identify the \emph{k-nearest-neighbors (kNN)} from a query point. To  process a \emph{kNN} query, one of the three approaches described below is generally taken: \cite{ChenIndexingmetricSpaceSurveyPaper}
\begin{enumerate}
    \item Range search is performed several times starting with a very small radius and then the search radius is increased gradually until \emph{k-nearest-neighbors} are found.
    
    \item The search radius is initially set to infinity and then objects in the indexes are visited in the order of increasing distance to the query point, where the search radius is gradually tightened. This is the most commonly taken approach.
    
    \item A set of candidate objects is determined and the distance from $q$ to its $k^{th}$ nearest neighbor in this set is determined. Then a range search with this distance is performed. 
\end{enumerate}

In the N-tree,  the third approach is used. Starting from a small radius, the radius is gradually increased until we find an approximate radius $r_{approx}$ from the query point within which it is guaranteed that all the \emph{k-nearest-neighbors} lie (possibly along with other points). Then \emph{range search} is employed only once with $r_{approx}$, from which the \emph{k-nearest-neighbors} are obtained. The \emph{kNN} technique is shown in Algorithm \ref{algo:findkNN}.

\begin{algorithm}[ht]
\caption{\ic{kNN}$(root, q, k, DE)$}
\label{algo:findkNN}
  \KwIn{$root$ -- the root node of the N-tree\;
  \q\q $q$ -- a query object\;
  \q\q $k$ -- the number of neighbors to find\;
  \q\q $DE$ -- the distance estimate DE to choose\;}
  \KwOut{$k$ nearest neighbors from $q$}
$approxRadius := \ic{getApproxRadius}(root, q, k, DE)$\;
$Res_1 := \ic{rangeSearch}(root, q, approxRadius)$\;
$Res_2 := \emptyset$\;
\For{$c_i \in Res_1$}{
    $dist_i := distance(q, c_i)$\;
    $Res_2 := Res_2 \cup (c_i, dist_i)$\;
}    
Sort $Res_2$ in increasing order of $dist$\;
$finalResult$ := first $k$ elements of $Res_2$\;
\Return $finalResult$ \;
\end{algorithm}

The backbone of the \emph{kNN} algorithm is to find the approximate radius within which it is guaranteed that all the \textit{k-nearest-neighbors} lie (Line 1). We propose algorithm \emph{getApproxRadius} (Algorithm \ref{algo:getApproxRadius}) to approximate the radius.

\begin{algorithm}[ht]
\caption{\ic{getApproxRadius}$(root, q, k, DE)$}
\label{algo:getApproxRadius}
  \KwIn{$root$ -- the root node of the N-tree\;
  \q\q $q$ -- a query point\;
  \q\q $k$ -- the number of neighbors to find\;
  \q\q $DE$ -- distance estimate DE to choose\;
  }
  \KwOut{radius within which all the k points are guaranteed to lie}
$Q := priorityQueue()$\;
$Q.enqueue((root, 0, true))$ /* $isInside$ is set to $true$ */\;  
$r_{approx} := -1$\;
$pointsVisited := 0$\;
\While {$Q \neq \emptyset$} {
    $(n, dist, isInside) := Q.dequeue()$\;
    \If {n is a data object} {
        $r_{approx} := max(r_{approx}, dist)$\;
        $pointsVisited := pointsVisited + 1$\;
        \If{pointsVisited = k}{
            \textbf{break}\;
        }
    }
    \Else (/* $n$ is an internal node or a leaf */) {
        let $n.C = \{c_1, ..., c_m\}$\;
        $(c_i, d_x) := \ic{chooseCenter}(n, q, isInside)$ \;
        $Q.enqueue((c_i, d_x, isInside))$ \;
        \If{$n$ is an internal node}{
            $Q.enqueue((node(c_i), d_x - r_i, isInside))$ \;
        }
        \For{$j \in \{1, ..., m\} \setminus \{i\}$}{
            $Q.enqueue((c_j, d_x + d_{ij}, false))$ \;
            \If{$n$ is an internal node}{
                $Q.enqueue((node(c_j), de(d_x, d_{ij}, r_j, DE), false))$\;
            }
        }
    }
}
\Return $r_{approx}$ \;
\end{algorithm}

The idea of the algorithm is as follows. A priority queue $PQ$ is maintained which keeps objects (centers within internal nodes or entries in leaves) as well as nodes ordered by some estimate of the distance from the query point $q$. $PQ$ is initialized with the root node, at distance 0.

Then in a loop, elements are removed from $PQ$ and possibly other elements added. In each step, the first element (with the minimal distance estimate) is removed (dequeued). 

If the element is a node (internal or leaf), then its entries (centers) are entered into $PQ$ with a distance estimate. If it is an internal node, then for each center also the related subtree (node) is entered into $PQ$ with another distance estimate.

If the element is an object (center or leaf entry), the object is counted (counting objects up to $k$) and the maximum distance estimate from $q$ of this and the previously encountered objects is maintained. When the $k$-th object is encountered, the current maximal distance estimate is returned by the algorithm.

The remaining question is how distances are estimated for objects or nodes.  Of course, we want to use the precomputed distances between centers as well as the radii of subtrees available in nodes.

Let us first consider objects, i.e., centers or leaf entries. The general strategy is to determine the distance $d_x$ from $q$ to one center $c_i$ precisely and to estimate the distance to other centers $c_j$ using the precomputed distance $d_{ij}$. For these other centers the true distance can lie between $d_x - d_{ij}$ and $d_x + d_{ij}$ (triangle inequality). For the estimate, it is necessary to use the \emph{maximal} distance 
$d_x + d_{ij}$ because the distances for objects determine the returned range query radius and we must ensure that the $k$ objects found lie within this radius.

For the selection of the one center we distinguish - as for range queries - the two cases (i) the query point lies inside the partitioning of this subtree, or (ii) it lies outside (Figures~\ref{fig:rangequery} and \ref{fig:rangequery2}). If the query point lies inside the partition, we select the closest center. If it lies outside, we randomly select some center. This is because in Section~\ref{sec:range} we have seen that using the closest center is not efficient in the ``outside'' case.

Regarding the distance estimates for subtrees, for the subtree belonging to the selected center $c_i$ we use estimate $d_x - r_i$. This is motivated by the fact that $d_x$ is a precise distance and we estimate elements of the subtree by their minimal possible distance.

For the other subtrees, it is not so clear. We need to combine the available information $d_x, d_{ij}$, and $r_j$. Distances of elements of the subtree $node(c_j)$ may lie between $|d_x - d_{ij}| - r_j$ and $d_x + d_{ij} + r_j$. It is not clear whether subtrees should be handled (dequeued) as early or as late as possible or at some ``most likely'' distance. We have finally set up the nine distance estimates shown in Table~\ref{tab:Distance Estimates} and evaluated the performance experimentally in \cite{ntree_technical_report}. It turned out that there are considerable differences in performance for the different estimates and that DE3 was the most efficient, that is, $max(d_x, d_{ij}) - r_j$.

Note that the choice of distance estimates for subtrees does not affect the correctness of the \emph{kNN} algorithm which only depends on the fact that $k$ objects are found that are guaranteed to lie in the range query radius $r_{approx}$. It only affects execution time.

\begin{table}[htb]
  \centering
  \caption{Distance Estimates $de(d_x, d_{ij}, r_i, DE)$}
  \label{tab:Distance Estimates}
  \scalebox{.9}{
    \begin{tabular}{lr}
    \hline
    \multicolumn{1}{c}{\textbf{DE0}} & \multicolumn{1}{c}{\textbf{$|d_x - d_{ij}| - r_j$}}\\ 

    \multicolumn{1}{c}{\textbf{DE1}} & \multicolumn{1}{c}{\textbf{$|d_x - d_{ij}|$}}\\ 

    \multicolumn{1}{c}{\textbf{DE2}} & \multicolumn{1}{c}{\textbf{$|d_x - d_{ij}| + r_j$}}\\ 

    \multicolumn{1}{c}{\textbf{DE3}} & \multicolumn{1}{c}{\textbf{$max(d_x, d_{ij}) - r_j$}}\\ 

    \multicolumn{1}{c}{\textbf{DE4}} & \multicolumn{1}{c}{\textbf{$max(d_x, d_{ij})$}}\\ 

    \multicolumn{1}{c}{\textbf{DE5}} & \multicolumn{1}{c}{\textbf{$max(d_x, d_{ij}) + r_j$}}\\ 

    \multicolumn{1}{c}{\textbf{DE6}} & \multicolumn{1}{c}{\textbf{$d_x + d_{ij} - r_j$}}\\ 

    \multicolumn{1}{c}{\textbf{DE7}} & \multicolumn{1}{c}{\textbf{$d_x + d_{ij}$}}\\ 

    \multicolumn{1}{c}{\textbf{DE8}} & \multicolumn{1}{c}{\textbf{$d_x + d_{ij} + r_j$}}\\ 

\hline
    \end{tabular}}
\end{table}

To control whether for a given node the closest center or a random one should be selected as $c_i$, Algorithm~\ref{algo:getApproxRadius} stores triples rather than pairs in the priority queue $PQ$ where the third component is a Boolean value called $isInside$. For the root node it is clear that $isInside$ is true, so this is entered in the triple for the root node. The parameter is passed into triples stored for subtrees in $PQ$. Exactly for one path from the root to a leaf, $isInside$ is true; for all other subtrees $false$ is passed into the triple. On dequeueing a node, Algorithm \ref{algo:chooseCenter} uses the $isInside$ parameter to select the closest or a random center.

\begin{algorithm}[htb]
\caption{\ic{chooseCenter}$(p, q, isInside)$}
\label{algo:chooseCenter}
  \KwIn{$p$ -- a node of the N-tree\;
  \q\q$q$ -- a query point\;
  \q\q $isInside$ -- a Boolean value identifying whether $q$ lies inside or outside of partition\;
  }
  \KwOut{a center $c_i$\;
    \q\q $d_x = distance(q, c_i)$ }
    \If{$isInside$}{
        $(c_i, d_x) := closestCenter(p.C, q, p.D, p.\{p_1, p_2\}, p.PD)$ \;
    }
    \Else{
        choose a random center $c_i$ from $p.C$ \;
        $d_x := distance(q, c_i)$ \;
    }
\Return $(c_i, d_x)$ \;
\end{algorithm}

\section{Experimental Evaluation}
 \label{sec:evaluation}

 \subsection{Overview}
 \label{subsection:evaluation_overview}
 
In this section, we evaluate both N-tree as well as the new distance measures, i.e., precise and approximated integral-based distance computations. The N-tree metric index is compared with two other popular metric indexes, GNAT~\cite{Brin1995Near} and MVPT~\cite{Bozkaya1999Indexing}. GNAT and MVPT have been found in a recent survey \cite{ChenIndexingmetricSpaceSurveyPaper} to be among the best performing main memory metric indexes. The exact integral-based distance function (denoted as \emph{DistanceAvg}) is compared with Hausdorff distance, which is one of the most commonly used metric distance measures for trajectories. In addition, we also perform a comparison between the exact and the approximate average distance functions.

We employ four real-world and one synthetic dataset as well as five different metric distance functions as shown in Table \ref{tab:Dataset_info}. 

\begin{table}[htb]
\caption{Datasets used in the Experiments    }
\label{tab:Dataset_info}
\begin{tabular}{lllrll}
\toprule
Dataset              & \begin{tabular}[l]{@{}l@{}}Distance \\ Function\end{tabular}    & Object Type & \begin{tabular}[l]{@{}l@{}}Number\\ of Objects\end{tabular} & \begin{tabular}[l]{@{}l@{}}(Avg.) \\ Object Size\end{tabular} & Total Size                                                              \\
\toprule
\emph{Trips}       & \begin{tabular}[l]{@{}l@{}}Hausdorff or \\ DistanceAvg\end{tabular} & Trajectory  & 550,841                                                     & 37.7 points                                                   & 20,790,341 points                                                       \\ \midrule
\emph{X-Rays}      & $L_1$-Norm                                                         & Image       & 55,000                                                      & 32 x 32 pixels                                                & 56,320,000 pixels                                                       \\ \midrule
\emph{Sentences}   & Jaccard                                                         & Text        & 158,914                                                     & 13.4 words                                                    & 2,127,821 words                                                         \\ \midrule
\emph{Buildings}   & Euclidean                                                       & 2D-points   & 3,257,397                                                   & 1 point                                                             &  3,257,397 points                                                      \\ \midrule
\emph{Hermoupolis} & \begin{tabular}[l]{@{}l@{}}Hausdorff or \\ DistanceAvg\end{tabular} & Trajectory  & 1,000                                                      & 5,000 points                                                & 5,000,000 points \\
\bottomrule
\end{tabular}
\end{table}

The \emph{Trips} dataset represents  550,841 trips of New York taxis. On average, each trip consists of around 38 points and the total number of points is over 20 million, where each point represents a 3D-coordinate of the form $(lat, long, time)$. The distance between two trips is measured using Hausdorff Distance or  \emph{DistanceAvg} depending on the evaluation criteria. 

The \emph{Trips} dataset is semi-real in the sense that it is based on original taxi trip data.\footnote{https://www1.nyc.gov/site/tlc/about/tlc-trip-record-data.page} The original data set includes for each trip pick-up and drop-off dates/times and pick-up and drop-off locations. From these data, we have created\footnote{http://newton2.fernuni-hagen.de/secondo/download/TripsCM200R.tar} for each trip a continuous trajectory of data type \tc{mpoint} (see Section~\ref{sec:distance}) by (i) applying a shortest path algorithm on the road network of New York to create a path from origin to destination location, (ii) determining the average speed given by distance traversed and time difference, and (iii) creating an \tc{mpoint} following this path using this average speed on all segments (units of the \tc{mpoint}).

\textit{X-Rays}\footnote{https://nihcc.app.box.com/v/ChestXray-NIHCC} represents 55,000 de-identified images of chest X-Rays \cite{chest-xray-dataset} in PNG format provided by the National Institute of Health (NIH) Clinical Center. The distance between two images is measured using $L_1$-norm. For indexing and application of the distance function images have been scaled down to a size of
32 x 32 pixels, thus totalling to over 56 million pixels in the dataset.

\emph{Sentences}\footnote{https://www.kaggle.com/datasets/hsankesara/flickr-image-dataset} represents around 159,000 annotated sentences as captions for about 32,000 images, where five captions were stored for each image. Each sentence is comprised of around 13 words on average, totalling to over 2 million words in the dataset. The distance between two sentences (or captions) is measured using Jaccard Distance (on the sets of words of the two sentences). 

\emph{Buildings} represents 2D locations of 3,257,397 buildings of the Niedersachsen state of Germany obtained from the Open Street Map dataset.\footnote{https://download.geofabrik.de/europe/germany/niedersachsen-latest-free.shp.zip} The motivation was to have a large real 2D point dataset, so the original polygonal shape of a building has been reduced to a point by taking the center of the bounding box. The distance between buildings is measured using Euclidean Distance.

\emph{Hermoupolis} is a pattern and semantic-aware synthetic trajectory generator, which is able to produce realistic semantic trajectory datasets \cite{Hermapoulis_Springer, Hermapoulis_ACM}. Using this dataset, we generated 1,000 trajectories, each having 5,000 points on an average. Similar to \emph{Trips} the distance between two trajectories is measured by either using Hausdorff Distance or \emph{DistanceAvg}. This dataset is used only to compare the performance of \emph{DistanceAvg} against Hausdorff as shown in Section \ref{subsection:avg_dist_vs_hausdorff_eval}. In the remaining experiments, all four real-world datasets are used.

The four datasets vary considerably in cardinality (about 50000 to 3 million), object size (2d point to trajectory), cost of distance function (Euclidean distance on points to Hausdorff or \emph{DistanceAvg}), and distance distributions (see the next subsection). Experiments on these datasets therefore provide a good overview of the behaviour of the index structures in different application scenarios.

We break the evaluation into several sections. First, we try to understand the geometry of all four of the real-world datasets by measuring their distance distributions. This is important because the performance of metric indexes highly depends on the data distribution. Second, for each of the metric index structures, we compare the time taken to build it over each of the datasets by using the corresponding distance functions. Third, we evaluate the performance of the N-tree  against all the other structures on \emph{range} and \emph{kNN search}. The performance metrics are the query execution time and the number of distance evaluations and we show that the N-tree outperforms the other structures in these metrics. 

In a separate experiment, we compare the time taken to evaluate  \emph{DistanceAvg} with that of Hausdorff, and show that our distance function takes much less time compared to Hausdorff. This is essential in scenarios where trajectories are long and thus have many points. Our distance function is well-suited in such scenarios as it runs in linear time. Throughout all experiments, the leaf node degree is always maintained to be 100. To identify the other parameters such as the internal node degrees of the various metric index structures that would lead to their best performance, we conducted extensive experiments, which are described in our technical report \cite{ntree_technical_report}. We found that the best performance is achieved with an internal node size of 4 for GNAT and MVPT, and with a node size of 36 for N-tree. Hence, these parameter values are kept constant for all experiments.
 
The three indexes and the associated similarity search algorithms were implemented in Java. Experiments were conducted on an Intel i5-1035G1 CPU having 16GB RAM. Each measurement we report is an average over 100 query points chosen randomly from the respective dataset.

Finally, in a different environment (the \Secondo\ system \cite{GAA+05}) we compare the exact and the approximate average distance function on the \emph{Trips} dataset and the N-tree, varying the approximation radii.

\subsection{Distance Distributions Of The Different Datasets}
\label{subsection:Eval_dist_distribution}

\begin{figure}[ht]
\centering
\subfigure[Trips (Hausdorff)]{%
\label{subfig:trips_hausdorff_data_distribution}%
\includegraphics[height=1.6in]{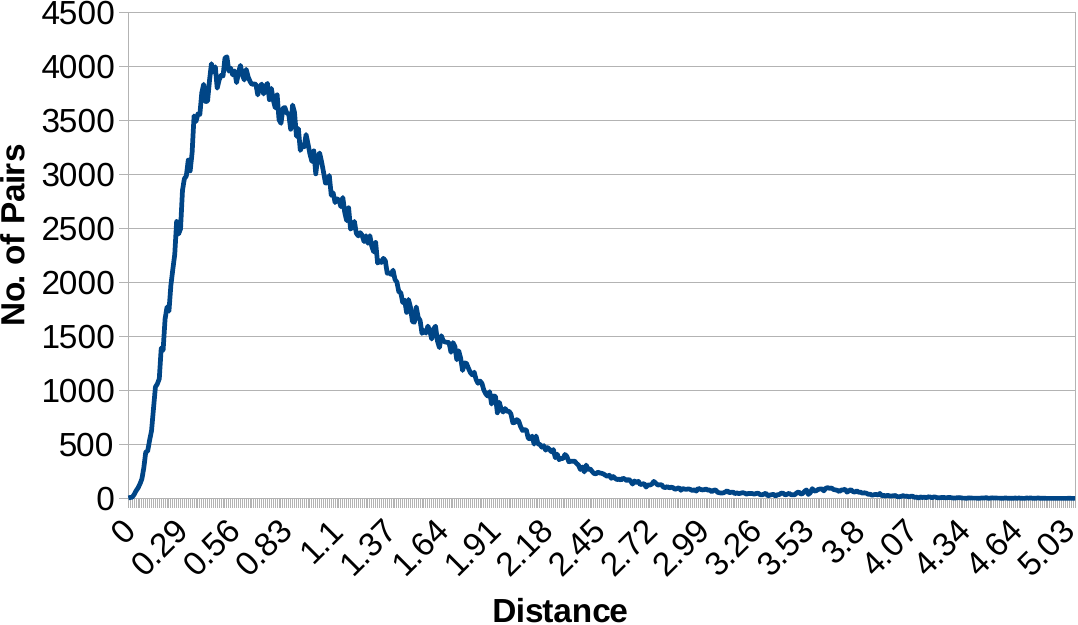}}%
\qquad
\subfigure[Trips (\emph{DistanceAvg})]{%
\label{subfig:trips_avg_dist_data_distribution}%
\includegraphics[height=1.6in]{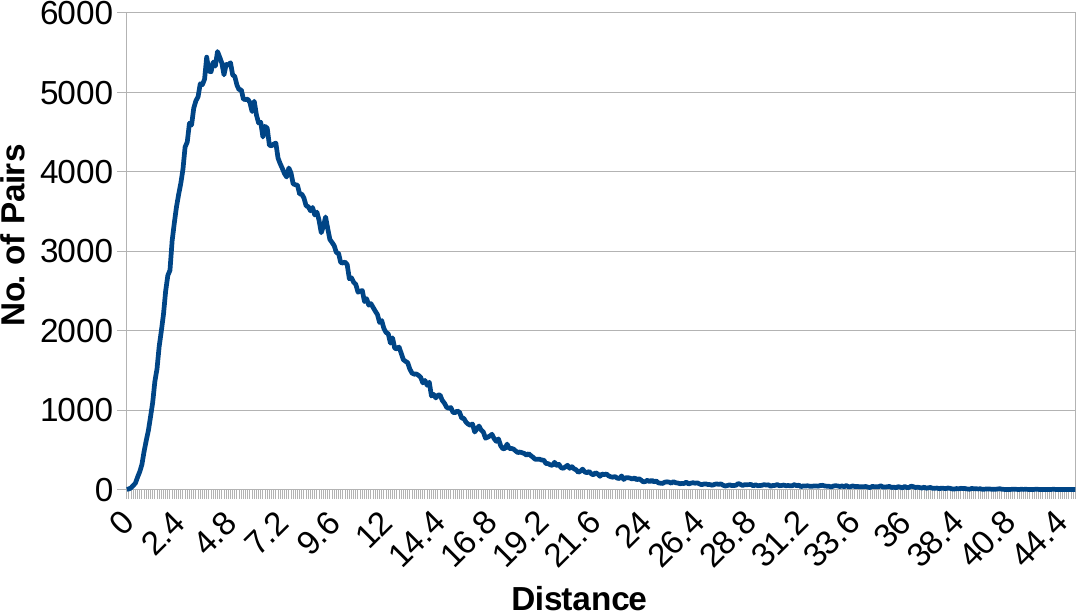}}%
\qquad
\subfigure[X-Rays]{%
\label{subfig:image_data_distribution}%
\includegraphics[height=1.6in]{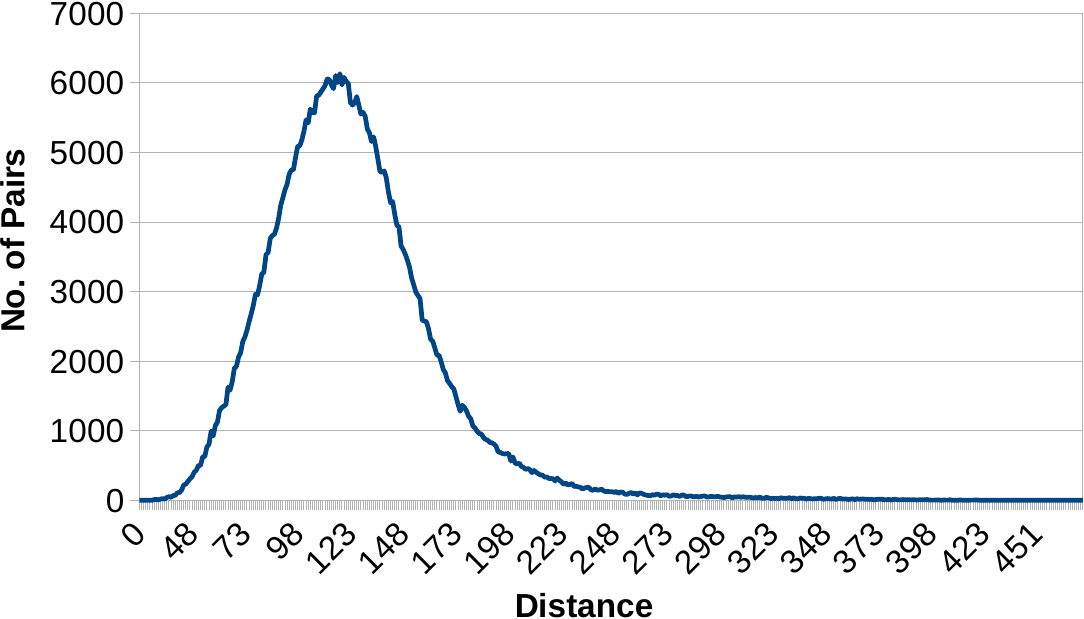}}%
\qquad
\subfigure[Sentences]{%
\label{subfig:sentence_data_distribution}%
\includegraphics[height=1.6in]{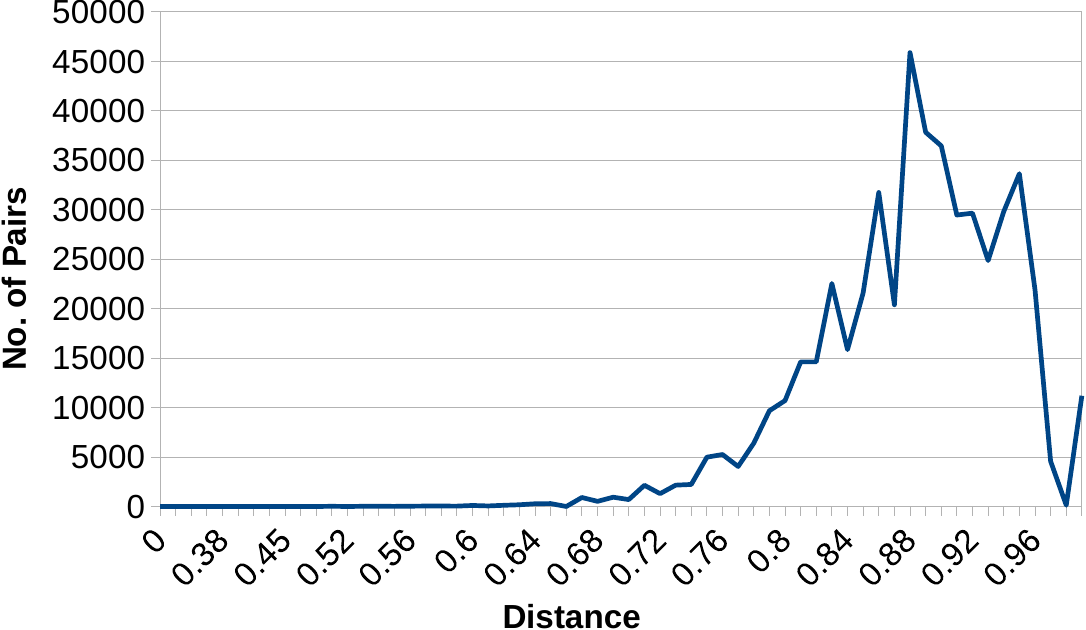}}%
\qquad
\subfigure[Buildings]{%
\label{subfig:buildings_data_distribution}%
\includegraphics[height=1.6in]{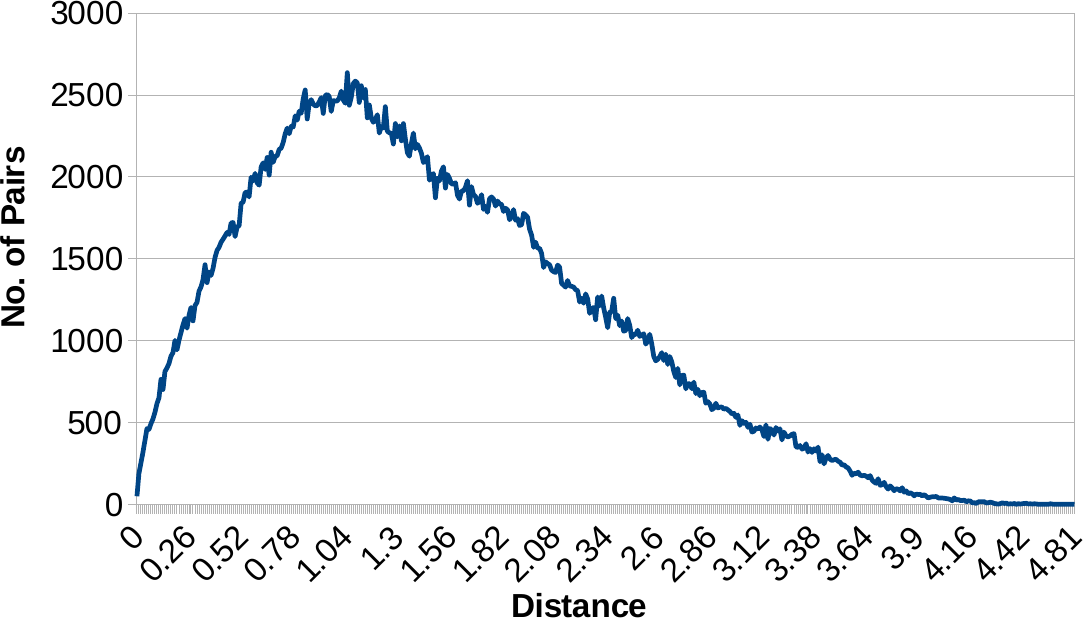}}%
\caption{Distance distributions of the different datasets   }
\label{fig:Data Distribution}
\end{figure}

In order to find the distance distributions of the different datasets, 500,000 unique pairs were randomly selected and the distances between them were calculated using the distance functions shown in Table \ref{tab:Dataset_info}. The distance distributions of the four different datasets obtained are shown in Figure \ref{fig:Data Distribution}. The Y-axis shows the number of data object pairs that have the corresponding distance value (X-axis). For the \emph{Trips} and \emph{X-Rays} dataset, the distributions are similar to a Gaussian curve. For \emph{Trips} the data distribution using both Hausdorff (Fig. \ref{subfig:trips_hausdorff_data_distribution}) and \emph{DistanceAvg} (Fig. \ref{subfig:trips_avg_dist_data_distribution}) are illustrated. Most of the data points are close to each other (since the ``bell'' is very close to the Y-axis) for both of the distance measures. For \emph{X-Rays}, the images are also close to each other, but not as close as in the \emph{Trips} dataset (the ``bell'' is slightly away from the Y-axis). For \emph{Sentences}, we see that most of the captions (or annotations) are highly different from each other (the ``ridges'' of the curve appear almost at the right end), which is different from the behavior that we observe for \emph{Trips} and \emph{X-Rays}. 
Lastly, for \emph{Buildings}, the distribution almost resembles a Gaussian curve.  Till a distance of about 1.3 units, the curve has a steep increase, after which the curve gradually slopes down as the distance increases. This possibly indicates numerous clusters of points within the dataset. Each cluster could indicate a city, town or a village of Niedersachsen. Thus we see that the data distributions for each of the datasets are  quite different from each other which will help us to understand how the indexes behave under different data distributions.

\subsection{Index Construction}
\label{subsection:Eval_index_construction}

\begin{table*}[]
\begin{minipage}[b]{0.5\textwidth}
\centering
	  \includegraphics[height=1.6in]{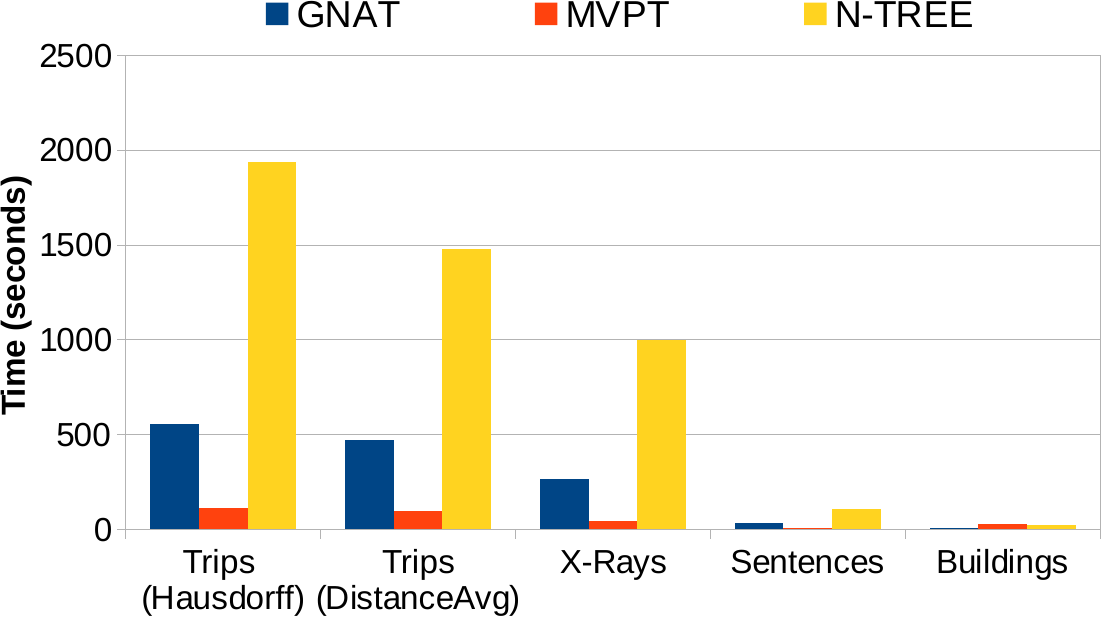}
   \captionof{figure}{Index Construction Times }
   \label{fig:index_build_time}
	\end{minipage}
 \hspace{4pt}%
\begin{minipage}[b]{0.5\textwidth}
\centering
\setlength\tabcolsep{1.5pt} % default value: 6pt
\begin{tabular}{c|c|c|c|}
\cline{2-4}
\multicolumn{1}{l|}{}                                                                        & GNAT & MVPT & N-tree \\ \hline
\multicolumn{1}{|c|}{\begin{tabular}[c]{@{}c@{}}Trips\\ (Hausdorff)\end{tabular}}   & 558.094       & 111.317       & 1939.835        \\ \hline
\multicolumn{1}{|c|}{\begin{tabular}[c]{@{}c@{}}Trips\\ (DistanceAvg)\end{tabular}} & 474.016       & 99.349        & 1479.605        \\ \hline
\multicolumn{1}{|c|}{X-Rays}                                                        & 265.262       & 45.133        & 998.044         \\ \hline
\multicolumn{1}{|c|}{Sentences}                                                     & 35.76         & 8.069         & 106.038         \\ \hline
\multicolumn{1}{|c|}{Buildings}                                                   & 5.848         & 29.757        & 21.6            \\ \hline
\end{tabular}
\caption{Construction Times [seconds]}
\label{tab:index_build_time}
\end{minipage}

\vspace{-14pt}
\end{table*}
In this section we compare the index construction times of all the metric indexes, i.e., GNAT, MVPT and N-tree over all the datasets using their corresponding distance functions. The plot comparing the index building time is shown in Fig. \ref{fig:index_build_time} 
and the actual values are shown in Table \ref{tab:index_build_time}. From the figure and the table, we can see that except for \emph{Buildings}, MVPT always requires the least time to build. GNAT always takes less construction time compared to the N-tree. The construction times for the\emph{Trips} and \emph{X-Rays} datasets are longer compared to the other datasets since the distance functions used for these two datasets (i.e. Hausdorff, \emph{DistanceAvg} and $L_1$-Norm) are expensive and thus require more time to evaluate compared to the Jaccard and Euclidean (for \emph{Sentences} and \emph{Buildings} datasets respectively) distances, which are simpler and faster to evaluate. In \emph{Trips}, the time taken to build  the indexes is less with \emph{DistanceAvg} compared to Hausdorff distance. In Section \ref{subsection:avg_dist_vs_hausdorff_eval}, we will show that the time taken to compute \emph{DistanceAvg} is much less compared to Hausdorff, due to which indexes take less time to build in the former scenario.

Except for \emph{Buildings}, the time taken by the N-tree is comparatively larger than that of GNAT and MVPT. One reason for this is 
its internal node degree. As mentioned before, the internal node degree for N-tree was 36, whereas those of GNAT and MVPT were set to 4. It takes comparatively more distance evaluations to select 36 centers and partition the dataset in 36 disjoint sets, compared to 4 centers (and partitions). When the distance function is expensive, this leads to higher index construction time, as observed in the N-tree. Another reason is the precomputation of pairwise distances of centers within nodes in the N-tree.

In the \emph{Buildings} dataset, whereas the distance function used is \emph{Euclidean} which is cheap to calculate, the index construction time for MVPT gets dominated by the fact that it requires to rearrange the dataset into spherical cuts and create partitions in respect to the vantage points.  This rearrangement is taking time due to the large dataset size compared to the other ones. This is why the N-tree takes  less time to build compared to MVPT in spite of having a large number of distance evaluations.

However, we will see later that in most of the cases, the \emph{range} and \emph{kNN} queries are answered quickly by the N-tree in terms of time taken and with less distance evaluations as it extensively uses the stored distances in each node. So, although the N-tree takes some time to be constructed, once built it can be more efficient in answering queries compared to the other indexes. Moreover, the issue of long construction times can be addressed by parallel computation and by saving and restoring an N-tree index as shown in Section~\ref{sec:distconst}.

\subsection{Comparison Among Different Structures}

 In this set of experiments, the performances of all the indexes were compared by executing \emph{range} and \emph{kNN} queries by varying radius and $k$. The search radius for \emph{range} queries  was varied as shown in Table \ref{tab:Search_Radius_Setting}. For \textit{range search}, the evaluation is further sub-divided into two parts:
 \begin{enumerate}
     \item Keeping the search radius low such that the selectivity is also quite low, i.e., the cardinality of the dataset returned after applying the \textit{range search} is within 750. For each dataset, the third column (Low Radius) of Table \ref{tab:Search_Radius_Setting} denotes the search radii for which the selectivity is low.
     
     \item Varying the search radius from low to high to compare the behavior of all the index structures. The radii are increased such that the  cardinality of the \textit{range search} result varies between 100 to nearly the entire dataset. The fourth column (Low to High Radius) of Table \ref{tab:Search_Radius_Setting} denotes such radii.
 \end{enumerate}
 
 For \emph{kNN} queries, the set of values chosen for $k$ were \{5, 10, 20, 50, 100\}. For both range search and \emph{kNN} search, we compare the running time and the number of distance functions evaluated for N-tree with other structures. Now let us look into the performance of the indexes for each dataset.

\begin{table}[ht]
\caption{Search Radius Settings   }
\label{tab:Search_Radius_Setting}
\begin{tabular}{llll}
\toprule
Dataset                                                                         & \begin{tabular}[l]{@{}l@{}}Distance \\ Function\end{tabular} & Low Radius              & Low to High Radius      \\ \toprule
\multirow{2}{*}{\emph{Trips}}                                                          & Hausdorff                                                             & \{0.02, 0.04, 0.06, 0.08, 0.1\} & \{0.1, 0.5, 1, 1.5, 2\}     \\ \cmidrule{2-4} 
& \emph{DistanceAvg}                                                & \{0.1, 0.2, 0.3, 0.4, 0.5\} [km]     & \{0.5, 5, 10, 15, 20\} [km]           \\ \midrule
\emph{X-rays}                                                                          & $L_1$-norm                                                               & \{10, 20, 30, 40, 50\}       & \{50, 100, 150, 200, 250\}      \\ \midrule
\emph{Sentences}                                                                       & Jaccard                                                               & \{0.2, 0.3, 0.4, 0.5, 0.6\} & \{0.6, 0.675, 0.75, 0.825, 0.9\} \\ \midrule
\emph{Buildings}                                                                       & Euclidean                                                             & \{1, 2, 3, 4, 5\}$\times10^{-3}$         & \{0.005, 1, 2, 3, 4\}    \\ \bottomrule
\end{tabular}
\end{table}

\begin{figure}[ht!]
\centering
\vspace{1em}
\subfigure[Range Search (Low Radius) - run time ]{%
\label{subfig:trips_range_search_exe_time_low}%
\includegraphics[height=1.5in]{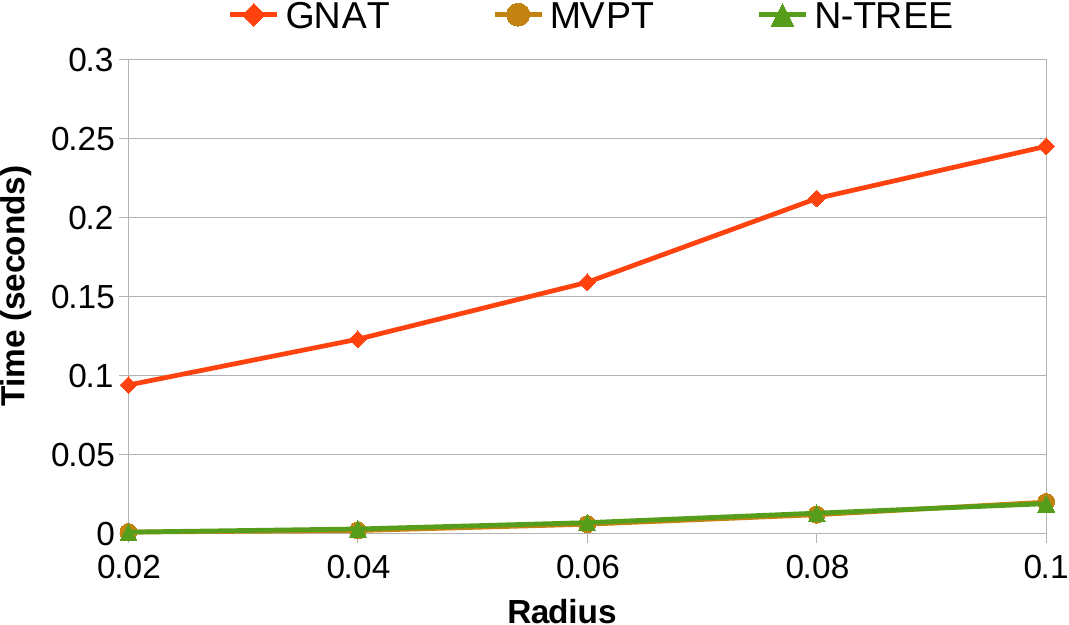}}%
\qquad
\subfigure[Range Search (Low Radius) - distance evaluations]{%
\label{subfig:trips_range_search_dist_eval_low}%
\includegraphics[height=1.5in]{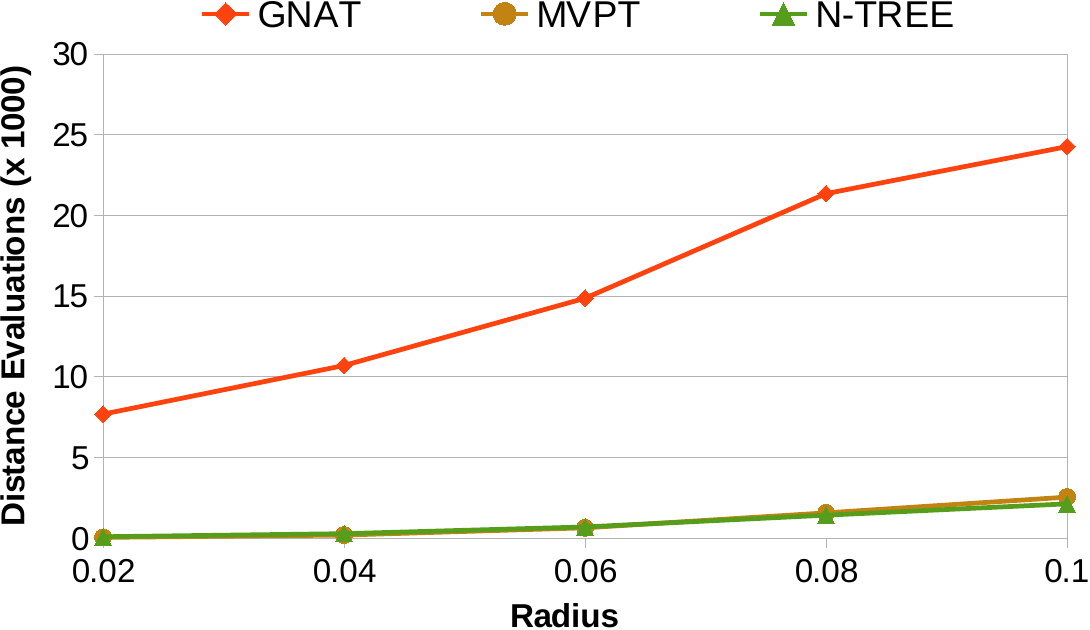}}%
\qquad
\subfigure[Range Search (Low to High Radius) - run time ]{%
\label{subfig:trips_range_search_exe_time_low_high}%
\includegraphics[height=1.5in]{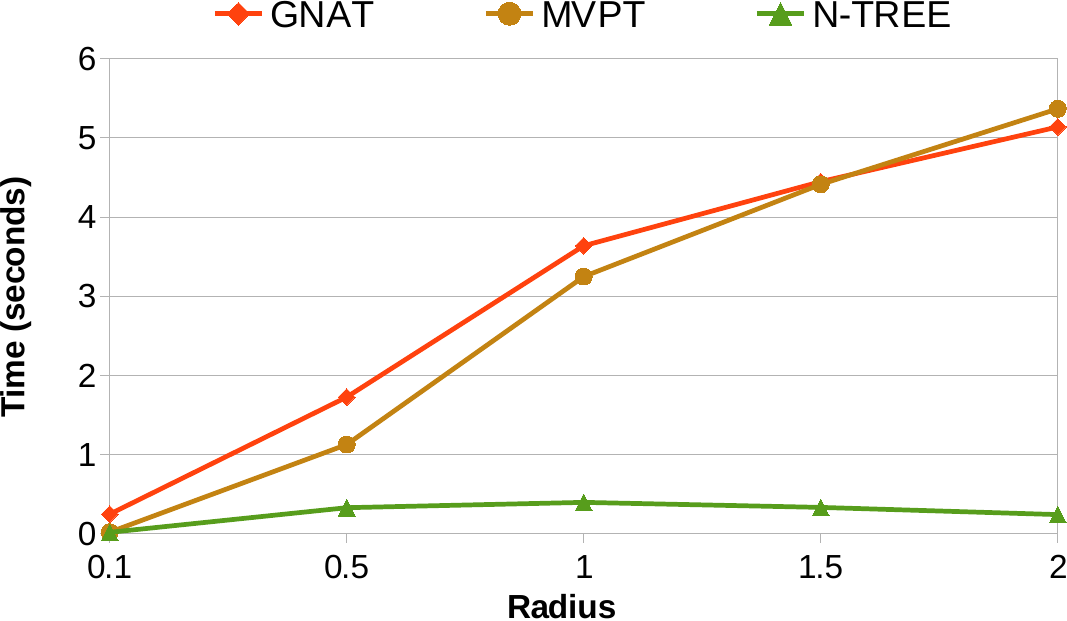}}%
\qquad
\subfigure[Range Search (Low to High Radius) - distance evaluations]{%
\label{subfig:trips_range_search_dist_eval_low_high}%
\includegraphics[height=1.5in]{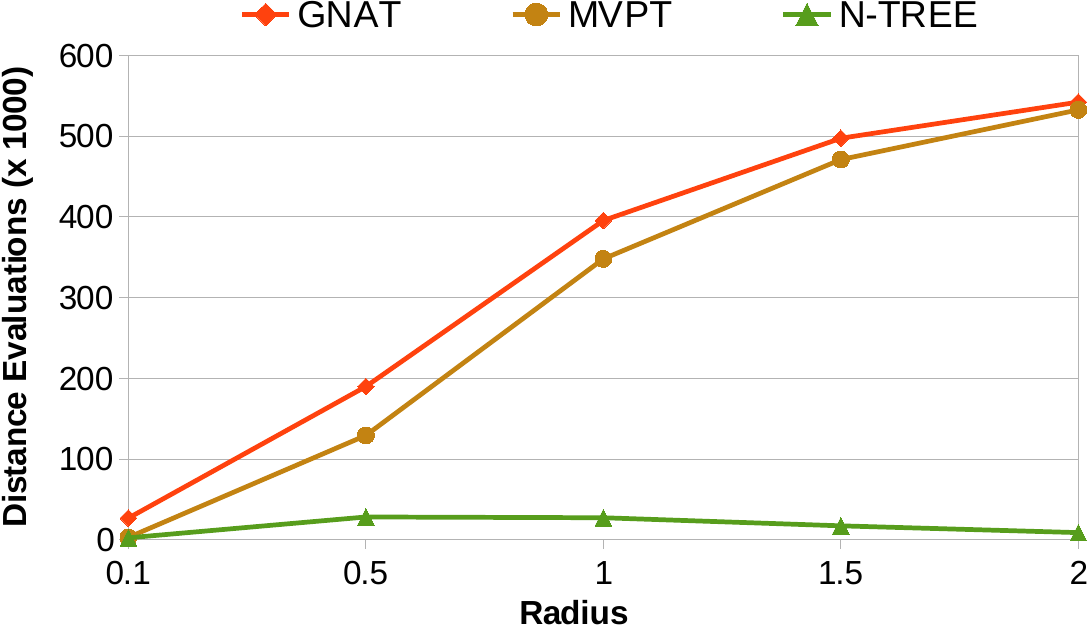}}%
\qquad
\subfigure[kNN Search - run time ]{%
\label{subfig:trips_knn_search_exe_time}%
\includegraphics[height=1.5in]{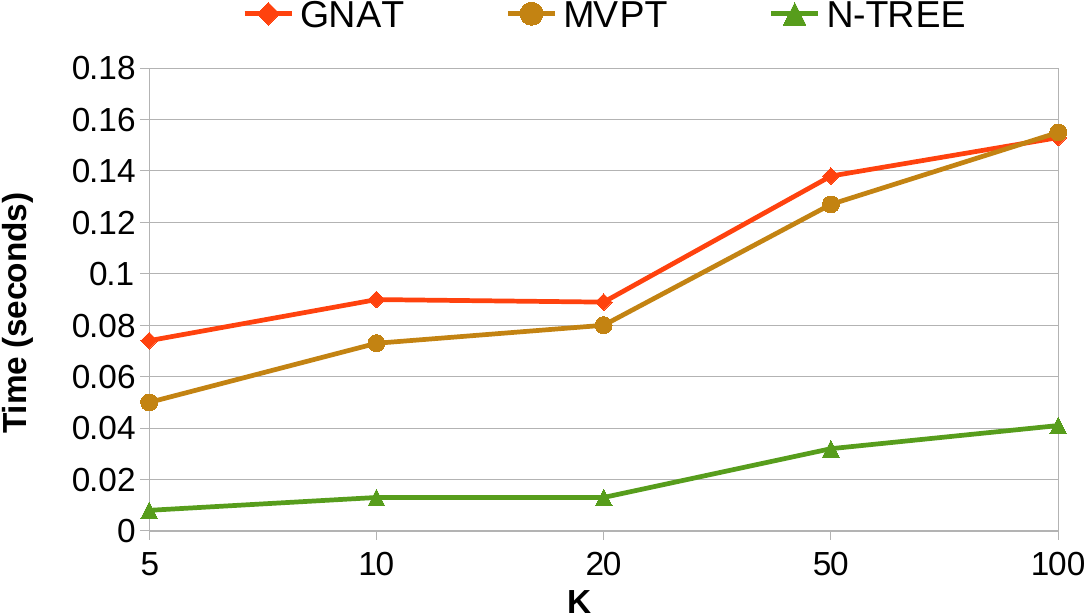}}%
\qquad
\subfigure[kNN Search - distance evaluations]{%
\label{subfig:trips_knn_search_dist_eval}%
\includegraphics[height=1.5in]{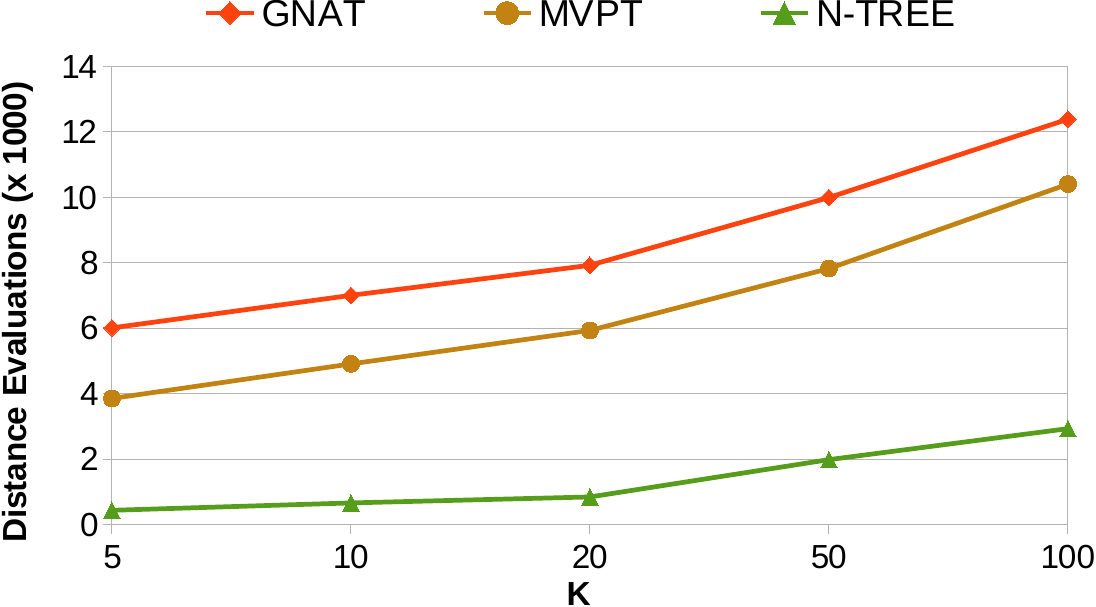}}%
\caption{Trips with Hausdorff Distance   }
\label{fig:trips_hausdorff_eval}
\end{figure}

\subsubsection{Baseline}
\label{sec:baseline}

As a baseline, we note the cost of performing a range query by sequential scan of the respective dataset. The times shown in Table~\ref{tab:baseline} denote the average of the times required for five random elements of the dataset as query objects.

\begin{table}[htb]
  \centering
  \caption{Range Query by Sequential Scan}
  \label{tab:baseline}
    \begin{tabular}{|l|r|r|r|r|r|}
\hline
& Trips (H) & Trips (DA) & X-Rays & Sentences & Buildings \\
\hline
Total time [seconds] & 4.802 & 4.669 & 3.849 & 0.479 & 0.0508 \\
Time per distance eval. [$\mu$s] & 8.717 & 8.476 & 69.981 & 3.014 & 0.015 \\
\hline 
    \end{tabular}
\end{table}

 \subsubsection{Trips with Hausdorff Distance}
 \label{subsubsection:Eval_trips_hausdorff}
 All the metric indexes are evaluated on the \emph{Trips} dataset using Hausdorff distance. The results of \emph{range} and \emph{kNN} search are shown in Fig. \ref{fig:trips_hausdorff_eval}. From Figs. \ref{subfig:trips_range_search_exe_time_low} and \ref{subfig:trips_range_search_dist_eval_low}, 
 which illustrate the effect on range search on low radius, it is observed that the performances of N-tree and MVPT are very similar in terms of both query execution time and distance evaluations, whereas GNAT performs the worst. 
 Looking at the actual numbers, for very low radii, MVPT performs slightly better than N-tree. Till radius 0.06 units, MVPT was the best candidate, after which at radius 0.08, N-tree took the lead. However, the differences are very small.
 
 But as we gradually increase the radius from low to high (Figs. \ref{subfig:trips_range_search_exe_time_low_high} and \ref{subfig:trips_range_search_dist_eval_low_high}), N-tree outperforms the other two indexes both in terms of query execution time and distance evaluations.  A closer look reveals
 that after radius of 0.5 units, the number of distance evaluations decreases with the increase in radius. At radius 0.5 units, the number of distance evaluations was around 28,000 whereas at radius 1 unit it gets reduced to around 27,000 distance evaluations. We call this the \emph{U-turn effect}. 
 
 Lastly, for the \emph{kNN} search, N-tree clearly outperforms the other two metric indexes both in terms of execution time and distance evaluations (Figs. \ref{subfig:trips_knn_search_exe_time} and \ref{subfig:trips_knn_search_dist_eval}).  MVPT performs well for range queries on very low radius if the data points in the dataset are very close to each other (like in the \emph{Trips} dataset) but slowly loses its performance advantage compared to N-tree as the radius increases. On the other hand, the performance of MVPT degrades when the data points are far away from each other, which we will highlight in the later experiments.
 
\begin{figure}[ht]
\centering
\subfigure[Range Search (Low Radius ) - run time ]{%
\label{subfig:trips_avg_dist_range_search_exe_time_low}%
\includegraphics[height=1.5in]{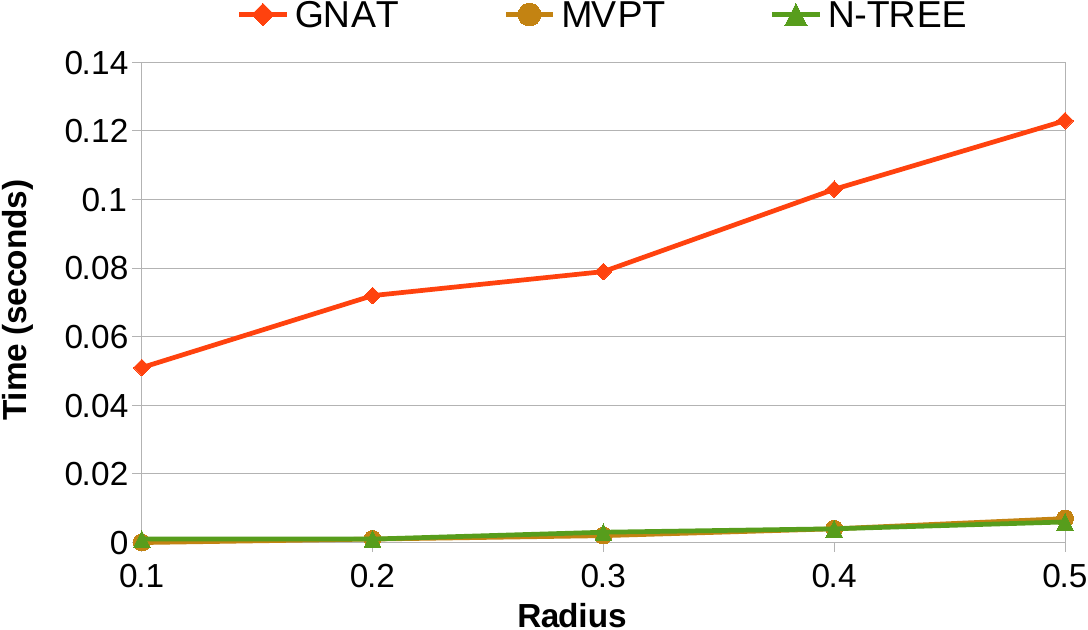}}%
\qquad
\subfigure[Range Search (Low Radius) - distance evaluations]{%
\label{subfig:trips_avg_dist_range_search_dist_eval_low}%
\includegraphics[height=1.5in]{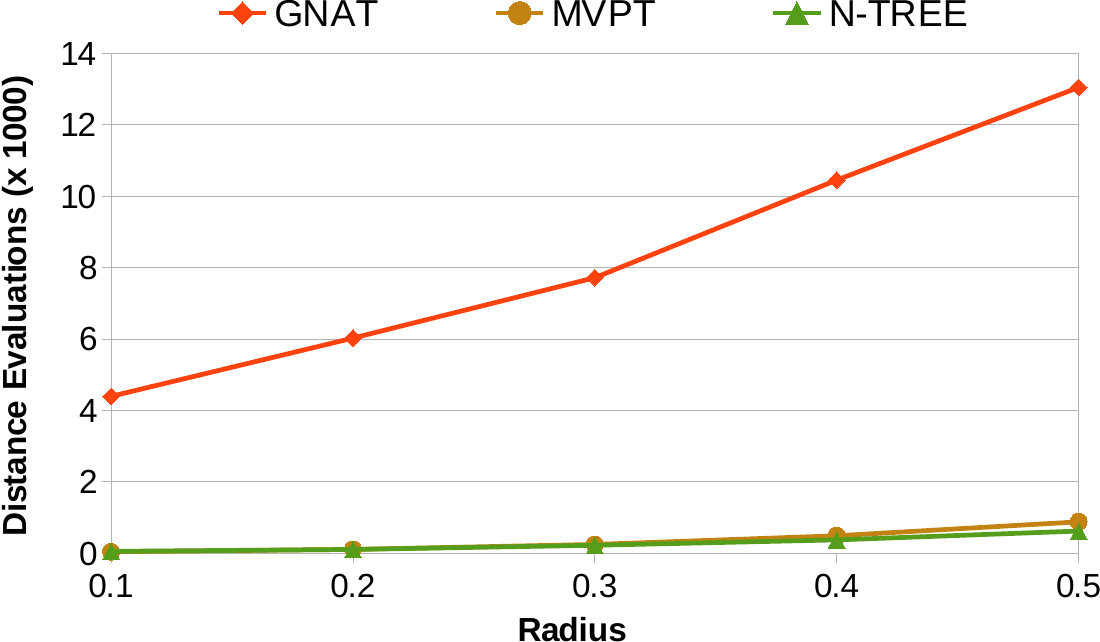}}%
\qquad
\subfigure[Range Search (Low to High Radius) - run time ]{%
\label{subfig:trips_avg_dist_range_search_exe_time_low_high}%
\includegraphics[height=1.5in]{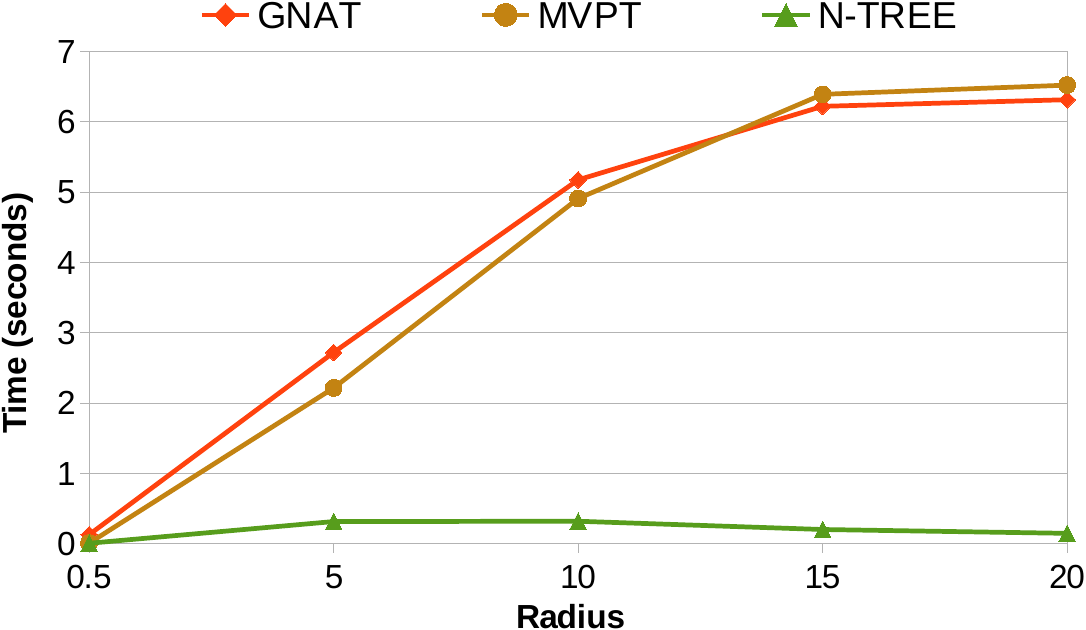}}%
\qquad
\subfigure[Range Search (Low to High Radius) - distance evaluations]{%
\label{subfig:trips_avg_dist_range_search_dist_eval_low_high}%
\includegraphics[height=1.5in]{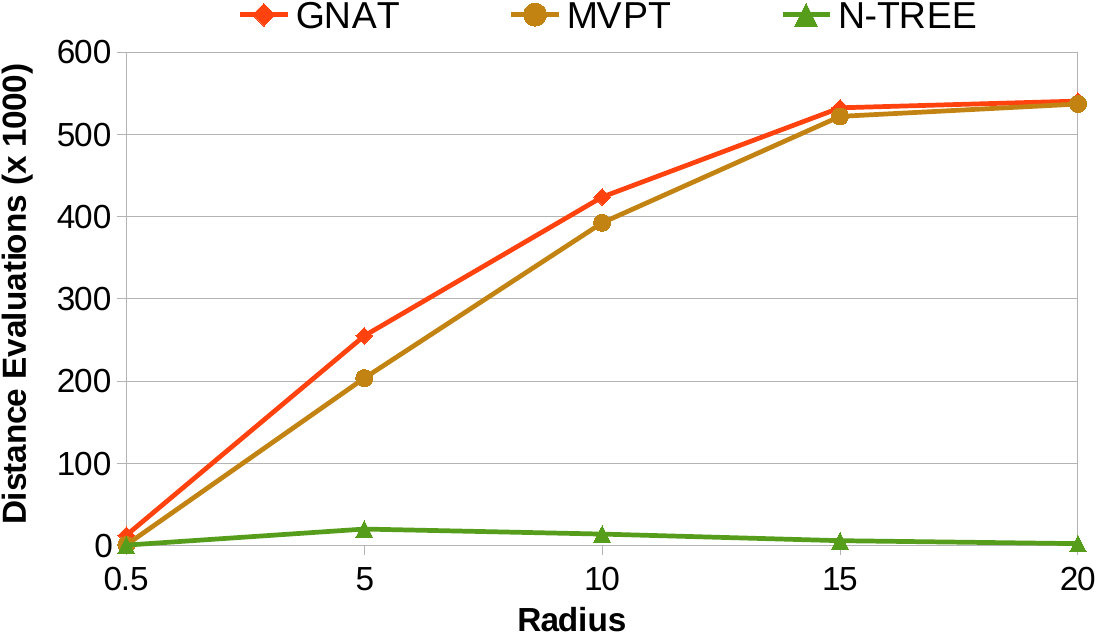}}%
\qquad
\subfigure[kNN Search - run time]{%
\label{subfig:trips_avg_dist_knn_search_exe_time}%
\includegraphics[height=1.5in]{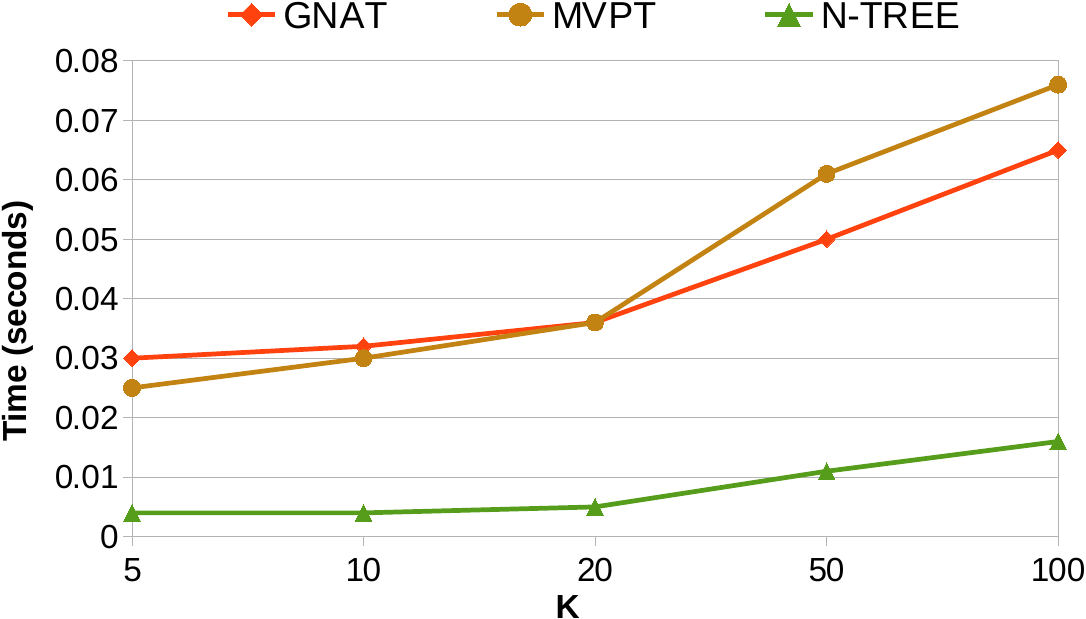}}%
\qquad
\subfigure[kNN Search - distance evaluations]{%
\label{subfig:trips_avg_dist_knn_search_dist_eval}%
\includegraphics[height=1.5in]{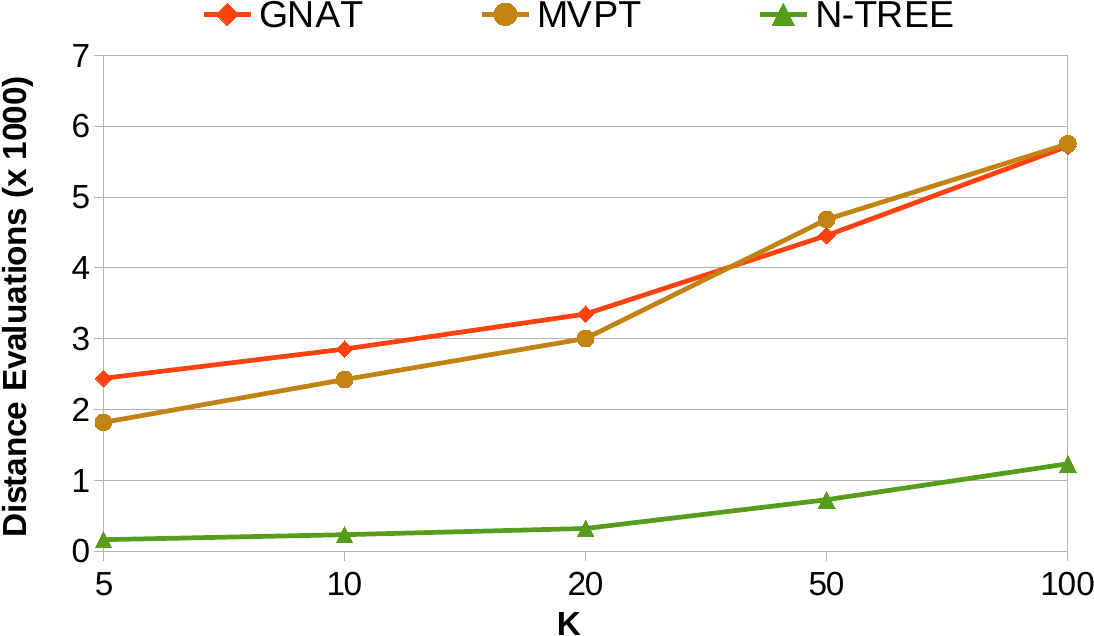}}%
\caption{Trips with Average Distance }
\label{fig:trips_avg_dist_eval}
\end{figure}

 \subsubsection{Trips with  \emph{DistanceAvg}}
 \label{subsubsection:Eval_trips_avg_dist}

 The evaluation of all the metric indexes over the \emph{Trips} dataset using our novel  distance function \emph{DistanceAvg} is shown in Fig. \ref{fig:trips_avg_dist_eval}. Here the units are meaningful and denote kilometers; so the low radius ranges from 100 to 500 meters and the low to high radius from 500 meters to 20 kms average distance of movements.
 
 Overall, we see that the behavior of index structures is similar to using Hausdorff distance as the distance measure, as seen in Section \ref{subsubsection:Eval_trips_hausdorff}. From Section \ref{subsection:Eval_dist_distribution}, we see that the data distribution of the  \emph{Trips} dataset using \emph{DistanceAvg} is similar to using Hausdorff distance. Due to this reason, the performances of all the indexes over both \emph{range} and \emph{kNN} search are similar to those in  the previous scenario. From Figs. \ref{subfig:trips_avg_dist_range_search_exe_time_low} and \ref{subfig:trips_avg_dist_range_search_dist_eval_low}, we see that for \emph{range} queries on low radius, the performance of MVPT is slightly better than N-tree till radius 300 meters, after which N-tree takes the lead.  For \emph{range} query on low to high radii (Figs. \ref{subfig:trips_avg_dist_range_search_exe_time_low_high} and \ref{subfig:trips_avg_dist_range_search_dist_eval_low_high}), N-tree outperforms the other metric indexes and it also exhibits the \emph{U-turn effect}, which is clearly visible from the plots. For \emph{kNN} queries too (Figs. \ref{subfig:trips_avg_dist_knn_search_exe_time} and \ref{subfig:trips_avg_dist_knn_search_dist_eval}), N-tree performs the best in terms of query execution time and the distance evaluations.

Whereas the behaviour of the index structures is similar for Hausdorff distance and \emph{DistanceAvg}, we note that in Fig. \ref{fig:trips_avg_dist_eval} 
the running times and numbers of distance evaluations are smaller, at least for low radius range queries and \emph{kNN} queries, than in Fig. \ref{fig:trips_hausdorff_eval}. As an example, the concrete numbers for \emph{kNN} queries at $k = 100$ are shown in Table \ref{tab:knncomparison}.
So \emph{DistanceAvg} cannot only be evaluated faster than Hausdorff but also requires less distance evaluations. This can possibly be explained by the smoother shape of the distance distribution (see Figures \ref{subfig:trips_hausdorff_data_distribution} and
\ref{subfig:trips_avg_dist_data_distribution}).

\begin{table}[ht]
  \centering
  \caption{\emph{kNN} queries at radius $k=100$ for Hausdorff and \emph{DistanceAvg}}
  \label{tab:knncomparison}
    \begin{tabular}{|l|r|r|r|r|}
\hline
 & \multicolumn{2}{c|}{Run Time [ms]}  & \multicolumn{2}{c|}{Distance Evaluations}  \\ \cline{2-5}
 Index & Hausdorff & \emph{DistanceAvg} & Hausdorff & \emph{DistanceAvg} \\
 \hline
 GNAT & 153 & 65 & 12389 & 5717 \\
 MVPT & 155 & 76 & 10403 & 5751 \\
 N-tree & 41 & 16 & 2928 & 1234 \\
 \hline
    \end{tabular}
\end{table}

When we consider the two contributions of this paper in combination, N-tree and \emph{DistanceAvg}, we can compare the previously best solutions for \emph{kNN} queries on trajectories with what is now available. We compare the run times of MVPT using Hausdorff distance with N-tree using \emph{DistanceAvg} in Table   \ref{tab:knncomparison2}.  The new techniques yield a speedup of more than an order of magnitude.

\begin{table}[h]
  \centering
  \caption{Run times for \emph{kNN} queries on MVPT with Hausdorff and N-tree with \emph{DistanceAvg}}
  \label{tab:knncomparison2}
    \begin{tabular}{|l|r|r|r|r|r|}
\hline
& \multicolumn{5}{c|}{Run time [ms]} \\
\hline
$k$ & 5 & 10 & 20 & 50 & 100 \\
\hline
MVPT with Hausdorff & 50 & 73 & 80 & 127 & 155 \\
N-tree with \emph{DistanceAvg} & 4 & 4 & 5 & 11 & 16 \\
 \hline
 Speedup & 12.5 & 18.2 & 16.0 & 11.5 & 9.7 \\
 \hline
    \end{tabular}
\end{table}

\begin{figure}[ht]
\centering
\subfigure[Range Search (Low Radius) - run time ]{%
\label{subfig:image_range_search_exe_time_low}%
\includegraphics[height=1.5in]{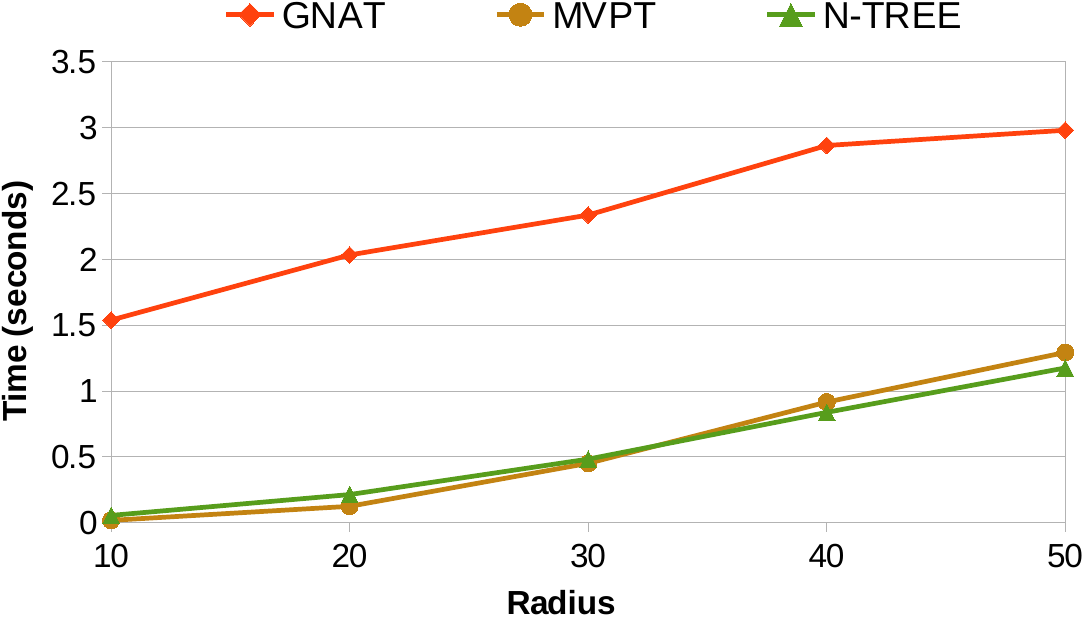}}%
\qquad
\subfigure[Range Search (Low Radius) - distance evaluations]{%
\label{subfig:image_range_search_dist_eval_low}%
\includegraphics[height=1.5in]{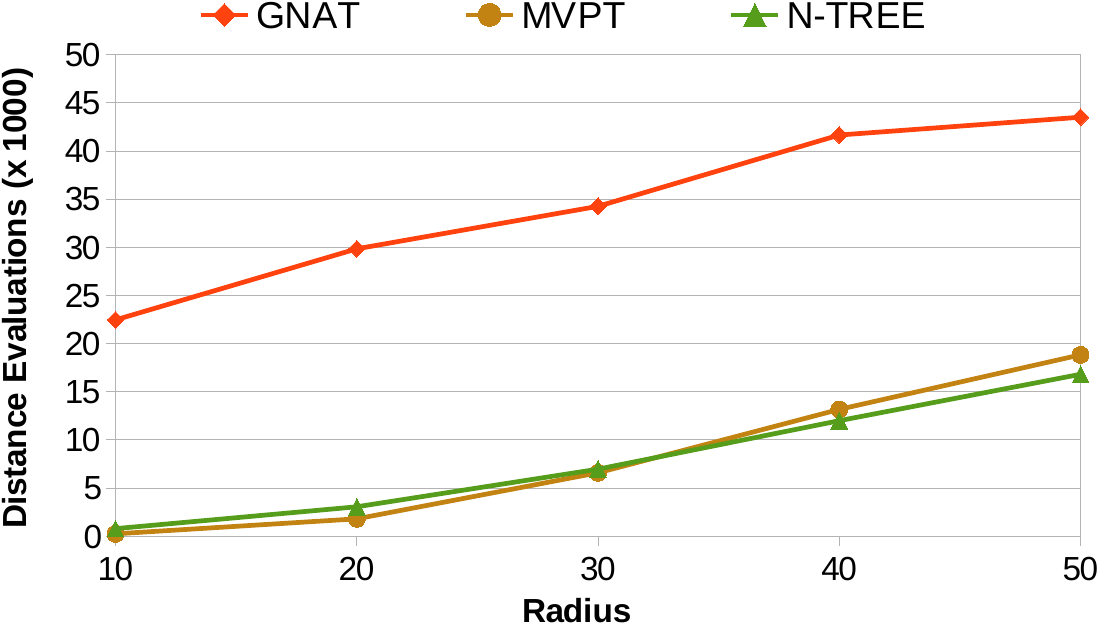}}%
\qquad
\subfigure[Range Search (Low to High Radius) - run time ]{%
\label{subfig:image_range_search_exe_time_low_high}%
\includegraphics[height=1.5in]{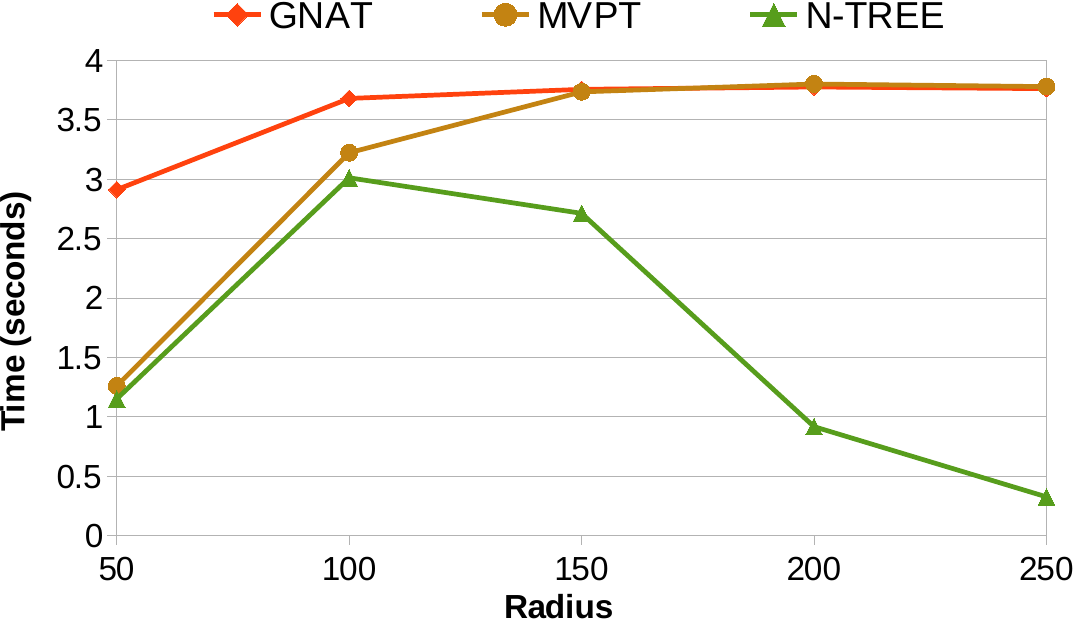}}%
\qquad
\subfigure[Range Search (Low to High Radius) - distance evaluations]{%
\label{subfig:image_range_search_dist_eval_low_high}%
\includegraphics[height=1.5in]{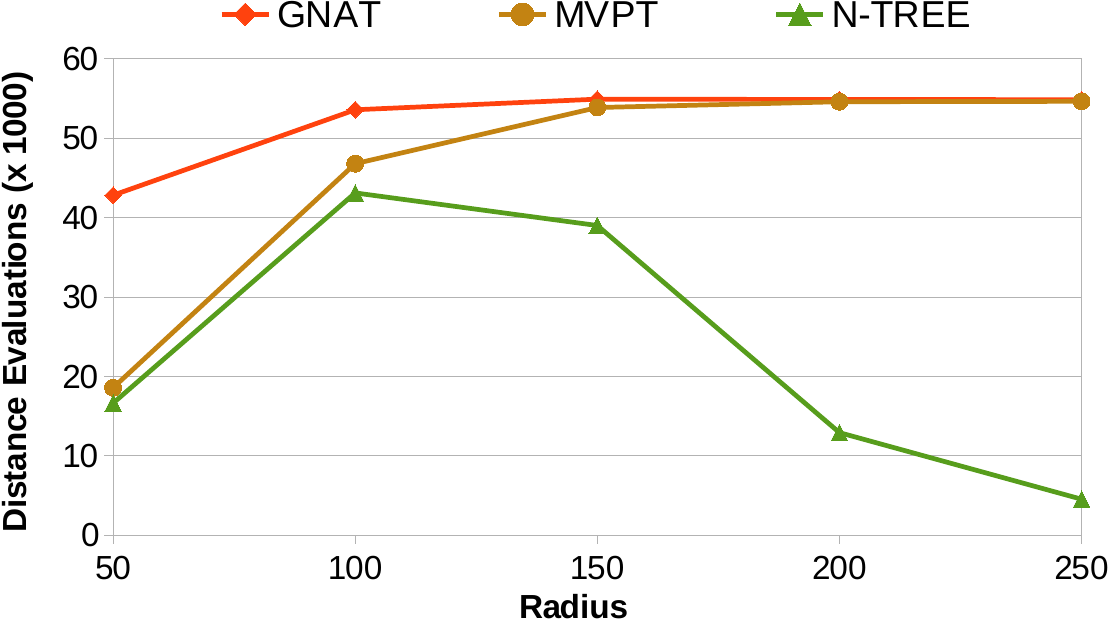}}%
\qquad
\subfigure[kNN Search - run time]{%
\label{subfig:image_knn_search_exe_time}%
\includegraphics[height=1.5in]{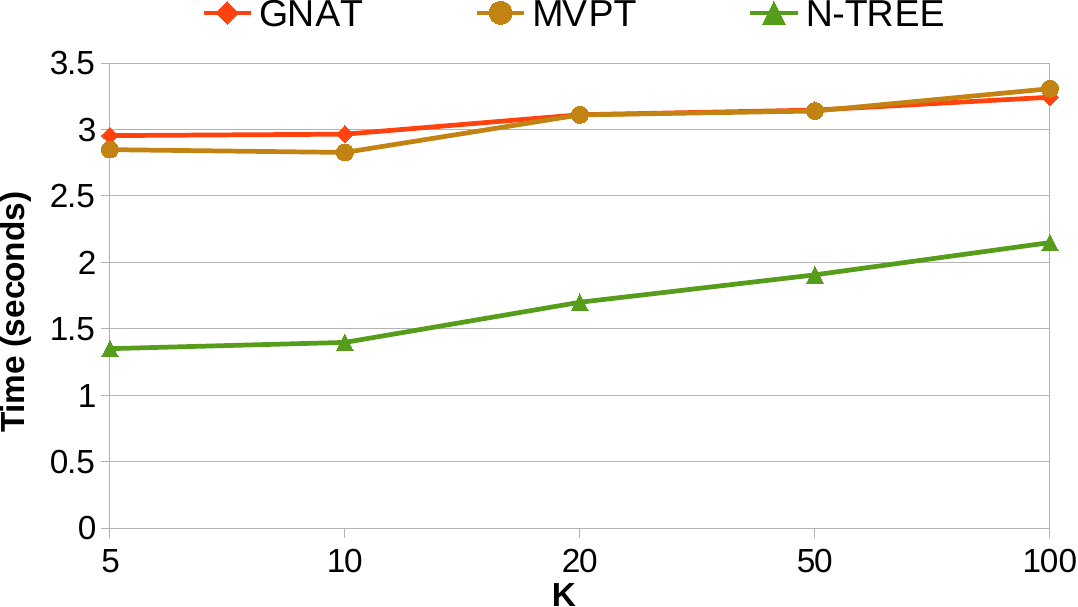}}%
\qquad
\subfigure[kNN Search - distance evaluations]{%
\label{subfig:image_knn_search_dist_eval}%
\includegraphics[height=1.5in]{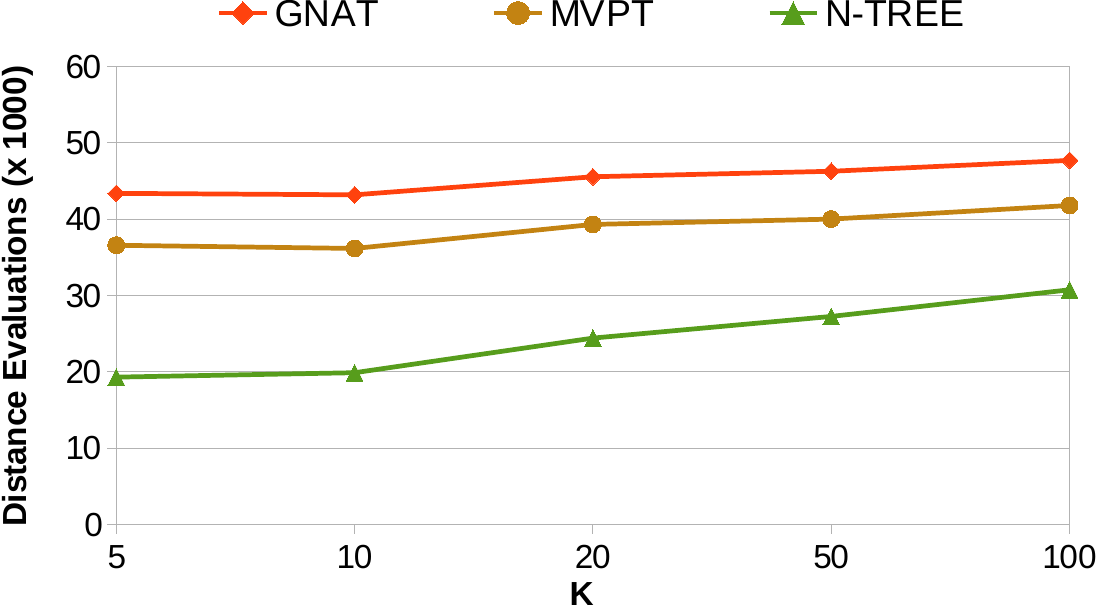}}%
\caption{X-Rays}
\label{fig:image_eval}
\end{figure}
 
 \subsubsection{X-Rays}
 \label{subsubsection:Eval_image}

 This section highlights the performance of all the metric indexes over the \emph{X-Rays} dataset as shown in Fig. \ref{fig:image_eval}. Similar to the \emph{Trips} dataset, from Figs. \ref{subfig:image_range_search_exe_time_low} and 
 \ref{subfig:image_range_search_dist_eval_low}, we observe that MVPT performs better on very small query radius for \emph{range} queries, after which N-tree starts performing better. The \emph{U-turn effect} is clearly observed in Figs. \ref{subfig:image_range_search_exe_time_low_high} and \ref{subfig:image_range_search_dist_eval_low_high}, where the number of distance evaluations (consequently, the execution time) decreases as we keep on increasing the search radius for \emph{range} query after 100 units. From Figs. \ref{subfig:image_knn_search_exe_time} and \ref{subfig:image_knn_search_dist_eval}, we see that N-tree outperforms MVPT  with \emph{kNN} queries both in terms of execution time and distance evaluations. Throughout all these experiments, the performance of GNAT was the worst, as can be seen from the plots.

\begin{figure}[ht]
\centering
\subfigure[Range Search (Low Radius) - run time ]{%
\label{subfig:text_range_search_exe_time_low}%
\includegraphics[height=1.5in]{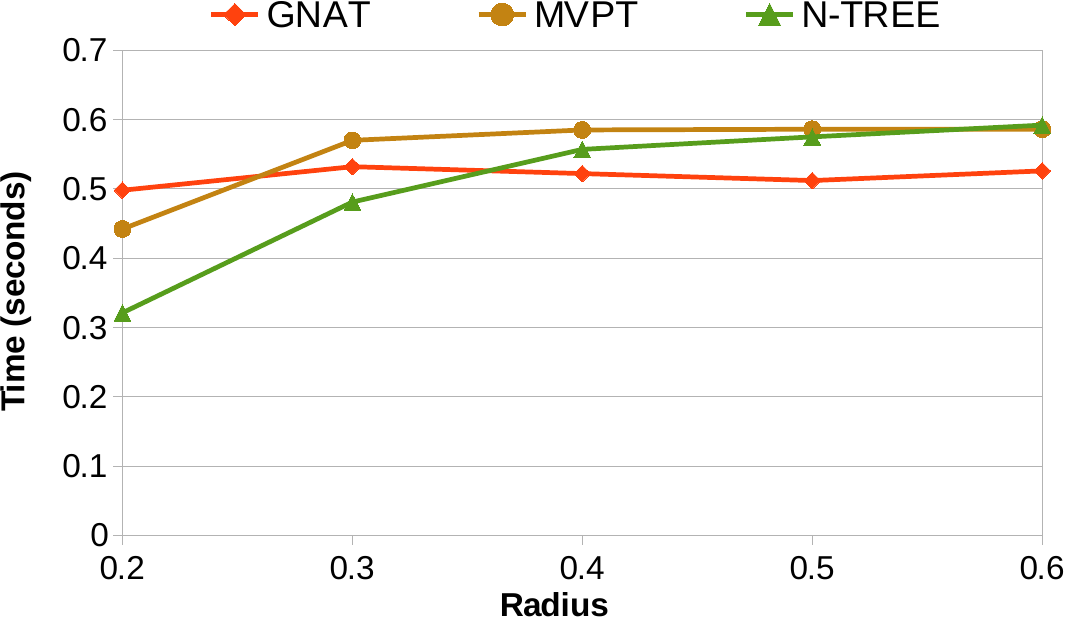}}%
\qquad
\subfigure[Range Search (Low Radius) - distance evaluations]{%
\label{subfig:text_range_search_dist_eval_low}%
\includegraphics[height=1.5in]{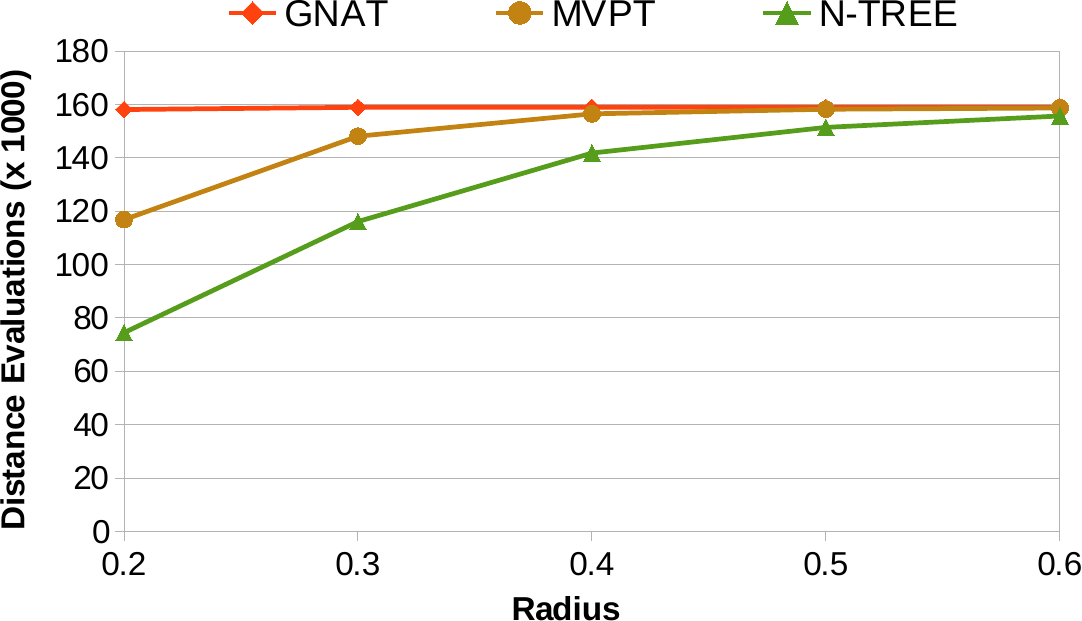}}%
\qquad
\subfigure[Range Search (Low to High Radius) - run time ]{%
\label{subfig:text_range_search_exe_time_low_high}%
\includegraphics[height=1.5in]{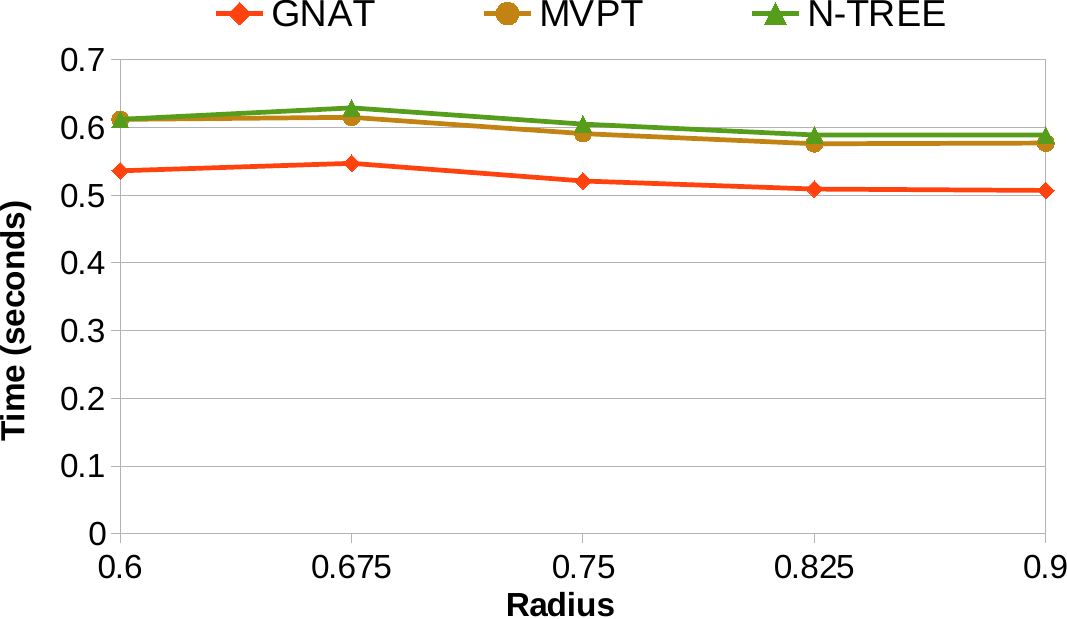}}%
\qquad
\subfigure[Range Search (Low to High Radius) - distance evaluations]{%
\label{subfig:text_range_search_dist_eval_low_high}%
\includegraphics[height=1.5in]{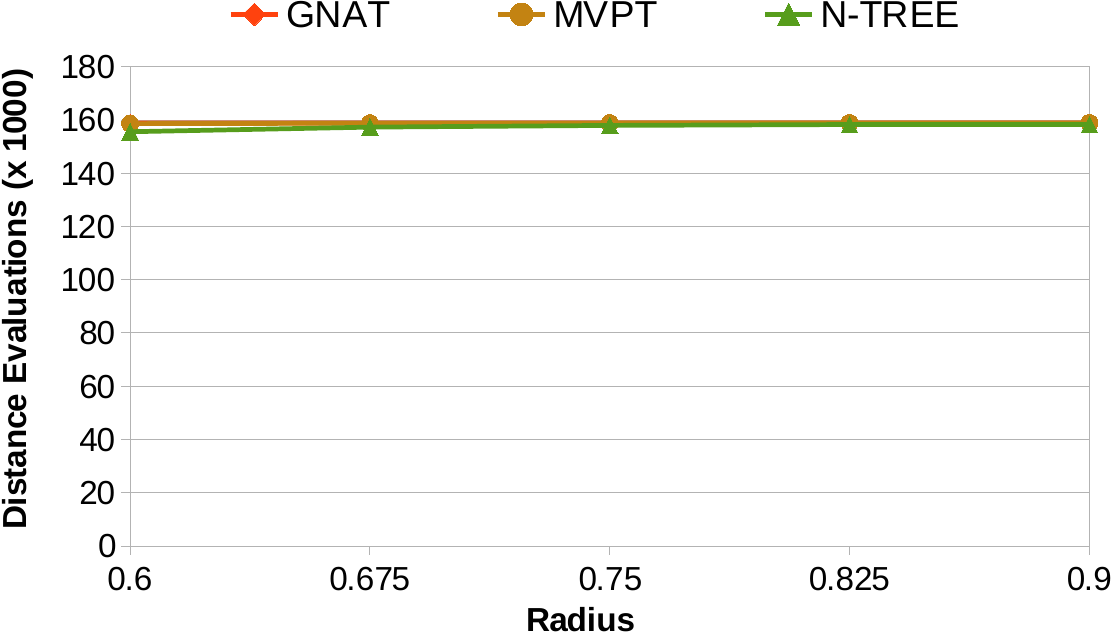}}%
\qquad
\subfigure[kNN Search - run time]{%
\label{subfig:text_knn_search_exe_time}%
\includegraphics[height=1.5in]{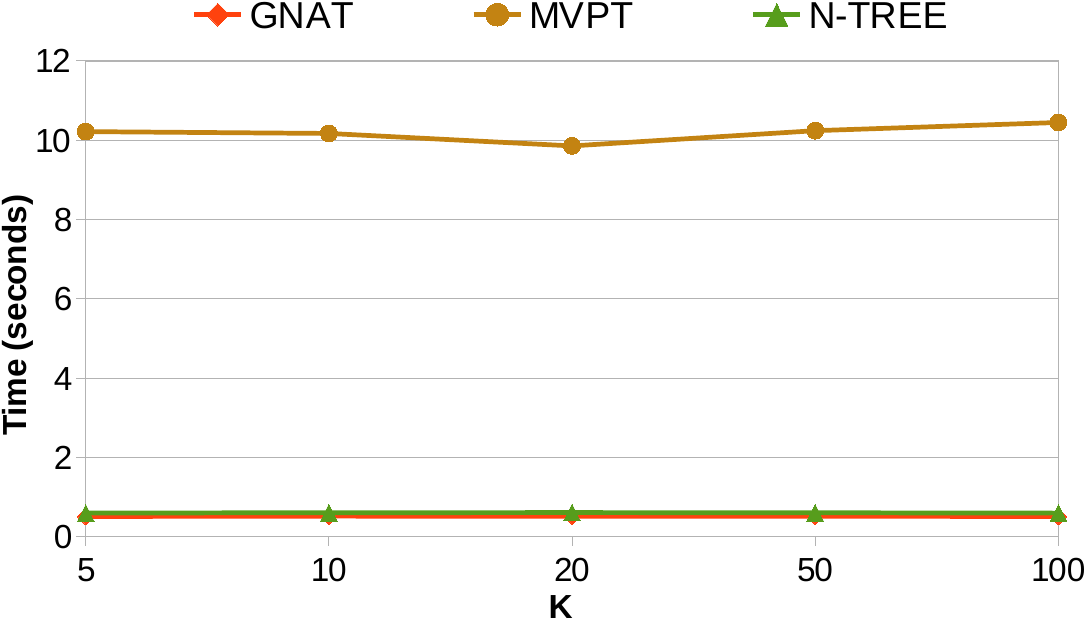}}%
\qquad
\subfigure[kNN Search - distance evaluations]{%
\label{subfig:text_knn_search_dist_eval}%
\includegraphics[height=1.5in]{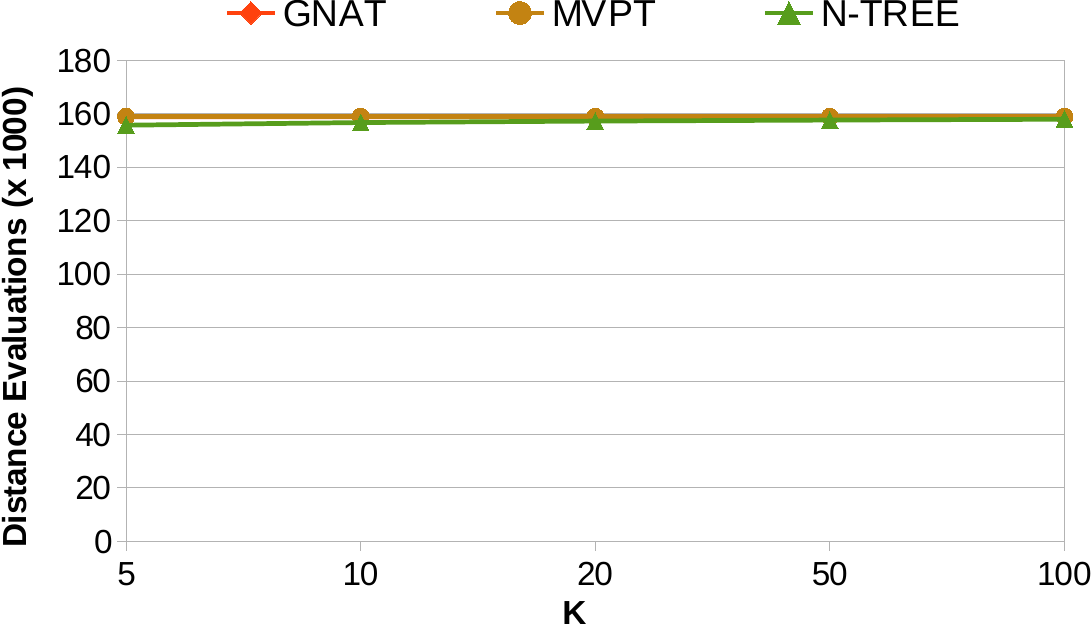}}%
\caption{Sentences }
\label{fig:text_eval}
\end{figure}
 
 \subsubsection{Sentences}
 \label{Eval_sentences}

This section shows our evaluation of all metric indexes over the \emph{Sentences} dataset, with plots presented in Fig. \ref{fig:text_eval}.
Considering first the numbers of distance evaluations (Figures 
\ref{subfig:text_range_search_dist_eval_low},
\ref{subfig:text_range_search_dist_eval_low_high}, and
\ref{subfig:text_knn_search_dist_eval})
 we note that for low to high radii range queries and \emph{kNN} queries all indexes require about as many distance evaluations (159000) as there are objects in the dataset. For low radii, the situation is not much better. Whereas GNAT needs the full number of distance evaluations even from the smallest query radius, N-tree and MVPT start with smaller numbers but also soon approach this maximal number with increasing radius.

This means that essentially all indexes fail as they offer hardly any advantage over exhaustive search. In Section~\ref{sec:baseline} we see that the time for exhaustive search is 0.479 seconds.

In the running times for range queries (Figures
\ref{subfig:text_range_search_exe_time_low} and 
\ref{subfig:text_range_search_exe_time_low_high}),
from a radius of 0.4 units on GNAT is a bit faster than MVPT and N-tree. Only for the smallest radii N-tree and MVPT need a bit less time due to their smaller numbers of distance evaluations. 
Note however, that for the smallest radii 0.2 and 0.3 the average numbers of results returned are 1.08 and 1.28, respectively. Given that one of the results is the query object itself, this means that hardly any close objects are found. So such range queries are not so relevant.

For \emph{kNN} queries, GNAT and N-tree have similar running times whereas for MVPT it is much higher.

This behaviour is entirely different from what we have seen in the previous three subsections. This confirms that it is important to consider the distance distribution of a dataset as the performance of a metric index highly depends on it. From Fig. \ref{subfig:sentence_data_distribution}, we see that the captions of the \emph{Sentences} datasets are mostly dissimilar from each other, in terms of Jaccard Distance. It appears that in such situations, when all objects are far apart from each other, indexes simply do not work.

\begin{figure}[ht!]
\centering
\subfigure[Range Search (Low Radius) - run time ]{%
\label{subfig:buildings_range_search_exe_time_low}%
\includegraphics[height=1.5in]{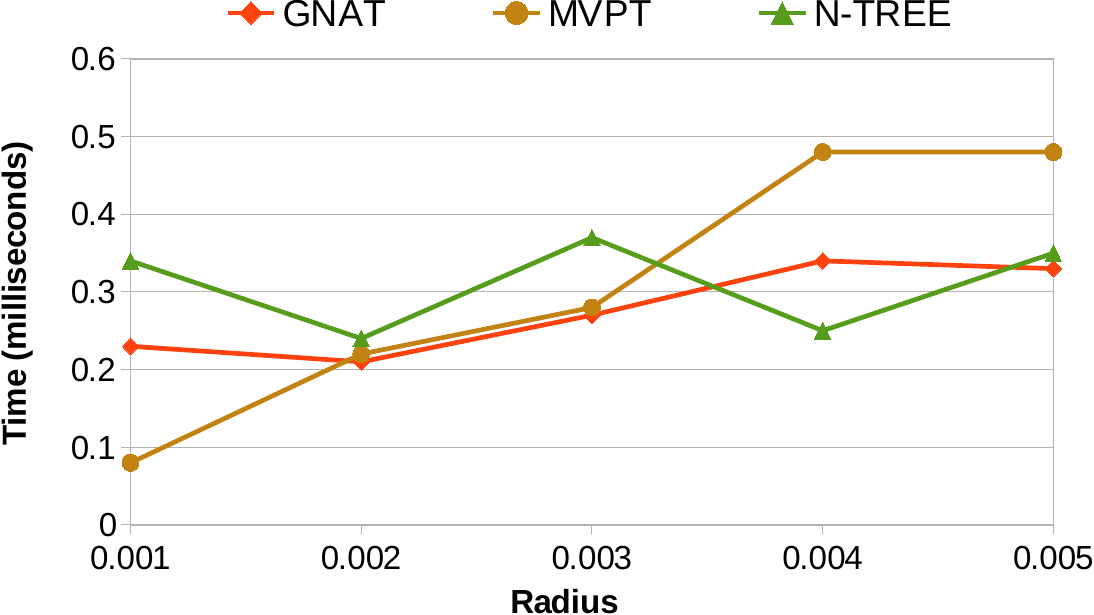}}%
\qquad
\subfigure[Range Search (Low Radius) - distance evaluations]{%
\label{subfig:buildings_range_search_dist_eval_low}%
\includegraphics[height=1.5in]{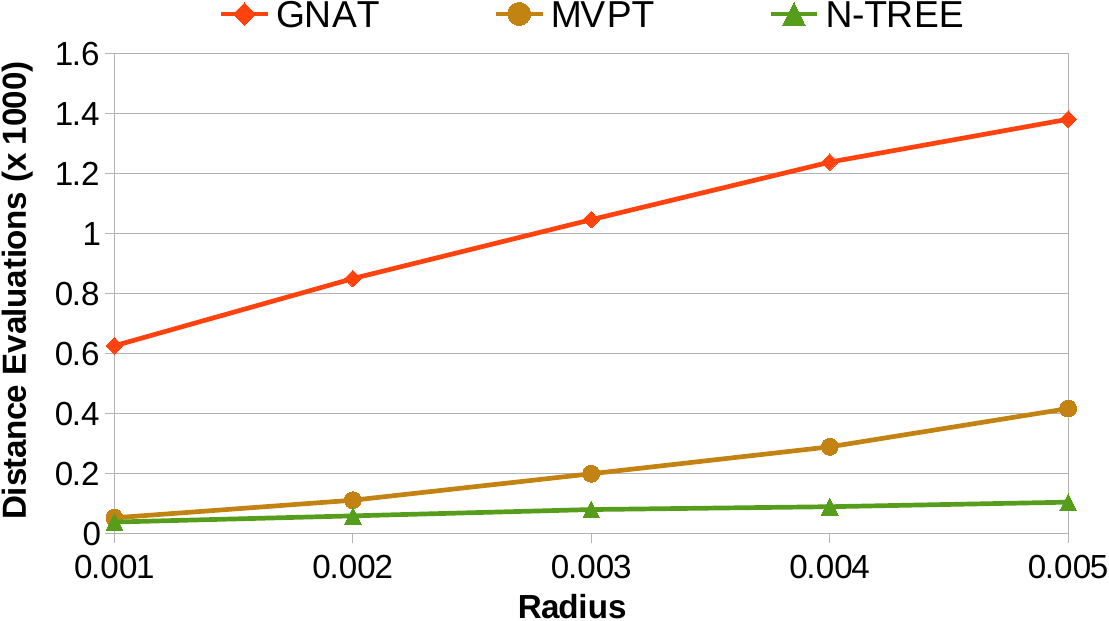}}%
\qquad
\subfigure[Range Search (Low to High Radius) - run time ]{%
\label{subfig:buildings_range_search_exe_time_low_high}%
\includegraphics[height=1.5in]{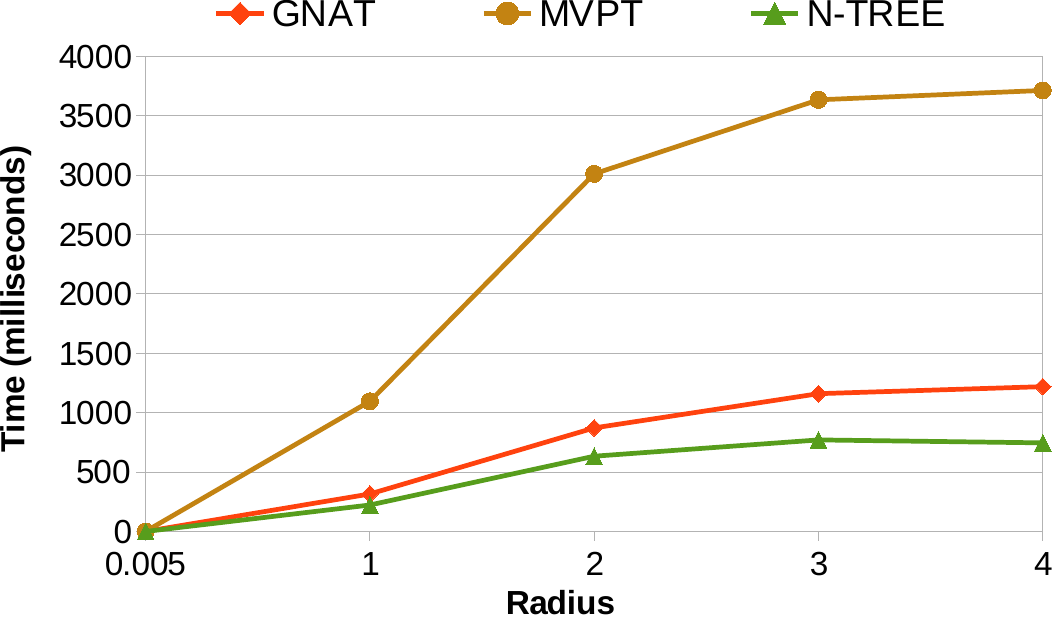}}%
\qquad
\subfigure[Range Search (Low to High Radius) - distance evaluations]{%
\label{subfig:buildings_range_search_dist_eval_low_high}%
\includegraphics[height=1.5in]{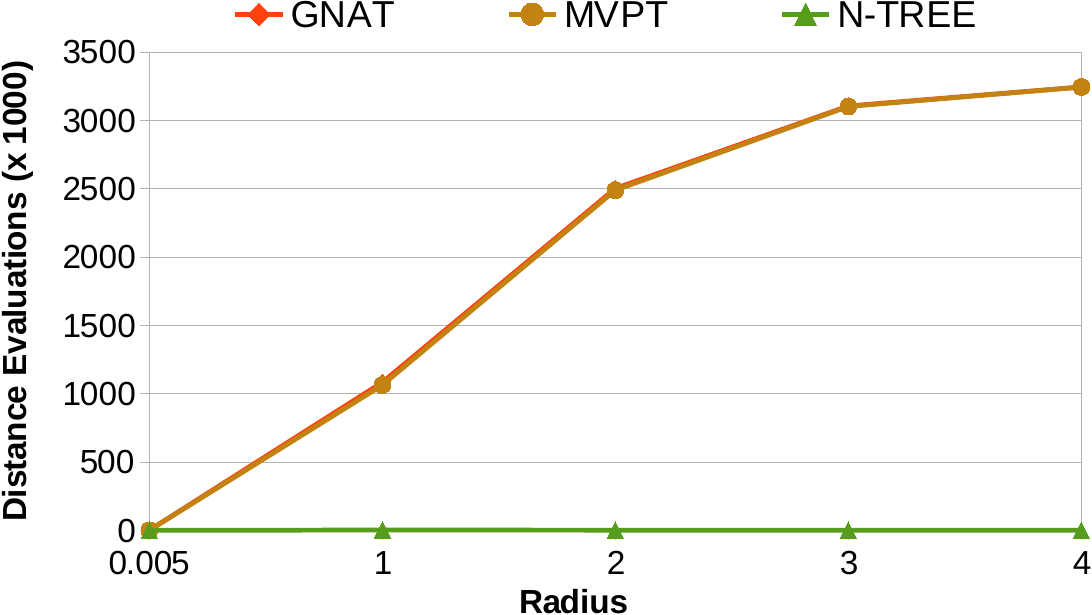}}%
\qquad
\subfigure[kNN Search - run time]{%text
\label{subfig:buildings_knn_search_exe_time}%
\includegraphics[height=1.5in]{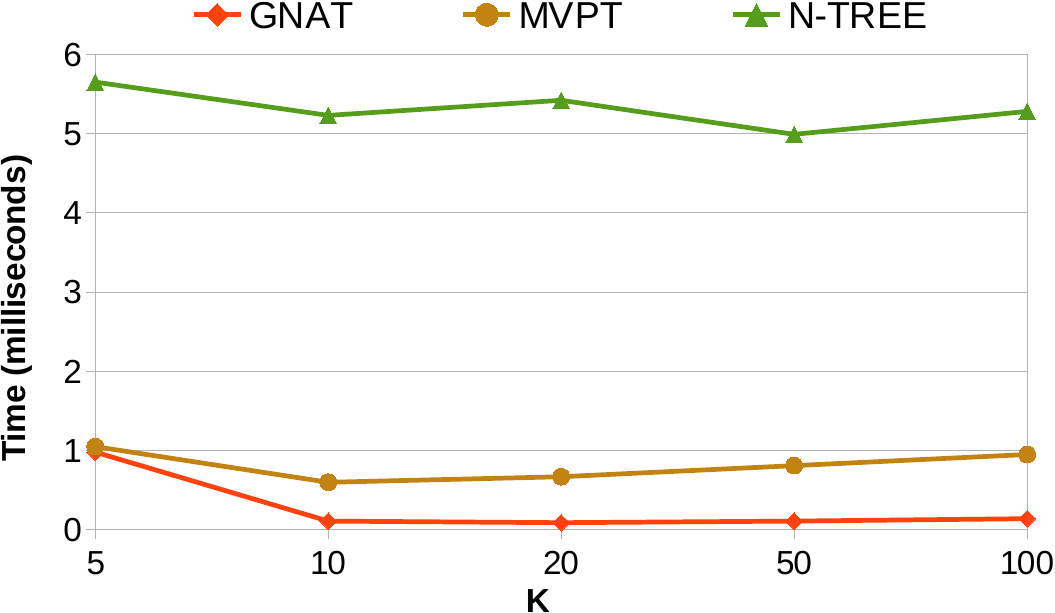}}%
\qquad
\subfigure[kNN Search - distance evaluations]{%
\label{subfig:buildings_knn_search_dist_eval}%
\includegraphics[height=1.5in]{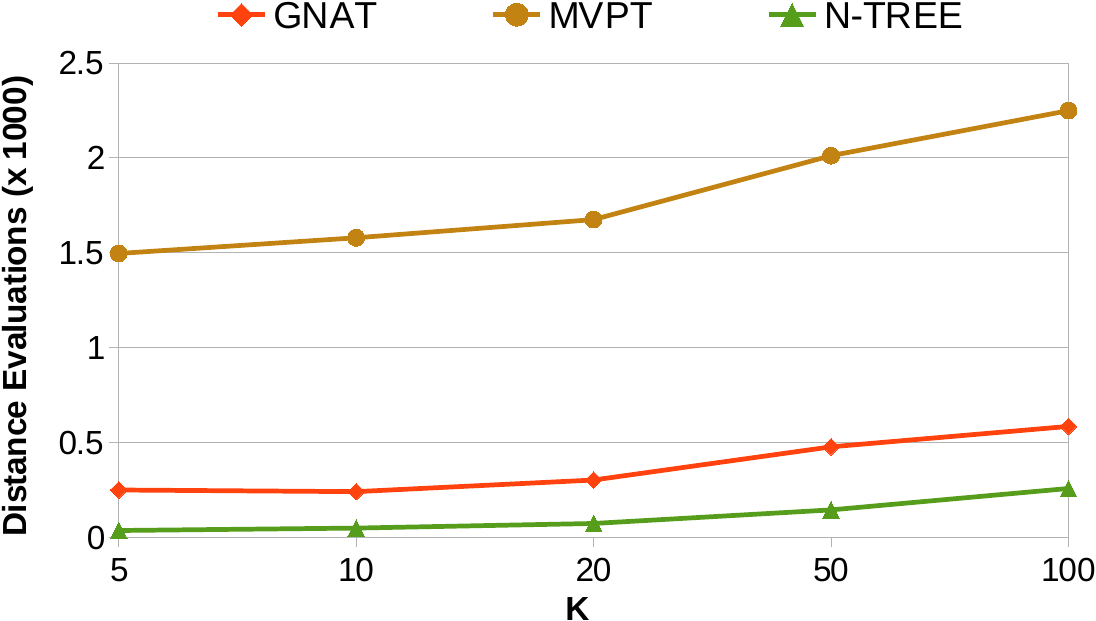}}%
\qquad
\subfigure[Fig. \ref{subfig:buildings_range_search_dist_eval_low_high} with N-tree only displaying \emph{U-turn effect}]{%
\label{subfig:buildings_ntree_only_range_search_dist_eval_low_high}%
\includegraphics[height=1.5in]{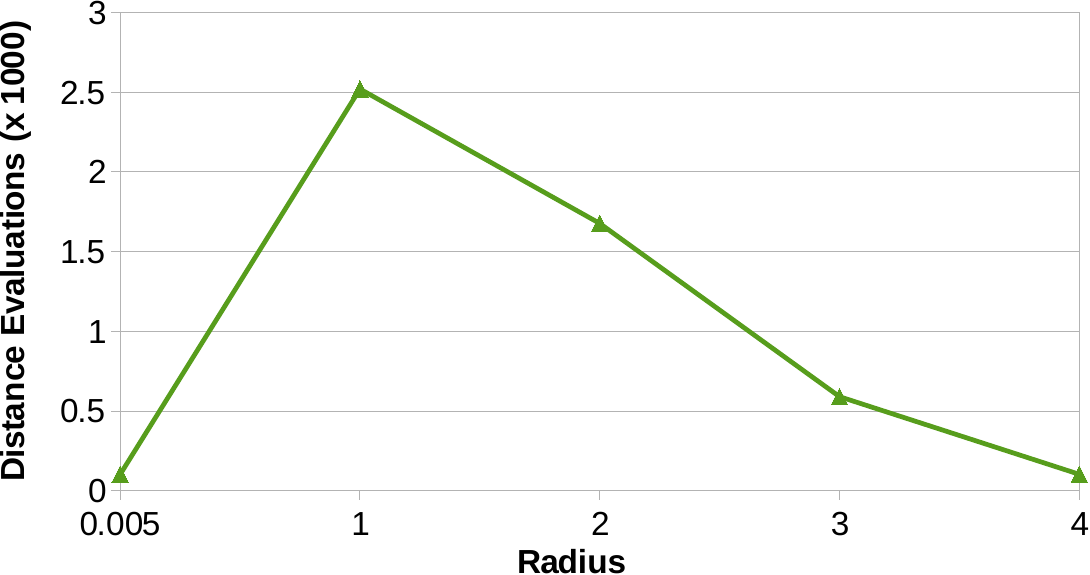}}%
\caption{Buildings  }
\label{fig:buildings_eval}
\end{figure}

\subsubsection{Buildings}

The performance evaluation of all the metric indexes using the \emph{Buildings} dataset is shown in Fig. \ref{fig:buildings_eval}. As discussed for Fig. \ref{subfig:buildings_data_distribution}, the dataset contains many clusters corresponding to towns or villages.
Due to this and the low evaluation cost for Euclidean distance, the behavior of the metric indexes  differs from the previous cases. For \emph{range} queries on low radii (Figs. \ref{subfig:buildings_range_search_exe_time_low} and \ref{subfig:buildings_range_search_dist_eval_low}), we see that GNAT performs the worst in terms of distance evaluations, whereas the performance of MVPT is similar to that of  N-tree when the search radius is very low (0.001 units in this case), after which the performance of MVPT starts to degrade. However, the query execution times varied and no relationship with the change in search radius was identified in this case. For higher search radii (Figs. \ref{subfig:buildings_range_search_exe_time_low_high} and \ref{subfig:buildings_range_search_dist_eval_low_high}), we see that the performance of GNAT and MVPT is almost the same in terms of distance evaluations, whereas N-tree  performs the best and requires far less distance evaluations compared to the other two structures.

Although it appears from Fig. \ref{subfig:buildings_range_search_dist_eval_low_high} that the number of distance evaluations for N-tree is constant irrespective of the radius,  that is not the case. The \emph{U-turn effect} occurs here as well, but that is not  clearly observable in the plot as the numbers of distance evaluations for the other two indexes are so high. If the other two indexes are removed from the plot keeping N-tree only, the \emph{U-turn effect} is clearly visible after radius 1 units, as shown in Fig. \ref{subfig:buildings_ntree_only_range_search_dist_eval_low_high}. MVPT requires the most time for query execution, whereas GNAT and N-tree almost take the same time to start with, and then gradually GNAT consumes more time compared to N-tree as the search radius grows.

For \emph{kNN} queries (Figs. \ref{subfig:buildings_knn_search_exe_time} and \ref{subfig:buildings_knn_search_dist_eval}), we see that N-tree requires the least number of distance evaluations, whereas MVPT requires the most, followed by GNAT. At the same time, N-tree exhibits the  highest execution time, followed by MVPT and GNAT. So for an extremely cheap distance function like Euclidean distance, the advantage of the N-tree in run time for \emph{kNN} queries, due to fewer distance evaluations, disappears.
In scenarios with expensive distance evaluation, the query execution time will be proportional to the number of distance evaluations. In such cases, less distance evaluations will lead to shorter query execution times.

\subsubsection{Summary}

To conclude, we can say that the N-tree performs better in almost all situations compared to GNAT and MVPT both in \textit{range} and \textit{kNN search}. 

We have evaluated N-tree, GNAT, and MVPT for five scenarios. We consider the first three (\emph{Trips} with Hausdorff and \emph{DistanceAvg} as well as \emph{X-Rays}) to be normal for the expected scope of applications of metric index structures, whereas the last two (\emph{Sentences} and \emph{Buildings}) have unusual properties. We first discuss the normal case and then the two special cases.

In the normal case, data sets have a ``reasonable'' distance distribution and the distance functions are somewhat complex and expensive to evaluate. In this case, we observe that the results and the ranking of index structures in terms of numbers of distance evaluations and query time are strongly correlated. For range queries at low radius, N-tree and MVPT perform about equally well whereas GNAT requires much more time and distance evaluations. For range queries at higher radii N-tree clearly outperforms the other two structures as it needs far less distance evaluations and consequently query time. The same holds for \emph{kNN} queries where N-tree is much faster than the other two.

The first special case occurs for the \emph{Sentences} dataset where the distance distribution is such that practically all objects are far from each other. We observe that all three index structures fail as they need a similar number of distance evaluations as sequential scan. It appears that the structure of indexes relies on distinguishing between objects that are close together and those that are far apart.

The second special case occurs for the \emph{Buildings} dataset where the cost of evaluating Euclidean distance is negligible. In this case, N-tree still has the smallest number of distance evaluations but this advantage does not play out as distance evaluations are so cheap. 

We also demonstrate the \emph{U-turn effect} where after a certain point, the number of distance evaluations decreases with the increase in search radius for \emph{range} search. Normally with larger query radius there are more data points and therefore more distance evaluations. However, our structure efficiently uses the $radius$ associated with each center $c_i$ within a node in the N-tree to include an entire partition into the final result set, without any distance evaluation. To the best of our knowledge, this property is not present in other metric indexes. 

\subsection{Comparison of Average Distance Function against Hausdorff}
\label{subsection:avg_dist_vs_hausdorff_eval}

In this experiment, we compare the time taken to execute our novel distance measure \emph{DistanceAvg} with the Hausdorff distance for trajectories, when used in an index. Here we consider trajectories with many points, varying the length. The \emph{Trips} dataset is not suitable for this purpose, as the average number of points on each trajectory is about 38, which is too low to conduct this experiment. 

So, we generated 1,000 trajectories using \emph{Hermoupolis} \cite{Hermapoulis_ACM}.
For the purposes of our experiment, each trajectory (having an average of 5,000 points) was divided into smaller trajectories with $M$ points each, where $M = \{100, 150, 200, 250, 300\}$. The experiments were conducted on 25,000 such smaller trajectories, by varying $M$ and executing \emph{kNN} ($k=50$) queries 100 times (randomly selecting a query point each time). For each $M$, we tabulated the time taken to build the N-tree index and the average time taken (for each run of 100 \emph{kNN} queries) to evaluate queries using
both distance measures.

\begin{figure}[ht]
\centering
\subfigure[N-tree build time]{%
\label{subfig:ntree_build_time_avg_with_hausdorff}%
\includegraphics[height=1.5in]{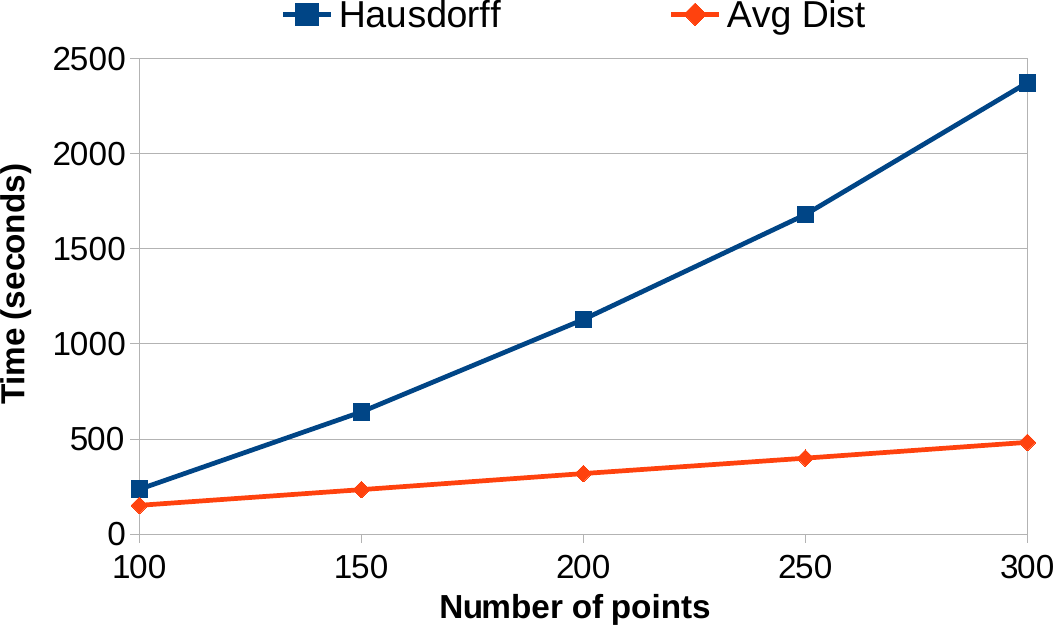}}%
\qquad
\subfigure[\emph{kNN} query evaluation time]{%
\label{subfig:dist_eval_avg_with_hausdorff}%
\includegraphics[height=1.5in]{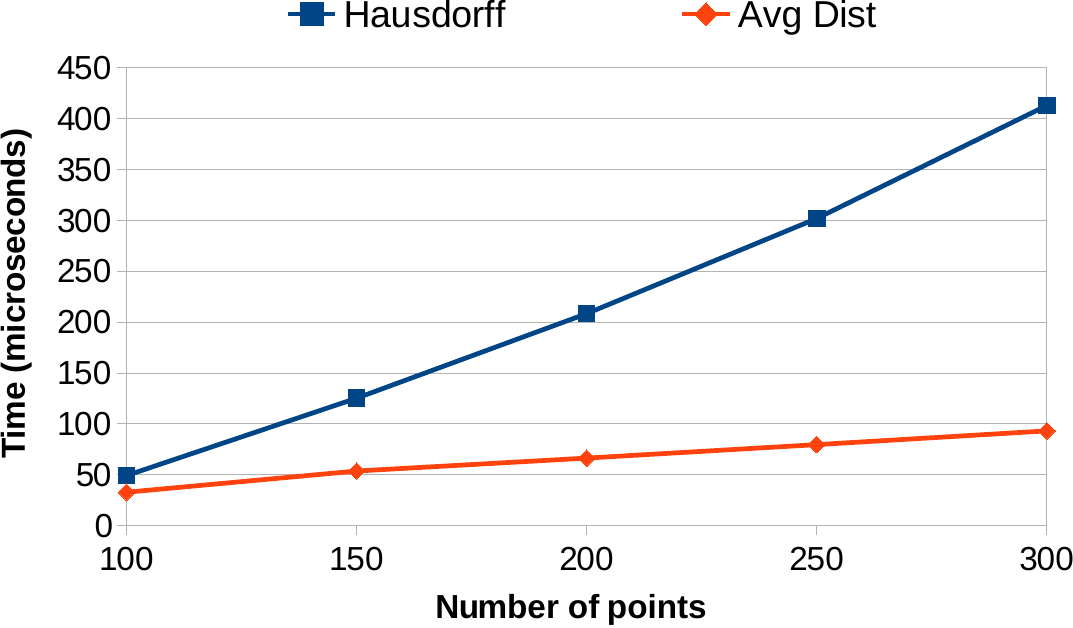}}%
\caption{Comparison of \emph{DistanceAvg} with Hausdorff}
\label{fig:avgdist_vs_hausdorff}
\end{figure}

Figs. \ref{subfig:ntree_build_time_avg_with_hausdorff} and \ref{subfig:dist_eval_avg_with_hausdorff} show the N-tree build times and query evaluation times, respectively, for both distance measures. We observe that, as the value of $M$  increases, it takes progressively more time to build the N-tree using Hausdorff compared to \emph{DistanceAvg}.  The query evaluation time for Hausdorff also increases at the same rate. When $M$ varies between 100 and 300, the average time taken to measure Hausdorff ranged from 49 to 412 microseconds, compared to a range from 32 to 93 microseconds for \emph{DistanceAvg}. Likewise, the N-tree index building time varied between 235 and 2374 seconds (i.e., around 4 to 40 minutes) for Hausdorff and between 150 and 482 seconds (i.e., around 2.5 to 8 minutes) for \emph{DistanceAvg}. Hence, our novel measure \emph{DistanceAvg} takes much less time to compute the distance between two trajectories compared to the well-known Hausdorff distance. As a consequence, both the index build time and query execution time can be reduced by applying \emph{DistanceAvg} instead of Hausdorff. 

\subsection{Filter-and-Refine}

In this section we evaluate the filter-and-refine technique for the \emph{DistanceAvg} function described in Section~\ref{sec:filterrefine}. For these experiments, we use the implementation in \Secondo{} and the New York Trips dataset as well as the computer described in Section~\ref{sec:distconsteval}.

\begin{figure}[ht]
\begin{center}
\begin{tabular}{cc}
\includegraphics[scale=0.8]{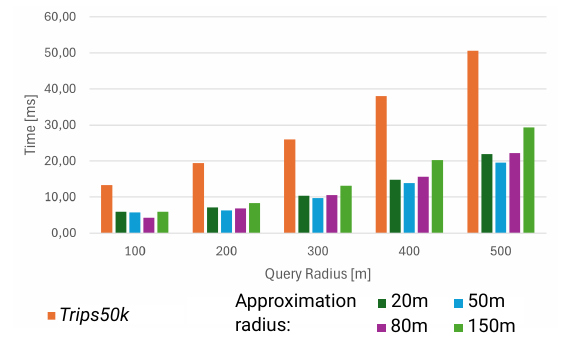}
& 
\includegraphics[scale=1.1]{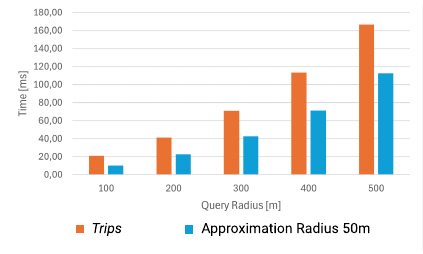} \\
(a) \footnotesize{\emph{Trips50k}}& (b) \footnotesize{New York Trips}
\end{tabular}
\end{center}
  \caption{ Range queries using filter-and-refine}
  \label{fig:rangeFR}
\end{figure} 

We evaluate the use of an N-tree index (Algorithm~\ref{alg:rangeSearchFR}). We use random subsets of the New York Trips dataset of cardinality 50,000 and of  1,000, called \emph{Trips50k} and \emph{Queries1000}, respectively. An N-tree index is built on \emph{Trips50k} and a range query is executed for each object in \emph{Queries1000}. We vary the approximation radius (20, 50, 80, 150 meters) and the query radius (100, 200, 300, 400, 500 meters) and report the average runtime for queries (Figure~\ref{fig:rangeFR} (a)).

The choice of the approximation radius has two opposing effects: A smaller radius yields longer cylinder approximations so that the evaluation of distances becomes more expensive. On the other hand, a finer approximation yields fewer results in the filter step that need to be examined further. Conversely, a coarse approximation (large radius) yields short cylinder approximations with cheap distance computations but a large result set in the filter step. This is shown in Table~\ref{tab:rangeFR} for queries with radius 300 m. Here the result sizes refer to the accumulated result of all 1000 queries.
\begin{table}[htp]
\small
\caption{Varying the approximation radius}
\begin{center}
\begin{tabular}{|l|r|r|r|r|r|}
\hline
& Exact & 20 m & 50 m & 80 m & 150 m \\
\hline
\hline
Index build time [sec] & 507.52 & 170.70 & 119.05 & 97.57 & 78.56 \\
Avg number of units & 36.91 & 11.78 & 7.78 & 6.62 & 5.25 \\
\hline
Query time 300 m [sec] & 26.02 & 10.35 & 9.71 & 10.57 & 13.14 \\
Result size filter step & & 6110 & 7932 & 10114 & 16829 \\
Distance $< q - r$ & & 4064 & 3008 & 2061 & 786 \\
No. exact evaluations & & 2046 & 4924 & 8053 & 16043 \\
Result size & 5001 & 5001 & 5001 & 5001 & 5001 \\
\hline
\end{tabular}
\end{center}
\label{tab:rangeFR}
\end{table}

In Figure~\ref{fig:rangeFR} one can observe that the three smaller approximation radii yield similar results; the 50 m approximation generally is the best. Radii above 100 meters are less efficient, especially for larger query radii. Results for the full New York Trips dataset, using the 50 m approximation, are shown in Figure~\ref{fig:rangeFR} (b).

\section{Distributed Construction of N-trees}
\label{sec:distconst}

The N-tree is a main-memory index, which for large datasets with expensive distance functions may have long construction times. To address this, we offer two techniques, namely (i) parallel construction and (ii) saving an index to and restoring it from persistent storage. 

The basis for both techniques is a relational representation of an index. We use relations rather than files to be able to use systems that offer distributed/parallel relational query processing (e.g. Apache Spark~\cite{ZXWD+16}). One such system is \Secondo\ and we have implemented the techniques in this environment.

\subsection{Exporting and Importing N-trees}

Suppose we have on a single node (the master) the relation to be indexed as $R(A_1, ..., A_n, X)$ where $X$ is the attribute to be indexed. The standard procedure is to load this relation into memory (call it $Rm$) and then to construct an N-tree index $Rm\_X\_ntree$ for it with entries of the form $(TID, X)$. $TID$ is the tuple identifier for $Rm$ which allows direct access to tuples.

We now assume we have operations \op{exportntree} and \op{importntree}, that allow one to export a given N-tree into four relations \emph{Nodes}, \emph{Distances}, \emph{Pivots}, and \emph{Info} and to build an N-tree from such a quadruple of relations, respectively.

We assume that in the N-tree each node has a unique \emph{NodeId}. The first relation \emph{Nodes} to represent the N-tree has the schema 

\begin{query}
(A1, ..., An, X, TID, NodeId: int, Entry: int, Subtree: int, MaxDist: real)
\end{query}

Here \emph{Entry} is the index of the respective center in an internal node or of the element in a leaf. The \emph{Subtree} field contains the \emph{NodeId} of the root of the respective subtree or 0 for a leaf.
Field \emph{MaxDist} contains the radius of the subtree.

The second relation \emph{Distances} has schema
\begin{query}
(NodeId: int, Entry1: int, Entry2: int, Distance: real)
\end{query}
and represents the precomputed distances within a node.

The third relation \emph{Pivots} represents the precomputed 2d points for all entries based on pivot distances (assuming 2d pivots) and has the schema:
\begin{query}
(NodeId: int, Entry: int, Pos: point, IsPivot: bool)
\end{query}
The last field allows one to identify the two pivot points.

The fourth relation \emph{TreeInfo} has only a single tuple which contains general information about the index such as the type of the indexed relation and paramaters $k, l$.

The \op{export} operator can be implemented in a single tree traversal, writing the three main relations, in linear time. Similarly the \op{import} operator can read the three relations in parallel and construct the tree in linear time, without any distance computations.

\subsection{Distributed Construction}

\subsubsection{One Level}
\label{sec:onelevel}

The basic idea is to partition the data set at the top level in the same way as it would be done in the root node construction. Partitions are then distributed to workers. Each worker builds for each partition that it handles an N-tree, corresponding to one subtree of the root. It then exports the N-tree into the four relations. These relations are then collected at the master and their union is formed. Further, the relations needed for the root node, e.g. the pairwise distances of centers, are computed directly
and added to the union. From this union, on the master an N-tree for the entire data set can be built fast, because all required information about placing data elements in nodes and about precomputed distances is available. 

It is necessary that in the global tree built from the local trees of workers all node identifiers are distinct. Gaps in the numbering are permitted. Therefore the \op{exportntree} operator has a parameter providing the start number for node identifiers exported. Then all local trees constructed by workers will export trees with distinct sets of node identifiers. Further, on import relations must be ordered by node identifiers and this order must correspond to a depth-first traversal of the tree, as it  is done on export.

\subsubsection{Two Levels}

In the approach described so far, a disadvantage is that the number of tasks for workers to construct N-trees is limited by the degree of the root node. For example, when we wish to construct N-trees with $k = 25$, we cannot employ more than 25 workers. Further, the sizes of the $k$ partitions may vary a lot and the runtime for parallel construction is determined by the largest partition.

For a good load balancing it is necessary to assign more than one task to each worker (corresponding to a slot in a distributed array\footnote{A distributed array has fields of the same type (called \emph{slots}) that may be stored on different computers and be processed by different workers. Each worker may process several slots sequentially.} in \Secondo). Experiments in previous work have shown that it is good to have a number of tasks (slots) corresponding to at least 4 to 6 times the number of workers \cite{GBN21}.

We therefore extend the approach to handle two levels, that is, to construct relations for the root node as well as the level 1 nodes below it directly (partly on the master, partly in parallel by workers) and to build N-trees for partitions and export them only at the next level 2. At this level we can create $k^2$ partitions and N-trees and have a reasonable number of tasks for $k = 25$ and a number of 30 or 80 workers, for example. Two levels should be sufficient for most applications.

The steps needed in the distributed construction, using the two-level approach, are listed in Table~\ref{tab:distconst}.

\begin{table}[ht]
  \centering
  \caption{Steps of distributed construction}
  \label{tab:distconst}
    \begin{tabular}{|r|l|r|r|}
\hline
No & Step & Time [sec] & Util [\%] \\
\hline \hline
1 & Distribute Relation $R$ & 57.0 & \\
2 & Select Centers for the Root Node & 4.3 & \\
3 & Distribution to Root Centers & 85.2 & \\
4 & Select Centers of Level 1 & 57.5 & \\
5 & Distribution to Level 1 Centers & 131.1 & 98.1\\
6 & Creating and Exporting N-trees at Level 2 & 262.0 & 98.7\\
7 & Collecting Exported Relations into DArrays & 57.9 & \\
8 & Computing Tree Relations for the Root Node & 6.9 & \\
9 & Computing Tree Relations for Level 1 Nodes & 70.7 & \\
10 & Merging Tree Relations on the Master & 505.7 & \\
11 & Importing the N-tree & 65.5 & \\
\hline
  & Total Time & 1303.8 & \\
\hline\hline
    \end{tabular}
\end{table}

We explain the steps for the concrete example of constructing an N-tree for the New York Trips dataset with 550841 tuples, using a computer with 32 workers. The N-tree has parameters $k=25, l = 50$. The precise \Secondo\ commands for these steps can be found at~\cite{distconstruction}.

In \textbf{Step 1}, the relation is randomly distributed from the master to the 32 workers, creating a distributed array with $4 * 32 = 128$ slots (partitions); each worker will handle 4 partitions. Since random distribution is balanced, each slot has 4303 or 4304 tuples (trips).

In \textbf{Step 2}, 25 centers for the root node are randomly selected. This is done by randomly selecting 1 element from each slot, transferring these to the master and selecting randomly 25 tuples from that set. The set of centers $C$ is copied to all workers.

In \textbf{Step 3}, each worker loads the relation $C$ into memory and builds an N-tree index over it, with a single node. It then processes each assigned slot executing for each tuple a \op{closestCenter} operation on the N-tree, extending the tuple by an attribute $N1$ with the index of that center and its distance $N1Dist$. 

In principle, we could now partition the resulting data set (for each slot) by the index $N1$ of its closest center. However, this would result in only 25 partitions, not enough to employ the workers. We therefore create a composite index with value $N1x = N1 * 20 + X$ where $X = slotindex \mod 20$ (slot indexes ranging from 0 to 127 in this example). Each tuple is also extended by attribute $N1x$. The result is a distributed array $Trips2501$, still with 128 slots, but the relations in the slots have the
$N1$ and $N1x$ values. From this we derive a data set partitioned by $N1x$ called $Trips2501b$ with 500 partitions.

The work in this step is well balanced, as all workers process the same number of partitions of the same size.

In \textbf{Step 4}, the goal is to find for each of the 25 level 1 nodes below the root a set of 25 centers.
For each slot of $Trips2501$, the relation is grouped by $N1$ and from each group one tuple is randomly selected as a center candidate. Hence for each $N1$ value (each level 1 node) we have 128 candidates. These are transferred to the master and there 25 centers are again selected randomly from the 128 candidates. The result is the set of centers for level 1 with 625 elements.

In the next step, we need to process the 500 partitions of $Trips2501b$ (20 partitions for each node of level 1) and assign each element to its closest center in its level 1 node. It is practical to have 500 sets of centers (slots) to be able to match the 500 partitions of $Trips2501b$. We therefore create 20 copies of each center tuple and extend each tuple by an $N1y$ attribute with value $N1y = N1 + Y$ with $Y = slotnumber \mod 20$, similar to Step 3. The result is distributed by $N1y$, resulting in a distributed array $Centers2501$ with 500 slots, each having the set of centers for one $N1$ value.

In \textbf{Step 5}, the 500 pairs of partitions from $Trips2501b$ and $Centers2501$ with the same index are processed. Note that partitions with the same index have the same $N1$ value, i.e., belong to the same level 1 node. For each pair of partitions, as in Step 3, a one-node N-tree index for the set of centers is built and for each trip the closest center is found by a \op{closestCenter} operation on that N-tree, extending the tuple by the index $N2$ of that center and $N2Dist$. The resulting distributed array $Trips2502a$ is partitioned by the combination of $N1$ and $N2$ values yielding partitions for level 2 nodes, resulting in $Trips2502$.

Note that the partition sizes in $Trips2501b$ may vary widely, resulting from partitioning by centers. We therefore use a load balancing technique available in \Secondo\ (see \cite{GBN21}). The \op{areduce} operator in \Secondo\ sorts partitions by size before processing them; it then assigns partitions by decreasing size to workers. When a worker finishes, it processes the next partition in this order. Processing many small partitions at the end allows one to have all workers finish at almost the same time.

In \textbf{Step 6}, the $k^2 = 625$ partitions of Trips2502 corresponding to level 2 subtrees are processed. For each partition, an N-tree is constructed and exported to relations using the \op{exportntree} operator. A numbering scheme for the start node numbers of exported N-trees is employed 
to fulfill the requirements mentioned in Section~\ref{sec:onelevel}. Again, the \op{areduce} operator is used to balance the work load for these partitions of widely varying size.

In \textbf{Step 7}, the $625 * 3$ relations \emph{Nodes}, \emph{Distances}, and \emph{Pivots} created in the previous step are collected in three distributed arrays to make them available for further processing. This step does not copy data; it just creates the summary information on the master making up the description of a distributed array.

In \textbf{Step 8}, the same three relations to represent the root node are constructed. As a preparatory step, for each of the 25 subtrees of the root, the radii are computed. This is done efficiently by determining for each $N1$ value the maximal $N1Dist$ value, where the $N1Dist$ values have already been computed at the time of determining closest centers, in Step 3, and are available in \emph{Trips2501}.

In \textbf{Step 9}, similarly the three relations for each of the 25 level 1 nodes are constructed. Again, the radii of the level 2 subtrees are computed beforehand from the $N2Dist$ values added to each element in Step 5 and available in \emph{Trips2502}. Node numbers for the level 1 nodes are assigned to lie between those of the subtrees, to respect the depth-first traversal order (the root has node number 0).

In \textbf{Step 10}, all relations of the three types are collected on the master and for each type their union is formed and sorted by node numbers. Further the relation \emph{TreeInfo} with one tuple is constructed.

In \textbf{Step 11}, the relation $R$ still present on the master is loaded into memory and an N-tree index over it is constructed from the four relations built in Step 10, using the \op{importntree} operator.

\subsubsection{Evaluation}
\label{sec:distconsteval}

We provide a brief experimental evaluation of this technique.
As a computing platform, we use a single computer with an AMD Ryzen 9 3950X 16-core processor (32 threads), 64 GB memory, and 4 disks. This is suitable to run up to 32 workers efficiently. We configure each worker to run with 1800 MB memory. The running times for an example run of the 11 steps with the dataset and configuration explained above are shown in Table~\ref{tab:distconst}. 

For the two steps 5 and 6 where highly unbalanced partitions are processed and \op{areduce} is used, Table~\ref{tab:distconst} shows the measure of utilization of workers. It is defined as follows. Let $W = \{w_1, ..., w_n\}$ be the observed running times of workers (in a step). Then $Util(W) = \frac{\sum_{i = 1, ..., n}w_i}{n \times \max_{i = 1, ..., n}w_i}$. In other words, this is the work workers have done compared to what they could have done if all had worked to the end (i.e. the running time of the worker finishing last).

\begin{figure}[thb]
\begin{center}
\begin{tabular}{ccc}
\includegraphics[scale=0.24]{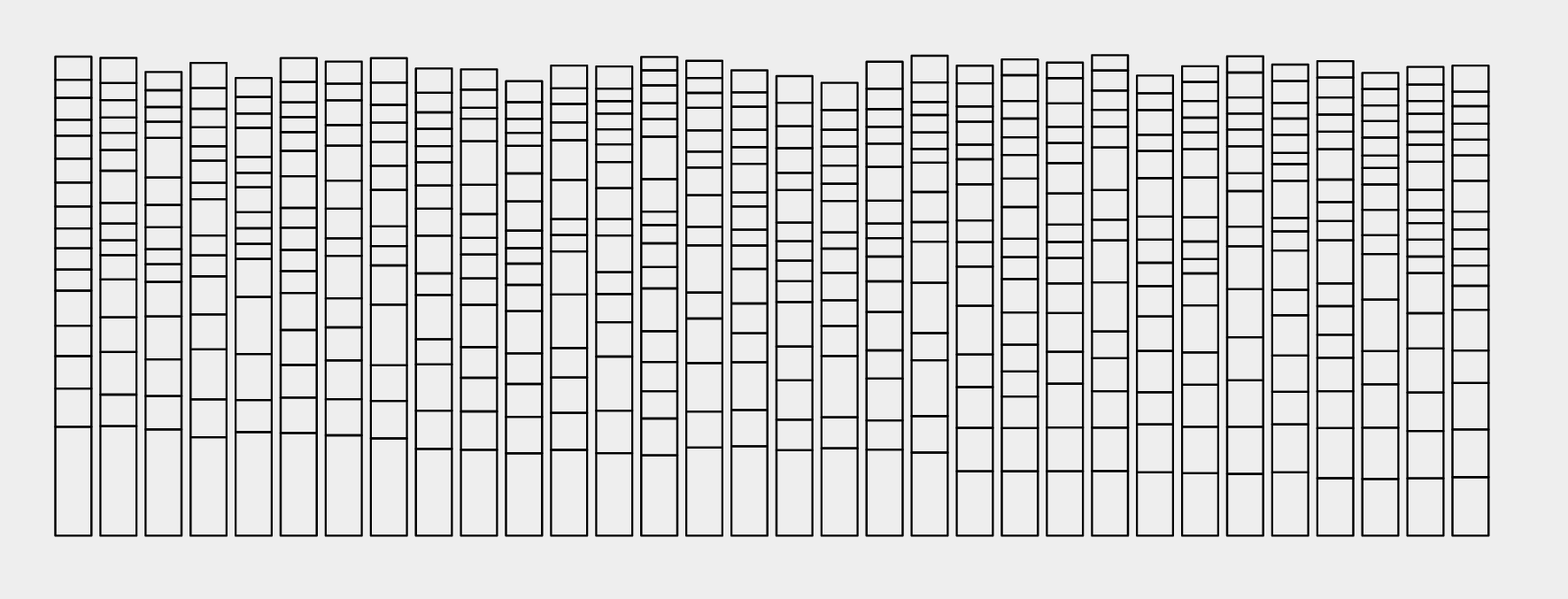}
& &
\includegraphics[scale=0.24]{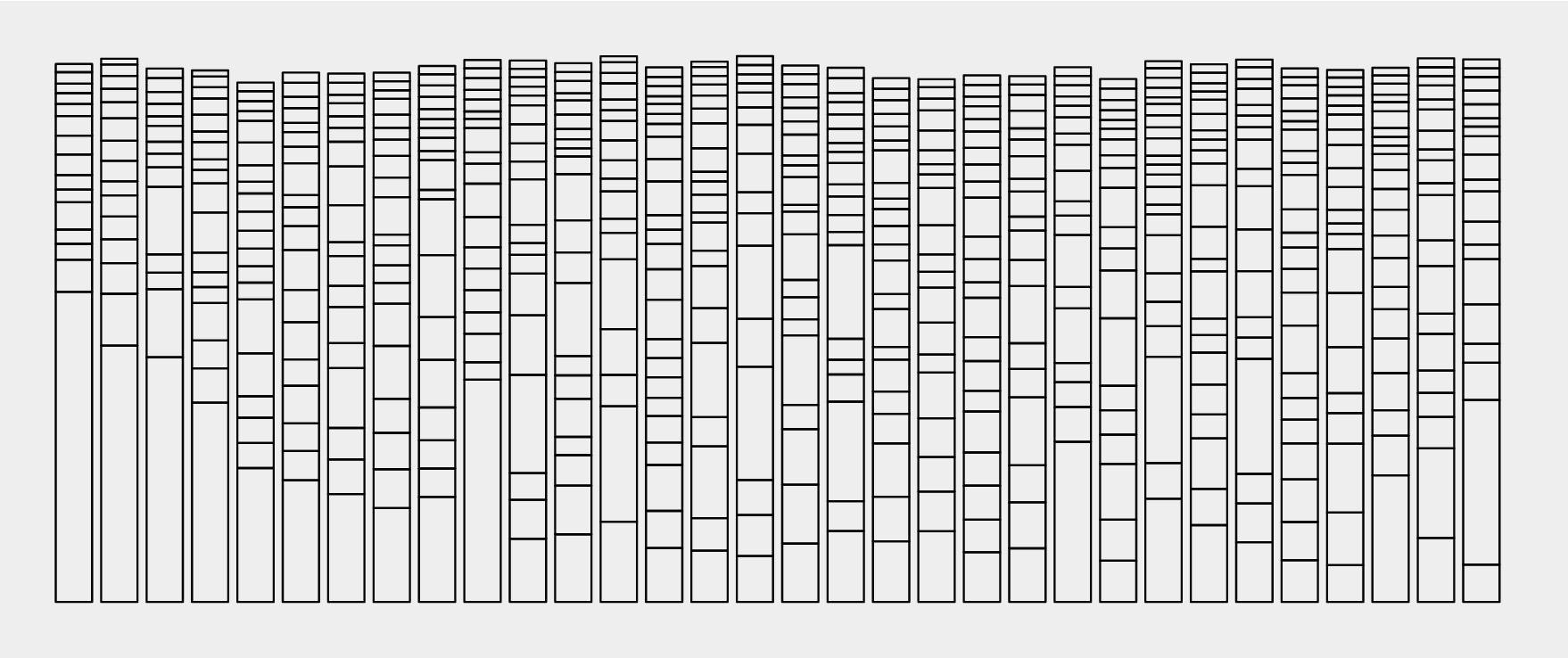} \\
(a) & & (b)
\end{tabular}
\end{center}
  \caption{ Load balancing in (a) Step 5 (b) Step 6. Horizontal: workers; vertical: tasks. }
  \label{fig:cost_trips}
\end{figure}

We can see that with values above 98\% utilization is excellent. The load balancing is also shown in Figure \ref{fig:cost_trips}.

In these figures, each vertical bar represents the running times of tasks performed by one worker, stacked vertically from bottom to top in the order of execution. One can observe that generally the running times decrease from left to right and bottom to top, reflecting the way \op{areduce} works. This is not precise as even for a given partition size the execution time can vary. But this is well captured by the adaptive behaviour of \op{areduce}.

We compare to the running time of constructing an index directly for the same dataset on the same machine. This takes 7109.8 seconds. With the time of 1303.8 seconds for the distributed/parallel construction we have a speedup of 5.4.

\subsection{Final Remarks}

The techniques of this section make the use of an N-tree main memory index entirely practical also for large data sets with expensive distance functions. On the one hand, construction time can be considerably reduced by parallel construction, even using a single 16-core computer, for example. On the other hand, it is not a problem to spend a long construction time once for a given dataset. After that, updates can be run on the main memory index. When the server with the index needs to be shut down, the index can be saved to relations in a few minutes, and it can be reconstructed in a similarly short time when the server is restarted. 

The running times for the New York Trips dataset are shown in Table~\ref{tab:distconst2}. When the index is restored from relations, \emph{cold} refers to the case that the relations used are not yet in the operating system's cache (e.g. the computer has just been started).

\begin{table}[ht]
  \centering
  \caption{Index Management for the New York Trips Dataset}
  \label{tab:distconst2}
    \begin{tabular}{|l|r|r|}
\hline
 & Time [sec] & Time [h:m:s] \\
\hline \hline
Sequential construction & 7109.8 & 1:58:29 \\
Parallel construction & 1303.8 & 0:21:53 \\
Saving to relations &  275.67 & 0:04:35 \\
Restoring from relations (warm) & 62.83 & 0:01:03 \\
Restoring from relations (cold) & 197.09 & 0:03:17 \\
\hline
    \end{tabular}
\end{table}

\section{Related Work}
\label{sec:related}

In this section, we first present existing work related to metric indexes. Then we discuss prior work on trajectory distance measures. 
Like the survey \cite{ChenIndexingmetricSpaceSurveyPaper}, we consider only work that provides exact results according to the given distance measure, i.e., a ``recall'' and ``precision'' of 100\%.

\subsection{Metric Indexes}
\label{sec:related_metric_index}

A significant body of previous work investigated the topic of indexing metric spaces. Chen et al.~\cite{ChenIndexingmetricSpaceSurveyPaper}  conducted a comprehensive survey of such indexing techniques. These approaches can be grouped into three categories: compact-partitioning based metric indexes (CMI), pivot-based metric indexes (PMI), and hybrid metric indexes (HMI) incorporating ideas from CMI and PMI.

\paragraph{Compact-partitioning based metric indexes} The indexes in the CMI category utilize the principle of partitioning the search space and they aim to attain this partitioning as compactly as possible. Such partitioning allows filtering out unqualified partitions during query execution. Three different kinds of partitioning techniques have been proposed, namely, ball partitioning, generalized hyperplane partitioning and hash partitioning. CMI indexes usually utilize one of these techniques, while a few use hybrid partitioning by leveraging both ball partitioning and hyperplane partitioning. The ball partitioning divides the search space into two subsets \textit{S1} and \textit{S2} using a  spherical cut~\cite{Zezula2006Metric}. Given an arbitrarily chosen point \textit{p} and radius \textit{r}, all the objects that are at a distance less than or equal to \textit{r} from \textit{p} belong to \textit{S1} and the rest of the objects are in \textit{S2}. The generalized hyperplane partitioning divides the search space into two subsets \textit{S1} and \textit{S2} using two arbitrarily chosen reference points \textit{p1} and \textit{p2}.  The objects are assigned to either \textit{S1} or \textit{S2} depending on their proximity to the reference points. The hash partitioning leverages a hash function for partitioning the search space. A popular hash function is the  $\rho$-split function that partitions the search space into three subsets \textit{S1},  \textit{S2} and \textit{S3}. This leaves out points near a particular threshold when determining membership in \textit{S1} and  \textit{S2}, whereas \textit{S3} includes the excluded points. This is why this technique is also known as excluded middle partitioning \cite{Yianilos99excluded}.

One of the earliest indexing approaches based on generalized hyperplane partitioning is the Bisector Tree (BST)~\cite{Kalantari1983ADS}. It is a binary tree that is built recursively using two reference  points, such that objects closer to the first reference  point belong to the first subtree and the objects nearer to the second reference  point are in the second subtree. The covering radii corresponding to each of the reference  points are maintained in the nodes. The Monotonous BST (MBST)~\cite{Noltemeier1992MonotonousBT} is a follow-up work of BST. It is based on the idea that every internal node (except for the root node) inherits one of the reference points from its parent node. The Voronoi Tree (VT)~\cite{Dehne1988Syntactic} is another extension of BST.  It attempts to pack the objects more compactly by decreasing the covering radii while moving downwards in the tree. The Generalized Hyperplane Tree (GHT) proposed by~\cite{Uhlmann1991Satisfying} is based on similar concepts as the BST. However, instead of using covering radii for pruning, it uses a criterion involving the hyperplane between the reference points to decide which subtrees to visit.

Among the ball partitioning based techniques, the M-tree~\cite{Ciaccia1997M-Tree} is perhaps the most well-known. It is a height-balanced index structure, and can support efficient external memory operations. In the M-tree, the internal nodes maintain pointers to the next level nodes, while the objects are kept at the leaf nodes. Each internal node entry also includes information regarding covering radius and parent distance. Several variants of M-tree have been proposed, such as the MM-tree~\cite{Pola07MM-Tree}, M$^+$-tree~\cite{Zhou2003MtreeA} and BM$^+$-tree~\cite{Zhou05BM+-Tree}. The List of Clusters (LC) index~\cite{Chavez2000effective} leverages a list of clusters in which each cluster is identified by a center and a radius. The main idea of LC is to store each object within a particular cluster whose distance to the center of that cluster is not larger than the radius. The LC index can be of two variants: fixed radius and fixed size. The Dynamic LC (DLC)~\cite{Navarro2016New} is an extension of LC, which is essentially a dynamic version of it.

The hash partitioning approaches include the MB$^+$-tree~\cite{MB+Tree00Ishikawa}. The MB$^+$-tree partitions the search space into two subsets using hash partitioning. Alternatively, it can also use generalized hyperplane partitioning recursively. The MB$^+$-tree utilizes two data structures: block tree and B$^+$-tree. The block tree is used to maintain partition information. For each object a key is generated by the MB$^+$-tree, which is indexed by the B$^+$-tree.

\paragraph{Pivot-based metric indexes (PMI)} The indexing approa\-ches in this category leverage pre-computed distance, since distance computation during query execution is an expensive operation. Sometimes, this distance pre-computation is performed by first selecting a set of pivots and then calculating the distance of every object to those pivots and storing those distances.

The Approximating and Eliminating Search Algorithm (AESA)~\cite{Ruiz1986Algorithm} is one of the earliest approaches in the PMI category. It maintains a table storing the distances between all pairs of objects. During query processing, the pre-computed distances can be used to filter out objects. Whereas AESA can improve search performance, a key drawback is that this requires scanning the pre-computed table to find a match. Another drawback is the space requirements to store the pairwise distances. To address these limitations with AESA, several approaches were subsequently proposed. Among them, the Reduced-Overhead AESA (ROAESA)~\cite{Vilar1995Reducing} sorts the pre-computed distances during the search and applies heuristics to potentially avoid unnecessary table scans. 

The Linear AESA (LAESA)~\cite{Figueroa2010Speeding} index utilizes a fixed number of pivot points and only stores pre-compu\-ted distances from objects to those pivot points.  However,  selecting suitable pivots becomes an issue. This can be addressed by other techniques, such as ~\cite{Mico1994LAESA}, which attempts to identify pivots that are as far away from each other as possible~\cite{Zezula2006Metric}. 

The idea of the Extreme Pivot Table (EPT)~\cite{Ruiz2013Extreme} approach  is to select ``extreme'' pivots and their associated objects so that the objects are likely to be eliminated with that pivot. In the EPT approach, a set of pivot groups are identified where each group contains $g$ pivots and the entire dataset is segmented into $g$ partitions. EPT uses a partition inclusion criterion for each object so that the absolute difference between the distance from the object to the pivot point and the expected distance between any object to the pivot point is greater than or equal to a threshold $\alpha$.
EPT maintains the data and pre-computed distance information in main memory. For larger datasets, CPT~\cite{Juraj11Clustered} proposes an I/O efficient approach. The objects are maintained on disk using an M-tree. The pre-computed distance table keeps pointers to the leaf node entries in the M-tree.

The Vantage Point Tree (VPT)~\cite{Uhlmann1991Satisfying} is based on partitioning a dataset into two subsets based on a vantage point or pivot. The pivot is chosen as the root node. All objects that are located within a distance from the pivot less than the median distance are kept in the left subtree, while the rest of the objects are maintained in the right subtree. The partitioning process is recursively applied to form a balanced binary tree. VPT is primarily targeted for continuous distance functions, and discrete distance functions can also be supported.  Instead of using the median distance, an alternative approach is to use the mean of distances from the pivot to all objects. This approach is known as \textit{middle point} in~\cite{Chavez01Searching}, which may perform better with high-dimensional data, however, it may result in an unbalanced tree. A dynamic variant of VPT, called DVPT~\cite{Fu2000DynamicVI}, was proposed that supports insertion and deletion operations. Another approach,  Multi Vantage Point Tree (MVPT)~\cite{Bozkaya1999Indexing},  extends the VPT approach. It utilizes multiple pivots (typically 2 or 3) to partition each node, rather than one as with VPT. On the other hand, with MVPT the children at the lower level leverage the same pivots, whereas with VPT there are different pivots at lower levels.

The Omni-family~\cite{Traina2007OmniFamily} of indexes utilizes the pivot mapping technique that is used to represent objects as vectors of their distances to pivots \cite{ChenIndexingmetricSpaceSurveyPaper}. Following this pivot mapping, the pre-computed distances with respect to the pivots are indexed using an existing exter\-nal-memory index. The OmniB$^+$-tree employs a B$^+$-tree, whereas the OmniR-tree utilizes an R-tree.

\paragraph{Hybrid metric indexes (HMI)} The approaches in the hybrid metric index category leverage both compact partitioning and pivots. The Geometric Near-Neighbor Access Tree (GNAT)~\cite{Brin1995Near} 
 utilizes $m$  pivots in each internal node. Based on the shortest distance of the objects to one of these pivots, the dataset is partitioned accordingly using generalized hyperplane partitioning (Voronoi partitioning). This process is applied recursively to build  an $m$-ary tree.  GNAT maintains  pre-computed distances from the objects to their corresponding pivots.

The Evolutionary Geometric Near-Neighbor Access Tree (EGNAT)~\cite{Navarro2011Fully} is an extension of GNAT and adap\-ted for external  memory. It can support insertion and deletion operations. The EGNAT index includes two types of nodes: buckets (leaves) and gnats (internal nodes), where gnats are similar to internal GNAT nodes. The index construction process is carried out by recursively selecting the closest pivot for a new object until reaching the leaf level. At the leaf level, objects are inserted into the buckets, without internal structure, which can result in reduction in storage compared to GNAT.   

The D-index~\cite{Dohnal04AnAccess} uses a combined hash partitioning and pivot mapping based approach. It utilizes several $\rho$-split functions, one at each level, to construct a multilevel structure. The Pivoting M-tree (PM-tree)~\cite{Skopal2004PMtreePM} is an extension of the M-tree that utilizes pivoting to reduce metric region volumes. The PM-tree first selects a set of pivots. For each inner node in the tree,  a routing entry is defined that includes an array of hyper-rings. Each hyper-ring  is the smallest interval encompassing  distances between the pivot and each of the objects stored in leaves of the subtree. Each leaf node in the PM-tree maintains an array of pivot distances.

\paragraph{Discussion}
Our approach, the N-tree, can be considered as a hybrid approach. It leverages both compact partitioning and pivoting techniques. The N-tree uses generalized hyperplane partitioning with many centers (Voronoi partitioning). Within each node, it maintains all pairwise distances between centers as well as distances to the two pivot elements. The N-tree uses various pruning criteria during query processing. 

The approach that appears to be closest to the N-tree is GNAT which also uses Voronoi partitioning. In the N-tree, an additional feature is precomputation of all distances between the entries of a node to enable pruning in range search and \emph{kNN} search. The use of precomputed distances also occurs in AESA. However, there it is only used at a global level (with problems of scale), not among the entries of a node in a multiway tree, as in the N-tree.

\subsection{Trajectory Distance Measures}
\label{sec:related_traj_dist}

A  survey on trajectory distance measures was provided by Su et al.~\cite{SuLZZZ20}. They categorized trajectory distance measures along two dimensions based on (i)  whether the measure considers spatial distance only or considers both spatial and temporal attributes, and (ii) whether the measure is defined  in a discrete or continuous manner. According to this, our proposed distance measure \emph{distanceAvg} belongs to the category of continuous spatiotemporal distance measures. Other measures in this category include Fréchet distance, STED, and STLIP. These approaches tend to compare the shape of the trajectories based on trajectory segments, rather than individual points.

Recently, Hu et al.~\cite{Hu2024Spatio} presented a comprehensive survey of trajectory similarity measures. In this, the non-learning measures are classified into (i) free-space based and (ii) road network based. The free-space based measures are designed for trajectories of objects moving freely. In contrast, the road network based measures take into account road network constraints. Each of these categories can be further sub-divided into standalone and distributed. Our proposed   \emph{DistanceAvg} falls under the  free-space based standalone category. Existing  distance measures in this category include ED, DFD~\cite{Alt1995Computing}, DTW~\cite{Byoung98Efficient}, EDR~\cite{Chen2005Robust}, EDwP~\cite{Ranu2015Indexing}, ERP~\cite{Chen2004Onthe},  Hausdorff~\cite{Rote1991Computing}, LCSS~\cite{Vlachos2002Discovering},  LIP~\cite{Pelekis2007Similarity}, and OWD~\cite{Frentzos2007Index}. However, among them only  ED, ERP and Hausdorff are metric distance measures~\cite{Hu2024Spatio}. Next, we discuss these three measures.  

Euclidean Distance (ED) is a popular metric distance measure that finds wide applications. It can be used to compute the distance between two sequences of the same length~\cite{Keogh2000Scaling}. This is not realistic for trajectory distance because it cannot be assumed that two trajectories are of the same length.  Edit Distance with Real Penalty (ERP) is a metric distance measure with time complexity of \textit{O(mn)}. Unlike ED, that requires every sample point in a trajectory $R$ to have a matching partner in the other trajectory $S$, ERP uses edit distance to match pairs of sample points. For a sample point in $R$ that has a matching partner in $S$, their $L_1$-norm is used to calculate the distance. Whereas, for a sample point in $R$ that matches a gap in $S$, $L_1$-norm from the point to a constant is calculated. Being a metric, ERP can leverage triangular inequality to  perform efficient pruning. Hausdorff distance is the largest of all the distances from each point in one sequence to its nearest point in the other sequence. It is a metric distance measure having time complexity of \textit{O(mn)}. Formally, the Hausdorff distance from a trajectory $R$ to a trajectory $S$ is function $h$, such that
\[  h(R, S) = \max_{r \in R} \mbox{ } \{ \min_{s \in S} \mbox{ } \{  d (r, s) \} \}\]
Here, $d (r, s)$ is any metric distance, such as ED.

%\paragraph{} 

\section{Conclusions and Future Work}
\label{sec:conclusions}

Motivated by similarity search on trajectories of mobile objects, we have presented the N-tree, a generic metric index supporting similarity search in many fields.

In our experimental evaluation we have compared the N-tree to the strongest competitors we could identify, MVPT and GNAT (ranked first and second in running time for \emph{kNN} queries according to \cite{ChenIndexingmetricSpaceSurveyPaper}). We have compared the three index structures for a diverse range of scenarios of different data types and distance functions. In normal situations, the N-tree clearly outperforms the competitors for \emph{kNN} queries and range queries with larger radius; only for range queries with low radius, MVPT performs about equally well. \emph{kNN} queries may be more relevant than range queries as it can be difficult to determine an appropriate query radius.

For the field of trajectory similarity search, our second contribution is a new similarity measure, the distance function \emph{DistanceAvg}. In contrast to the well-known metric distance functions for trajectories, having time complexity $O(mn)$, it has only linear complexity $O(m+n)$. Experiments show that it performs better than Hausdorff distance; this effect increases with the length of trajectories. We have also proposed a  cylinder approximation technique for trajectories that enables a filter-and-refine process for range query evaluation.

Combining the two contributions N-tree and \emph{DistanceAvg}, we observe an improvement of performance for \emph{kNN} queries on trajectories of an order of magnitude.

The N-tree has a property that we did not observe in other index structures: with increasing range query radius, at some point the number of distance evaluations begins to \emph{decrease} so that for very large query radii very few distances evaluations are needed (the U-turn effect).

The N-tree trades higher construction time for faster execution of queries. To remedy this, we have proposed a distributed/parallel technique for construction of a main-memory index. We are not aware of such work in the literature.

In this paper we have examined the N-tree as a main memory index structure. Due to the fact that the performance is equally good for large node sizes, we expect it to be also well suited as an external memory index. An experimental evaluation of this case is a subject for future work.

In designing the index structure, we have observed a strong duality between the use of a Voronoi partitioning in structuring an index and for parallel/distributed computation. Based on the range distribution property, a similarity join can easily be  implemented in a distributed manner. One possibility is to implement the join in each partition as a nested loop join using an index such as the N-tree. Partitioning the data set can be done in parallel and efficiently using the \emph{closestCenter} algorithm of the N-tree. This is another area for future work.

\paragraph{Acknowledgement} Thanks to Catherine Higgins for pre\-pa\-ring the New York trips data set and her support in the early stages of this project.

%\input{similarityJoin}
%\input{appendix}

% some previous version

%\input{abstract}
%\input{intro}
%\input{related}
%\input{preliminaries}
%\input{ntree}
%\input{range}
%\input{knn}
%\input{optimizations}
%\input{evaluation}
%\input{similarityJoin}
%\input{conclusions}

\bibliographystyle{plain}
\bibliography{bibliography}

\end{document}